\documentclass[11pt,tightenlines,eqsecnum,floats,aps,amssymb,nofootinbib,prd,shownopacs,floatfix]{revtex4-1}
\usepackage[english]{babel}
\usepackage[pdftex,dvipsnames]{xcolor}
\usepackage[utf8]{inputenc}
\usepackage{amsmath,amsfonts,extarrows,nicefrac}
\allowdisplaybreaks[1] 
\usepackage{marvosym} 
\usepackage[Symbolsmallscale]{upgreek}
\usepackage{graphicx}
\usepackage{subcaption}

\usepackage{marvosym} 
\usepackage[hidelinks]{hyperref} 
\usepackage{bbm}
\usepackage{pdfpages}
\usepackage{makecell}
\usepackage{mathtools}
\usepackage{mathrsfs}
\usepackage{ stmaryrd } 
\usepackage[scr=dutchcal]{mathalfa}
\usepackage{textgreek}
\usepackage{titlesec} 
\usepackage{tikz}
\usetikzlibrary{arrows,shapes}
\usetikzlibrary{positioning}

\usepackage[
    font=small,
    labelfont=bf,
    justification=RaggedRight,
    format=hang,
    width=0.8\textwidth
]{caption} 

\usepackage[T1]{fontenc}
\newcommand*{\dittostraight}{-----\textquotedbl-----} 

\usepackage{tikz}
\usetikzlibrary{fit}
\usetikzlibrary{patterns}
\usetikzlibrary{decorations.pathreplacing}
\tikzset{%
  highlight/.style={rectangle,rounded corners, fill=white, pattern color=gray, pattern=crosshatch dots, fill opacity=0.4,draw,thin,inner sep=0pt}
}

\newcommand{\nocontentsline}[3]{}
\newcommand{\tocless}[2]{\bgroup\let\addcontentsline=\nocontentsline#1{#2}\egroup}
\renewcommand{\thesection}{\Roman{section}}
\renewcommand{\thesubsection}{\Alph{subsection}}
\renewcommand{\thesubsubsection}{\arabic{subsubsection}}

\makeatletter
\renewcommand{\p@subsection}{\thesection.} 
\renewcommand{\p@subsubsection}{\thesection.\thesubsection.} 
\AtBeginDocument{\let\LS@rot\@undefined} 
\makeatother

\let\originalleft\left
\let\originalright\right
\renewcommand{\left}{\mathopen{}\mathclose\bgroup\originalleft}
\renewcommand{\right}{\aftergroup\egroup\originalright}
\newcommand{\lr}[1]{\left(#1\right)} 
\newcommand{\lrabs}[1]{\left|#1\right|} 
\newcommand{\I}{\mathcal{I}}
\newcommand{\ofI}{\lr{\I}} 
\newcommand{\ofIn}{\lr{\mathcal{I}_n}} 
\newcommand{\norm}[1]{\left\lvert\!\left\lvert #1 \right\rvert\!\right\rvert} 
\newcommand{\pb}[2]{\left\{#1,#2\right\}} 
\newcommand{\comm}[2]{\left[#1,#2\right]} 
\newcommand{\e}[1]{\ensuremath{\mathrm{e}^{#1}}} 
\newcommand{\ex}[1]{\exp\left(#1\right)} 
\renewcommand{\d}{\mathrm{d}}
\newcommand{\pdif}[2]{\ensuremath{\frac{\partial#1}{\partial#2}}}

\DeclareMathOperator{\tr}{tr} 
\newcommand{\inv}{{}^{-1}} 
\newcommand{\pt}{\partial_\theta} 
\newcommand{\Obs}[1]{\mathcal{O}_{#1}} 
\newcommand{\vm}{v^-} 
\newcommand{\vp}{v^+} 
\newcommand{\vxi}{v_\xi} 
\newcommand{\vpp}{v^{++}} 
\newcommand{\vmm}{v^{--}}  

\newcommand{\ev}{e_v} 
\newcommand{\evm}{e_{\vm}} 
\newcommand{\evmm}{e_{\vmm}} 
\newcommand{\evp}{e_{\vp}} 
\newcommand{\evpp}{e_{\vpp}} 
\newcommand{\graph}{\Gamma} 
\newcommand{\aqggraph}{\alpha} 
\newcommand{\halfalpha}{\frac{r}{2}}
\newcommand{\halfbeta}{\frac{r_2}{2}}
\newcommand{\PT}{{\cal P}(\epsilon)} 
\newcommand{\Sd}{{\cal S}^1} 

\renewcommand{\i}{\ensuremath{\mathrm{i}}} 
\newcommand{\immirzi}{\beta_{\mathrm{BI}}} 
\DeclareMathOperator{\V}{{\hat V}} 
\newcommand{\lp}{l_{\mathrm{P}}} 
\renewcommand{\pi}{\text{\textpi}} 
\newcommand{\Ham}{C} 
\newcommand{\euclHam}{\Ham_{\mathrm{eucl}}} 
\newcommand{\lorHam}{\Ham_{\mathrm{lor}}} 
\newcommand{\HH}{\mathrm{H}} 
\newcommand{\euclH}{\HH_{\mathrm{eucl}}} 
\newcommand{\lorH}{\HH_{\mathrm{lor}}} 
\DeclareMathOperator{\hatHH}{\hat{\operatorname{\HH}}} 
\DeclareMathOperator{\euclHhat}{\hatHH_{\mathrm{eucl}}} 
\DeclareMathOperator{\euclHhatI}{\hatHH_{\mathrm{eucl},\textit{v}}} 
\newcommand{\lorHhat}{\hatHH_{\mathrm{lor}}} 
\newcommand{\lorHhatpart}[1]{\hatHH_{\mathrm{lor}}^{(#1)}} 
\newcommand{\lorHhatI}{\hatHH_{\mathrm{lor},v}} 
\newcommand{\lorHhatIpart}[1]{\hatHH_{\mathrm{lor},v}^{(#1)}} 
\newcommand{\Diff}{C_\theta} 
\newcommand{\Diffeo}{C} 
\newcommand{\Gau}{G_3} 
\newcommand{\Gauss}{G} 
\newcommand{\Ett}{\ensuremath{E^{\theta}_3}} 
\newcommand{\Ex}{E^x} 
\newcommand{\Ey}{E^y} 
\newcommand{\Fx}[1]{\hat{\mathcal{F}}_{x,#1}} 
\newcommand{\Fy}[1]{\hat{\mathcal{F}}_{y,#1}} 
\newcommand{\Feta}[1]{\mathcal{F}_{\eta,#1}} 
\newcommand{\hatFeta}[1]{\hat{\mathcal{F}}_{\eta,#1}} 
\newcommand{\Att}{\ensuremath{A_{\theta}^3}} 
\newcommand{\bbar}{{\bar \beta}} 
\newcommand{\abar}{{\bar \alpha}} 
\newcommand{\Pbar}{{\bar P}} 
\newcommand{\A}{\mathcal{A}} 
\newcommand{\E}{\mathcal{E}} 
\newcommand{\Pxi}{P^\xi} 
\newcommand{\Peta}{P^\eta} 
\newcommand{\Hphys}{{\mathrm H}_{\mathrm{phys}}} 
\newcommand{\hatHphys}{\hatHH_{\mathrm{phys}}} 
\newcommand{\dustmanifold}{\mathcal{S}} 
\newcommand{\dustcoords}{\sigma} 
\renewcommand{\I}{\mathcal{I}} 
\newcommand{\holOp}{\hat{h}} 
\newcommand{\OOp}[2]{\hat{\operatorname{O}}^{#1}_{#2}} 
\newcommand{\Z}{Z} 
\newcommand{\Za}{Z_{r}} 
\newcommand{\ZaHatI}[1]{\hat{\operatorname{\Z}}_{r,#1}} 
\newcommand{\ZbHatI}[1]{\hat{\operatorname{\Z}}_{r_2,#1}} 
\newcommand{\keucl}{\kappa_{\text{eucl}}} 
\newcommand{\klorOne}{\kappa_{\text{lor},1}} 
\newcommand{\klorTwo}{\kappa_{\text{lor},2}} 
\newcommand{\Coeff}[3]{C_{\ifx&#1&%
    k   
    \else
    #1   
    \fi,
    \ifx&#2&%
    \mu   
    \else
    #2   
    \fi,
    \ifx&#3&%
    \nu   
    \else
    #3   
    \fi}
    } 
\newcommand{\kCoeff}[1]{C_{\ifx&#1&%
    k   
    \else
    #1   
    \fi}
    } 
\newcommand{\muCoeff}[1]{C_{\ifx&#1&%
    \mu   
    \else
    #1   
    \fi}
    } 
\newcommand{\nuCoeff}[1]{C_{\ifx&#1&%
    \nu   
    \else
    #1   
    \fi}
    } 
\newcommand{\state}[3]{\big\vert{\ifx&#1&%
    k   
    \else
    #1   
    \fi},
    {\ifx&#2&%
    \mu   
    \else
    #2   
    \fi},
    {\ifx&#3&%
    \nu   
    \else
    #3   
    \fi}
    \big\rangle}  
\newcommand{\stackpp}{\genfrac{}{}{0pt}{3}{+}{+}} 
\newcommand{\stackmm}{\genfrac{}{}{0pt}{3}{-}{-}} 
\newcommand{\stackpm}{\genfrac{}{}{0pt}{3}{+}{-}} 
\newcommand{\stackmp}{\genfrac{}{}{0pt}{3}{-}{+}} 
\newcommand{\singleCoeff}[1]{c_{#1}} 

\begin{document}
\titleformat{\section}[block]{\large\bfseries\filcenter}{\thesection.}{1em}{\MakeUppercase} 
\titleformat{\subsection}[block]{\bfseries\filcenter}{\thesection.\thesubsection}{1em}{} 
\titleformat{\subsubsection}[block]{\filcenter}{\thesection.\thesubsection.\thesubsubsection}{1em}{\textit} 

\title{{A reduced phase space quantisation of a model in Algebraic Quantum Gravity with polarised \texorpdfstring{$\mathbbm{T}^3$}{T3} Gowdy symmetry}}

\author{Kristina Giesel}
\thanks{kristina.giesel@gravity.fau.de}
\author{Andreas Leitherer}
\thanks{andreas.leitherer@gmail.com}
\author{David Winnekens}
\thanks{david.winnekens@fau.de}
\affiliation{Institute for Quantum Gravity, Friedrich-Alexander-Universit\"at Erlangen-N\"urnberg, \mbox{Staudtstr.~7}, 91058 Erlangen, Germany}

\begin{abstract}
We consider a reduced phase space quantisation of a model with $\mathbbm{T}^3$ Gowdy symmetry in which gravity has been coupled to Gaussian dust. We complete the quantisation programme in reduced loop quantum gravity (LQG) as well as algebraic quantum gravity (AQG) and derive a Schr\"odinger-like equation with a physical Hamiltonian operator encoding the dynamics. Due to the classical symmetries of the physical Hamiltonian, the operators are quantised in a graph-preserving way in both cases --- a difference to former models available in the literature. As a first step towards applications of the model in AQG, we consider an ansatz that we use to first construct zero volume states as specific solutions of the Schr\"odiger-like equation. We then also find states with a vanishing action of the Euclidean part of the physical Hamiltonian and investigate the degeneracies these states experience via the action of the Lorentzian part of the physical Hamiltonian. The results presented here can be taken as a starting point for deriving effective models as well as analysing the dynamics numerically in future work.
\end{abstract}
\maketitle

\newpage
\tableofcontents 
\newpage
\section{Introduction}\label{sec:Intro}

Loop quantum gravity provides a framework in which an analogue of the classical Einstein's equations can be formulated in the quantum theory. In the canonical approach one either considers solving the constraints in the quantum theory in the context of a Dirac quantisation~\cite{Dirac} or one solves the constraints already at the classical level by means of constructing suitable Dirac observables and subsequently quantises the physical phase space only. In full loop quantum gravity both approaches yield quantum Einstein's equations that are very complex and whose general solutions are not known \cite{thiemann_2007,rovelli_2004,rovelli_vidotto_2014,First30YearsLQG}. This is not too surprising because already at the classical level  the Einstein's equations without further assumptions are highly complex and constructing exact solutions is a very non-trivial task. However, exact solutions can be constructed in simpler setups where additional symmetry assumptions are implemented such as for instance in the context of cosmology or black holes. If we consider symmetry reduction in the context of a quantum gravity theory one can either symmetry reduce already at the classical level and then quantise or one can quantise full general relativity and afterwards access the symmetry reduced sector in the quantum theory. While the latter strategy is presumably the one that is able to capture more of the quantum nature of the symmetry reduced models \cite{Engle:2007qh,Brunnemann:2007du,Dapor:2017rwv,Kaminski:2020wbg}, it is also technically more involved than first symmetry reducing at the classical level and quantising only afterwards. 

For this reason we follow the first approach in this article and consider a symmetry reduction of classical models that experience a Gowdy symmetry  \cite{Gowdy:1971jh}. We furthermore specialise to the polarised case where the two commuting Killing vectors are orthogonal. Compared to other symmetry reduced homogeneous models in cosmology, such models similar to spherically symmetric models have the property that they are still a field theory with a non-trivial spatial diffeomorphism constraint and thus are closer to the situation we face in full general relativity. Hence, understanding these models allows to investigate properties of these models that might be absent in the homogeneous cosmological models in general. As we still have to deal with a constrained theory after symmetry reduction, we have again the option to either apply a Dirac or reduced quantisation of the symmetry reduced model where we follow the latter in this work. The quantisation of Gowdy models has been extensively discussed in the existing literature \cite{Misner:1973zz,BERGER1982394,Berger:1975kn,Berger:1984iya,Husain:1987am} starting after Gowdy's seminal paper \cite{Gowdy:1971jh}. Further work in terms of (complex) Ashtekar variables can be found in \cite{Husain:1989qq,Husain:1994dc,Husain:1995fd,MenaMarugan:1997us}. The quantisation programme could be completed in the context of a gravitational wave quantisation in geometrodynamics in  \cite{Pierri:2000ri}, see also \cite{Torre:2002xt} for a further extension of this model. A modified quantisation of the model in \cite{Pierri:2000ri} was later considered  in order to ensure that the dynamics is unitary \cite{Corichi:2002vy,Cortez:2005th,Corichi:2005jb,Corichi:2006xi,Corichi:2007ht}. All these models have in common that even if some of them start with Ashtekar variables the final quantum model does not involve a quantisation inspired from loop quantum gravity but considers techniques from geometrodynamics instead after gauge fixing the models. As a consequence, these models fail to resolve the singularity present in the classical Gowdy model. In \cite{Banerjee:classical,Banerjee:quant}, a loop quantisation of the polarised $\mathbbm{T}^3$ Gowdy has been introduced in the framework of a Dirac quantisation. However, due to the complicated form of the constraint operators, the quantisation programme could not be completed for that model. Progress in this directions was obtained using a hybrid quantisation procedure where the loop quantisation is applied to the homogeneous sector and a Fock quantisation to the inhomogeneous one \cite{Martin-Benito:2008eza,MenaMarugan:2009dp,Martin-Benito:2010dge,Garay:2010sk,Martin-Benito:2010vep}. A uniqueness result for the chosen Fock quantisation exists \cite{Corichi:2006zv} if one demands unitary implementation of the dynamics as well as invariance under the group of constant translations on the circle. This approach has turned out to be also useful in the context of cosmological perturbations, see for instance \cite{Fernandez-Mendez:2012poe,Gomar:2014faa,CastelloGomar:2017kbo} and \cite{ElizagaNavascues:2020uyf} for a recent review on the hybrid quantisation approach.

Because we will apply a reduced phase space quantisation for which we choose Gaussian dust as the reference matter \cite{Kuchar:1990vy,Giesel:2012rb}, we cannot consider the usual Gowdy solution that is a vacuum solution of Einstein's equations. Spacetimes with Gowdy symmetry coupled to matter have been considered in the literature, see for instance \cite{BarberoG:2007qra} for a coupling to a massless scalar field. For a corresponding quantum model see \cite{Martin-Benito:2010vep} and \cite{Andreasson:1998xr} for work on Einstein--Vlasov spacetimes with Gowdy symmetry extending former results for the vacuum case  \cite{Moncrief:1980pq,Berger:1997jd}. An introduction to the Einstein--Vlasov system can be found in \cite{Rendall:1996gx} and references therein. As discussed in \cite{Rendall:2010en}, dust can be understood as a distributional solution of the Vlasov equation and in this sense can be embedded in these systems. However, the specific characteristic properties of the matter component matter as for instance the results in \cite{Isenberg:1997re} show where the properties of the spacetimes are different if we couple generic Vlasov matter or dust respectively in the context of finding a global foliation of the spacetime. For the purpose of this work we consider general relativity coupled to Gaussian dust and then impose a Gowdy symmetry on the total system including the geometric as well as the matter degrees of freedom.

This setup allows us to construct Dirac observables associated with the geometric degrees of freedom in the framework of the relational formalism along the lines of \cite{Rovelli:1990ph,Rovelli:1990pi,Vytheeswaran:1994np,Dittrich:2004cb,Dittrich:2005kc,Thiemann:2004wk,Pons:2009cz,Pons:2010ad} that play the role of the elementary phase space variables in the reduced phase space. Their dynamics is generated by a so-called physical Hamiltonian whose Hamiltonian density in the Gaussian dust model is just given by the geometric contribution to the Hamiltonian constraint that is itself a Dirac observables and non-vanishing in the physical sector of the model \cite{Kuchar:1990vy,Giesel:2012rb}. Further reference matter models as well as their applications in the classical theory can for instance be found in \cite{Brown:1994py,Giesel:2007wi,Giesel:2007wk,Giesel:2009jp,Ali:2015ftw} and applications in the quantum theory are for example discussed in \cite{Ashtekar:2006rx,AQG4,Domagala:2010bm,Husain:2011tk,Giesel:2012rb,Alesci:2015wla,Giesel:2016gxq,Giesel:2020raf,Han:2019vpw}. Dirac observables for vacuum Gowdy spacetimes have for instance also been discussed in \cite{Torre:2005cn}.  In the context of the relational framework, this approach can be understood as choosing so-called geometrical clocks (or reference fields) constructed from purely geometric degrees of freedom. In contrast, we choose matter clocks in this work instead. As a consequence, we start with additional degrees of freedom compared to \cite{Torre:2005cn}, where in the end one independent Dirac observables exists, while here we end up with three independent ones. Furthermore, the construction of \cite{Torre:2005cn} is based on ADM variables, whereas here we will work with Ashtekar--Barbero variables in order to be able to apply a loop quantisation to the model later on. 

Loop quantisations that do not apply a hybrid approach of vacuum polarised $\mathbbm{T}^3$ Gowdy spacetimes in the context of a Dirac quantisation have been considered in \cite{Banerjee:quant,deBlas:2017goa}, where the latter model assumes a further rotational symmetry that simplifies the setup compared to \cite{Banerjee:quant}. The difference to the work here is that we consider a reduced phase space quantisation in the context of LQG as well as the Algebraic Quantum Gravity (AQG) framework for the Gaussian dust model. In both cases the physical Hamiltonian needs to be quantised in a graph-preserving manner in order to respect the classical symmetries of the physical Hamiltonian. This yields a different regularisation of the operator compared to the one presented in \cite{Banerjee:quant}, with in general different properties accordingly. Possible graph-preserving quantisations have been discussed in \cite{Bojowald:2005cb} in the context of spherically symmetric models and have also been mentioned in the final discussion of \cite{Banerjee:quant} as possible alternative regularisations. However, since in both works one uses Dirac quantisation with the corresponding constraint algebra in these models, a graph-modifying quantisation is motivated for the same reason we have in full LQG. 

Because we quantise the physical Hamiltonian in the AQG framework, a detailed discussion on how Gowdy states can be represented in the AQG framework is needed, allowing to implement the action of the physical Hamiltonian operator on this class of states properly. The dynamics of the physical states is encoded in a Schr\"odinger-like equation and finding its generic solution is far beyond the scope of this article. Nevertheless, the quantisation programme can be completed in this model here in the sense that the quantum dynamics is formulated at the level of the physical Hilbert space. The purpose of this work is to present how spacetimes with a Gowdy symmetry can be formulated in the AQG framework. The results presented here can be taken as the starting point for deriving effective models directly from the quantum theory because the graph-preserving regularisation chosen here is advantageous if semiclassical computations are to be performed as already existing semiclassical techniques can be directly used and need not be adapted to graph-modifying operators --- which is still an open and difficult question in full generality. As first steps towards applying the model, we compute the explicit form of the Schr\"odinger-like equation in the AQG framework and discuss a possible ansatz for the solution that can be considered for graph-preserving operators but will not work for graph-modifying ones as used in \cite{Banerjee:quant}. We further discuss how such an ansatz can be used to obtain zero volume eigenstates. 

The paper is structured as follows:
After the introduction in Section \ref{sec:Intro}, we briefly review the Gaussian dust model and dynamics of the corresponding Dirac observables in Subsection \ref{sec:GaussDust}, where we are very brief and refer for the main, already existing results in the case of general relativity to the work in \cite{Kuchar:1990vy,Giesel:2012rb}. We then rather focus on the symmetry reduction to the Gowdy symmetry in Subsection \ref{subsec:SymmRed}. Afterwards, in Section \ref{sec:QuantRedLQG}, we discuss the quantisation of the reduced model with polarised \texorpdfstring{$\mathbbm{T}^3$}{T3} Gowdy symmetry within reduced LQG. For this, we introduce in Subsection \ref{sec:PhysHSLQG} the physical Hilbert space of the model and its basic operators. In Section \ref{sec:DynRedLQG}, we present the quantisation of the corresponding physical Hamiltonian and discuss the difference in the regularisation and final properties of the operataor compared to the work in \cite{Banerjee:quant}. A quantisation of the classical model with polarised \texorpdfstring{$\mathbbm{T}^3$}{T3} Gowdy symmetry in the framework of AQG is discussed in Section \ref{Sec:AQGQuant}, where we introduce the physical Hilbert space in the AQG case in Subsection \ref{sec:PhysHSAQG} and the quantisation of the dynamics in Subsection \ref{sec:DynAQG}. We also emphasise the differences and similarities in the reduced LQG and AQG Gowdy model. As first steps towards applying the model, we consider the Schr\"odinger-like equation encoding the quantum dynamics of the model in Section \ref{sec:FirstStepsAppl} and compute the action of the physical Hamiltonian on the basic states of the AQG Gowdy model. Furthermore, we consider a solution ansatz for the Schr\"odinger-like equation that takes advantage of the fact that the physical Hamiltonian involved in the Schr\"odinger-like equation is quantised in a graph-preserving way and discuss how this can be used to construct zero volume states. Finally we summarise our result and conclude in Section \ref{sec:Conclusion}.


\section{Classical Setup: formulation of the model with polarised \texorpdfstring{$\mathbbm{T}^3$}{T3} Gowdy symmetry}
\label{sec:ClassSetup}
In this section, we briefly review the Gaussian dust model as well as its symmetry reduction to a model having a \texorpdfstring{$\mathbbm{T}^3$}{T3} Gowdy symmetry.
\subsection{Brief review of the classical reduced phase space using Gaussian dust}
\label{sec:GaussDust}

In this work, we aim at quantising the reduced phase space of general relativity formulated in terms of Ashtekar variables and symmetry reduced to the polarised Gowdy model. For this purpose, we choose as a first step some reference matter that we dynamically couple to gravity and that allows us to construct the corresponding elementary Dirac observables in the reduced phase space. For the reference matter we choose the Gaussian dust model \cite{Kuchar:1990vy} that was for instance considered in \cite{Giesel:2012rb} in the context of loop quantum gravity. Within the Gaussian dust model, one couples eight additional scalar fields to GR, leading to a system that involves second class constraints. As shown in \cite{Kuchar:1990vy,Giesel:2012rb}, after a reduction with respect to the second class constraints one obtains a first class system and next to gravity four additional dust fields --- denoted by $T$ and $S^j$ with $j=1,2,3$ --- that can be used as reference fields for the Hamiltonian and spatial diffeomorphism constraints respectively. In this model, these constraints take the form
\begin{eqnarray*}
C^{\rm tot} & =& C +\frac{P-\frac{E^a_jE^b_k\delta^{jk}}{\det(E)}T_{,a}C_b }{\sqrt{1+\frac{E^a_jE^b_k\delta^{jk}}{\det(E)}T_{,a}T_{,b}}},\quad 
C^{\rm tot}_a =C_a +PT_{,a}+P_jS^j_{,a} .
\end{eqnarray*}
Here, $C$ and $C_a$ denote the gravitational contribution to the total Hamiltonian and spatial diffeomorphism constraints in terms of the Ashtekar variables $A^j_a, E^a_j$, while $P, P_j$ are the momenta conjugate to $T,S^j$. Following \cite{Giesel:2012rb}, one solves $C^{\rm tot}$ for $P$ and $C^{\rm tot}_a$ for $P_j$ and then writes down an equivalent set of constraints that now is Abelian. The latter allows to directly apply the known observable map \cite{Vytheeswaran:1994np,Dittrich:2004cb,Dittrich:2005kc} in the framework of the relational formalism \cite{Rovelli:1990ph,Rovelli:1990pi} and construct the corresponding Dirac observables for the gravitational degrees of freedom denoted as $\Obs{A^j_a}$ and $\Obs{E^a_j}$. The algebra of the observables is given by
\begin{align}
    \{\Obs{A^j_a}(\tau,\dustcoords), \Obs{E_k^b}(\tau,\dustcoords^\prime)\}=\frac{\kappa}{2}\delta^b_a\delta^j_k \delta^{(3)}(\dustcoords - \dustcoords^\prime),
\end{align}
where we use $\kappa\coloneqq 16\pi G_{\mathrm{Newton}}$.

Using the properties of the observable map as shown in \cite{Thiemann:2004wk}, for a phase space function $f(A,E)$ we further have 
\begin{align}
    \Obs{f\lr{A,E}\lr{\tau,\dustcoords}} \approx f\lr{\Obs{A}\lr{\tau,\dustcoords},\ldots,\Obs{E}\lr{\tau,\dustcoords}},
\end{align}
where in general only a weak equality holds --- i.e.~one that only holds on the so-called constraint surface, the hypersurface on which all constraints are fulfilled. Hence, it is sufficient to construct Dirac observables for the elementary geometric phase space variables. To obtain their explicit form, one chooses $T$ as the temporal reference field for the Hamiltonian constraint and $S^j$ as spatial reference fields for the diffeomorphism constraint. These observables depend on physical temporal and spatial coordinates $\tau$ and $\dustcoords^j$ respectively and $\Obs{f}(\tau,\dustcoords^j)$ has the interpretation that it returns the values of the phase space function $f$ when the reference fields $T,S^j$ take the values $\tau,\dustcoords^j$ underlying the relational formulation of the model. 

Because all Dirac observables $\Obs{f}$ by construction commute with the canonical Hamiltonian, their dynamics in $\tau$ is generated by a so-called physical Hamiltonian $\Hphys$ via
\begin{align}
    \pdif{\Obs{f}\lr{\tau,\dustcoords}}{\tau} = \pb{\Obs{f}\lr{\tau,\dustcoords}}{\Hphys},
\end{align}
where for the Gaussian dust model the physical Hamiltonian reads
\begin{align}
   {\Hphys}=\int_{\cal S}{\rm d}^3\dustcoords\, \Obs{C} \label{eq:physHamObs}
\end{align}
and the integral is taken over the manifold coordinatised by the spatial dust reference fields, also called the dust space ${\cal S}$. This setup will be our starting point for the symmetry reduction in Subsection \ref{subsec:SymmRed}. In order to keep a more compact notation, we continue using $\lr{\A,X,Y,\E,\Ex,\Ey}$, $\lr{\Diffeo_\theta,\Ham}$ and $\lr{\theta,x,y}$ instead of $\lr{\Obs{\A}, \Obs{X}, \ldots, \Obs{\Ey}}$, $\lr{\Obs{\Diffeo_{\theta}},\Obs{\Ham}}$ and $\lr{\dustcoords^\theta, \dustcoords^x,\dustcoords^y}$ in the remaining part of this work.

\subsection{Brief review of the symmetry reduction to a model with polarised \texorpdfstring{$\mathbbm{T}^3$}{T3} Gowdy symmetry}
\label{subsec:SymmRed}

We start by introducing the basic elementary observables of the Gowdy model in loop quantum gravity along the seminal work of~\cite{Husain:1989qq,MenaMarugan:1997us,Bojowald} carried over to the reduced phase space considered here. We will also closely follow~\cite{Banerjee:classical,Banerjee:quant} concerning notation and to compare our results to the already existing ones for Dirac quantisation. For the derivation of the physical Hamiltonian in the reduced phase space we further follow \cite{MA:Andreas}.

Denoting the two Killing vector fields by $\pdif{}{x}$ and $\pdif{}{y}$ and the remaining cyclic variable by $\theta$, we decompose the connection and its conjugate momentum accordingly:
\begin{align}
    A & = A^i_\theta\lr{\theta}\tau_i \d\theta + A^i_\rho \tau_i \d x^\rho, \\
    E & = E^\theta_i\lr{\theta} \tau_i \pt + E^\rho_i\lr{\theta}\tau_i \partial_\rho.
\end{align}

Therein, we sum over $x$ and $y$ via $\rho$ and $x^x\coloneqq x, x^y\coloneqq y$, while $i$ takes the values 1, 2 and 3. Additionally, $\tau_i = -\frac{\i}{2}\sigma_i$ are the generators of $\mathfrak{su}(2)$, with the Pauli matrices $\sigma_i$.

The unpolarised Gowdy model, where the Killing vector fields are not demanded to be orthogonal, is then obtained via the choice
\begin{align}
    E^\theta_I & = E^\rho_3 = 0 \ \text{and} \nonumber\\
    A^I_\theta & = A^3_\rho = 0,
\end{align}
where the capital $I$ is now representing 1 and 2. With this set of variables, the Gauß constraints $\Gauss_1$ and $\Gauss_2$ are trivially satisfied. 
The same holds for the geometric contributions to the spatial diffeomorphism constraints  $\Diffeo_x$ and $\Diffeo_y$. Within the relational formalism this is taken into account by the fact that we couple only two additional dust fields $T,S^\theta$ in the symmetry reduced sector, both depending on the $\theta$ coordinate only and thus there is no contribution from $P_x,P_y$ in the total diffeomorphism constraints. The remaining geometric contributions to the total constraints involving gravity and dust at this stage read
\begin{align}
    \Gau = & \frac{4\pi^2}{\kappa \immirzi} \lr{\pt \Ett + \epsilon_{3J}{}^K A^J_\rho E^\rho_K} \eqqcolon \frac{1}{\kappa' \immirzi} \lr{\pt \Ett + \epsilon_{3J}{}^K A^J_\rho E^\rho_K},\\
    \Diff = & \frac{1}{\kappa'\immirzi} \lr{E^\rho_I \lr{\pt A^I_\rho} + \epsilon_{3J}{}^K A^J_\rho E^\rho_K \Att - \kappa\immirzi\Att\Gau},\\
    \Ham = & \frac{1}{2\kappa' \sqrt{\det E}} \left(2\Att\Ett A^J_\rho E^\rho_J + A^J_\rho E^\rho_J A^K_\sigma E^\sigma_K - A^K_\rho E^\rho_J A^J_\sigma E^\sigma_K - 2\epsilon_{3J}{}^K\lr{\pt A^J_\rho} E^\rho_K\Ett\right.\nonumber\\
    & \left. \hphantom{\frac{1}{2\kappa' \sqrt{\det E}}} \ - \lr{1+\immirzi^2}\lr{2K^3_\theta \Ett K^J_\rho E^\rho_J + K^J_\rho E^\rho_J K^K_\sigma E^\sigma_K - K^K_\rho E^\rho_J K^J_\sigma K^J_\sigma E^\sigma_K}\right), \label{eq:C}
\end{align}
where we have introduced $\det E \coloneqq \Ett\lr{E^x_1 E^y_2 - E^x_2 E^y_1}$ and $\kappa' \coloneqq \frac{\kappa}{4\pi^2}$. The latter absorbs an additional factor of $4\pi^2$ that stems from smearing over the two variables $x$ and $y$ the model does not depend on anymore. In that sense, we solved already two of the integrals in \eqref{eq:physHamObs}.

We then proceed towards our final description via two canonical transformations. For the first one, we perform a polar decomposition of the $A^I_\rho$ and $E^\rho_I$ according to
\begin{align}
     A^1_x \eqqcolon A_x \cos\lr{\alpha + \beta},\quad & A^1_y \eqqcolon -A_y \sin\lr{\abar + \bbar},    & E^x_1 \eqqcolon E^x \cos{\beta},\quad & E^y_1 \eqqcolon - E^y \sin{\bbar}, \nonumber\\
     A^2_x \eqqcolon A_x \sin\lr{\alpha + \beta},\quad & A^2_y \eqqcolon A_y \cos\lr{\abar + \bbar}, & E^x_2 \eqqcolon E^x \sin{\beta},\quad & E^y_2 \eqqcolon E^y \cos{\bbar},
\end{align}
and as a result define
\begin{align}
    X & \coloneqq A_x \cos\alpha,  & P^\beta & \coloneqq -E^x A_x \sin\alpha, \nonumber\\
    Y & \coloneqq A_y \cos\abar, & \Pbar^\beta & \coloneqq -E^y A_y \sin\abar, \nonumber\\
    \A & \coloneqq \frac{1}{\immirzi}\Att,  & \E & \coloneqq \Ett.
\end{align}

Then, the second canonical transformation reads
\begin{align}
    \xi & \coloneqq \beta - \bbar, & \Pxi & \coloneqq \frac{P^\beta-\Pbar^\beta}{2}, \nonumber\\
    \eta & \coloneqq \beta + \bbar, & \Peta & \coloneqq \frac{P^\beta + \Pbar^\beta}{2},
\end{align}
resulting in the five pairs of canonically conjugate variables $(\A,\E)$, $(X,\Ex)$, $(Y,\Ey)$, $(\eta,\Peta)$ and $(\xi,\Pxi)$. Since the remaining diffeomorphism constraint $C_\theta^{\rm tot}$ and the Hamiltonian constraint $C^{\rm tot}$ are reduced at the classical level by means of constructing Dirac observables, the only left first class constraint is the Gauss constraint $G_3$. The latter reduced two degrees of freedom in phase space such that for the unpolarised Gowdy model we end up with four physical degrees of freedom.

The polarised $\mathbbm{T}^3$ Gowdy model can now be constructed by taking a look at the line element up to this point,
\begin{align}
    \d s^2 = \frac{\Ex\Ey}{\Ett} \cos\xi \d\theta^2 + \frac{\Ett \Ey}{ \Ex \cos\xi}\d x^2 + \frac{\Ett\Ex}{\Ey \cos\xi}\d y^2 - 2\Ett\frac{\sin\xi}{\cos\xi}\d x\;\! \d y,
\end{align}
and demanding the $\d x\;\! \d y$-term to vanish. This can be realised by imposing the constraints
\begin{align}
    \xi\lr{\theta} & \approx 0 \quad \text{and}\\
    \dot{\xi}\lr{\theta} & \approx 0,
\end{align}
where the latter guarantees the stability of the former. We get
\begin{align}
    \chi\lr{\theta} \coloneqq \dot{\xi}\lr{\theta} = 2\Pxi + \Ett \: \pt \ln \frac{\Ey}{\Ex},
\end{align}

which fixes the conjugate momentum $\Pxi$ and also results in $\dot{\chi}\lr{\theta}\approx 0$ with no further ado. These polarisation constraints together with the Gau\ss{} constraint
\begin{align}
    \Gau = & \frac{1}{\kappa' \immirzi} \lr{\pt\E + \Peta} \label{eq:GaussConstraint}
\end{align}
complete the set of constraints.
The symmetry reduced physical Hamiltonian then has the form
\begin{align}
   {\Hphys}=\int_{{\Sd}}{\rm d}\theta\, \Ham(\theta) ,
\end{align}
where $\Sd$ denotes the symmetry reduced dust space. The geometric contributions to the Hamilton constraint in terms of the Dirac observables  now read
\begin{align}
    \Ham = & -\frac{1}{\kappa' \sqrt{\det E}} \Bigg( \frac{1}{\immirzi^ 2}\lr{X\Ex Y\Ey + \A\E\lr{X\Ex+Y\Ey} + \E\pt\eta\lr{X\Ex+Y\Ey}} \nonumber\\
    &  \quad \hphantom{-\frac{1}{\kappa' \sqrt{\det E}}} +\frac{1}{4}(\pt\E )^2- \frac{1}{4}\lr{\E\pt\ln\frac{\Ey}{\Ex}}^2 \Bigg) + \frac{1}{\kappa'} \pt\lr{\frac{\E\pt\E}{\sqrt{\det E}}} \nonumber \\
    &- \frac{\kappa'}{4}\frac{G^2_3}{\sqrt{\det E}} - \frac{\immirzi}{2}\pt\frac{G_3}{\sqrt{\det E}}, \label{eq:Hamiltonconstraint}
\end{align}
where $\det E = \E\Ex\Ey$. Note that the other two integrations were already dealt with before, in \eqref{eq:C}. Note that \eqref{eq:Hamiltonconstraint} is in agreement with the result in \cite{Banerjee:classical} with the only difference that here only $C^{\rm tot}$ is required to vanish due to the presence of the dust. For completeness, we also present the geometric contribution to the diffeomorphism constraint in the symmetry reduced Gowdy model that takes the form
\begin{align}
 \Diff = & \frac{1}{\kappa'\immirzi} \lr{E^\rho_I \lr{\pt A^I_\rho} + \epsilon_{3J}{}^K A^J_\rho E^\rho_K \Att - \kappa\immirzi\Att\Gau}   
\end{align}
but does not contribute to the physical Hamiltonian in the case of the Gaussian dust model. Also $\Diff$ agrees with the results in \cite{Banerjee:classical} and likewose only $C_\theta^{\rm tot}$ is a constraint, but not $\Diff$ individually in general.

When it comes to the Gauß constraint, we may solve it already at this (classical) level. Noticing that $\eta$ is just translated via the action of the Gauß constraint, we can impose $\eta \approx 0$. With the Gauß constraint $\Gau$ and $\eta$ being second class, we proceed with the Dirac bracket and make the two constraints vanish strongly. For all other quantities that do not depend on $\eta$, the Dirac bracket reduces to the Poisson bracket and nothing more has to be done.\footnote{Note that $\Gau = 0$ allows to solve $\Peta$ for variables independent of $\eta$ and $\Peta$.} Following this route, we end up with three independent pairs of elementary Dirac observables: $(\A,\E)$, $(X,\Ex)$ and  $(Y,\Ey)$. Alternatively, as shown in \cite{Bojowald,Bojowald:2005cb}, the Gauß constraint can also easily be solved at the quantum level. In this case, operators associated with $(\eta,\Peta)$ will be involved in the kinematical Hilbert space and after solving the Gau\ss{} constraint the subspace of the kinematical Hilbert space no longer contains these quantum degrees of freedom.

\section{Quantisation of the reduced LQG model with  polarised \texorpdfstring{$\mathbbm{T}^3$}{T3} Gowdy symmetry}
\label{sec:QuantRedLQG}

\subsection{The physical Hilbert space in reduced LQG}
\label{sec:PhysHSLQG}
As discussed in the former section, the physical phase space of the polarised Gowdy model involves three independent pairs of canonically conjugate Dirac observables after solving the Gau\ss{} constraint:  $(\A,\E)$, $(X,\Ex)$ and $(Y,\Ey)$. Imposing the polarisation condition eliminated $(\xi,\Pxi)$ and, accordingly, fulfilling the Gauß constraint made $(\eta,\Peta)$ vanish. Since the algebra of these Dirac observables is given by the standard Poisson bracket we can use the same representations that was used in \cite{Banerjee:quant,Bojowald:2005cb} for the kinematical Hilbert space for the physical Hilbert space:
\begin{equation*}
{\cal H}_{\rm phys} = L_2(\overline{\cal A}_{\Sd\times {T}^2},\mu_0),
\end{equation*}
where $\overline{\cal A}_{\Sd\times {T}^2}$ denotes the space of generalised connections on $T^3\simeq \Sd\times {T}^2$ and $\mu_0$ is the analogue of the Ashtekar--Lewandowski measure in full LQG. $\overline{\cal A}_{\Sd\times {T}^2}$ is constructed as follows: We consider ${\cal A}_{\Sd}$ and its projective limit over graphs $\graph$ in $\Sd$, which are just non-intersecting unions of edges $e_n$ that correspond to arcs here. A graph $\graph$ is then given by $\graph=\cup_i e_i$. We denote by $V(\graph)$ the graph's set of vertices, which is just the union of all end points of the $e_i$, and by $E(\graph)$ its set of edges. For a given graph $\graph$ we can understand the space ${\cal A}^\graph_{\Sd}$ as a set of maps from $E(\graph)$ to $U(1)^{|E(\graph)|}$, that is one copy of U$(1)$ for each edge of the graph. For a fixed edge we have 
\begin{align}
{\cal A}_{\Sd}:\graph\to {\rm U(1)}, e\mapsto   h^{(k_e)}_{e}\lr{\A} \coloneqq \ex{\i\frac{k_e}{2}\int_{e}\A}, \label{eq:holonomy}
\end{align}{}
where the ${\rm U(1)}$ charges $k_e \in \mathbbm{Z}$ and, for later convenience, a factor $\frac{1}{2}$ is introduced. Using that the set of graphs is a partially ordered directed set and introducing the projections $P_{\graph\graph'}:{\cal A}^\graph_{\Sd}\to{\cal A}^{\graph'}_{\Sd}, {\cal A}\mapsto P_{\graph\graph'}({\cal A}^\graph)\coloneqq {\cal A}^\graph\big|_{\graph'}$ for $\graph'\leq \graph$ one can derive the set of generalised connections $\overline{{\cal A}}_{\Sd}$ as the projective limit over graphs in $\Sd$, that is 
\begin{equation*}
\overline{\mathcal{A}}_{\Sd}=\lim _{\substack{\longleftarrow \\ \graph\subset \Sd}} \mathcal{A}_{\Sd}^{\graph}.
\end{equation*}
$X$ and $Y$, in turn, are scalar fields and in order to still obtain a similar description, we follow ~\cite{Bojowald,Banerjee:quant} and define so-called point holonomies~\cite{Thiemann:1997rq}
\begin{align}
    h^{(\mu_v)}_{v} \lr{X} &\coloneqq \ex{\i\frac{\mu_v}{2}X\lr{v}} \quad \text{and} \label{eq:xholonomy}\\
    h^{(\nu_v)}_{v} \lr{Y} &\coloneqq \ex{\i\frac{\nu_v}{2}Y\lr{v}}  \label{eq:yholonomy}
\end{align}{}
sitting on the graph's vertices $v$ with corresponding charges $\mu_v,\nu_v\in\mathbb{R}$ and with $X\lr{v},Y\lr{v}\in \mathbb{R}$. For each fixed vertex $v$, the space $C(\overline{\mathbb{R}}_{\rm Bohr})$ of continuous almost periodic functions on the Bohr compactification of the real line is used. The space of generalised connections $\overline{\cal A}_{{T}^2}$ can be obtained again as a projective limit, this time over the vertex set $V(\graph)$. For a fixed graph $\graph$, the space ${\cal A}^\graph_{{T}^2}$ involves maps from $V(\graph)$ to $(\overline{\mathbb{R}}_{\rm Bohr}\times \overline{\mathbb{R}}_{\rm Bohr})^{|V(\graph)|}$. For a fixed vertex $v$, we have $A_{{T}^2}: V(\graph)\to\overline{\mathbb{R}}_{\rm Bohr}\times \overline{\mathbb{R}}_{\rm Bohr}$ with $v\mapsto (X(v),Y(v))$. Then we have 
$\overline{\mathcal{A}}_{\Sd\times T^2}=\lim\limits_{\substack{\longleftarrow \\ \graph\subset \Sd}} \mathcal{A}_{\Sd}^{\graph}\otimes \mathcal{A}_{T^2}^{\graph}$.

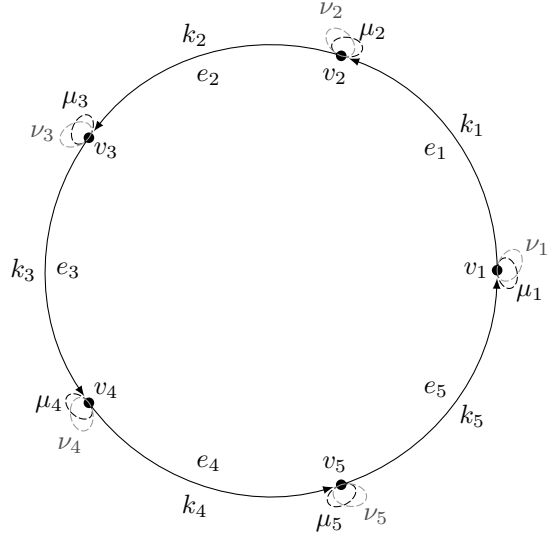
\begin{figure}
\begin{tikzpicture}

\def \n {5}
\def \radius {3cm}
\def \smallradius {0.2}
\def \margin {0} 
\def \halfmargin {2} 

\foreach \s in {1,...,\n}
{
  \node at ({360/\n * (\s - 1)}:\radius-8pt) {$v_\s$};
  \fill ({360/\n * (\s - 1)}:\radius) circle (2pt);
  \draw[-latex] ({360/\n * (\s - 1)+\margin}:\radius) 
    arc ({360/\n * (\s - 1)+\margin}:{360/\n * (\s)-\halfmargin}:\radius);
  \node at ({180/\n+(\s-1)*360/\n}:\radius*1.1) {$k_\s$};
  \node at ({180/\n+(\s-1)*360/\n}:\radius*0.9) {$e_\s$};
  \draw  [dash pattern=on 3pt off 1pt, black, rotate around y=-30, rotate around z=180, rotate around={{360/\n * (\s - 1)}:({360/\n * (\s - 1)}:\radius)}] ({360/\n * (\s - 1)}:\radius) arc (0:360:\smallradius);
  \node at ({360/\n * (\s - 1)-5}:{\radius+0.45cm}) {$\mu_\s$};
  \draw  [dash pattern=on 3pt off 1pt, black!40, rotate around y=60, rotate around x=0, rotate around z=180, rotate around={{360/\n * (\s - 1)}:({360/\n * (\s - 1)}:\radius)}] ({360/\n * (\s - 1)}:\radius) arc (0:360:\smallradius);
  \node at ({360/\n * (\s - 1)+5}:{\radius+0.55cm}) {\color{black!70}$\nu_\s$};
}
\end{tikzpicture}
    \caption{An embedded five-valent graph with charges $k_e$ on the edges and point holonomies labelled by $\mu_v, \nu_v$ on the vertices, serving as a basis element for Gowdy states. To keep the notation more compact, we used $k_{e_{v_I}}\eqqcolon k_I, \mu_{v_I}\eqqcolon\mu_I$ and $\nu_{v_I}\eqqcolon\nu_I$.}
    \label{fig:embeddedGraph}
\end{figure}

The basis states of ${\cal H}_{\rm phys}$ are then labelled by a graph $\graph$ --- defining the sets of the vertices $V\lr{\graph}$ and the edges $E\lr{\graph}$ ---, the U(1)-charges $k_e$ (collected in $k$) as well as the point holonomies' charges $\mu_v$ and $\nu_v$ (collected in $\mu$ and $\nu$ respectively)~\cite{Banerjee:quant}:
\begin{align}
    |\graph,k,\mu,\nu\rangle \coloneqq \prod_{e\in E\lr{\graph}} \ex{\i\frac{k_e}{2}\int_{e}\A} \prod_{v\in V\lr{\graph}} \ex{\i\frac{\mu_v}{2}X\lr{v}} \ex{\i\frac{\nu_v}{2}Y\lr{v}}. \label{eq:state}
\end{align}
We now use Figure~\ref{fig:embeddedGraph} --- showing exemplarily a five-valent Gowdy state in reduced LQG where we work with embedded graphs --- to introduce the states' composition and notation. The dashed miniature lines in Figure~\ref{fig:embeddedGraph} visualise the fact that the point holonomies are actually not along edges.

The physical Hilbert space ${\cal H}_{\rm phys}$ can also be written as a direct sum of the Hilbert spaces ${\cal H}_\graph$ associated to each graph $\graph\in\Sd$:
\begin{align}
\label{eq:DecHphys}
{\cal H}_{\rm phys}=\bigoplus\limits_{\graph}{\cal H}_\graph.    
\end{align}
The holonomy operators act on the basis states~\eqref{eq:state} via multiplication:
\begin{align}
 \hat{h}^{(k_{0})}_{e_I}\lr{\A}|\graph,k,\mu,\nu\rangle &=  \ex{\i\frac{k_0}{2}\int_{e_I}\A} |\graph,k,\mu,\nu\rangle = |\graph,k_{e_I}+k_0,\mu,\nu\rangle, \\
  \hat{h}^{(\mu_{0})}_{v_I} \lr{X}|\graph,k,\mu,\nu\rangle &=    \ex{\i\frac{\mu_{0}}{2}X\lr{v_I}} |\graph,k,\mu,\nu\rangle = |\graph,k,\mu_{v_I}+\mu_0,\nu\rangle, \\
 \hat{h}^{(\nu_0)}_{v_I} \lr{Y}|\graph,k,\mu,\nu\rangle &=    \ex{\i\frac{\nu_{0}}{2}Y\lr{v_I}} |\graph,k,\mu,\nu\rangle =|\graph,k,\mu,\nu_{v_I}+\nu_0\rangle.
\end{align}
Therein, we used the abbreviation $k_{e_I}+k_0 \eqqcolon k \lvert_{k_{e_I}=k_{e_I}+k_0}$ within the state and likewise for $\mu$ and $\nu$.

The according flux operators are implemented as follows. First, we have~\cite{Banerjee:quant}
\begin{align}
    \hat\E\lr{\theta}|\graph,k,\mu,\nu\rangle = -\i\immirzi\lp{}^2\frac{\delta}{\delta\A\lr{\theta}}|\graph,k,\mu,\nu\rangle = \frac{\immirzi\lp{}^2}{2}\frac{k_{e^{+}\lr{\theta}}+k_{e^{-}\lr{\theta}}}{2}|\graph,k,\mu,\nu\rangle, \label{eq:ActionFlux}
\end{align}
where $k_{e^{+}\lr{\theta}}$ is the U(1)-charge of the edge that is outgoing at $\theta$ and $k_{e^{-}\lr{\theta}}$ the one of the incoming edge. If $\theta$ does not coincide with a vertex, the two are the same and the factor $\frac{1}{2}$ vanishes.

For the $x$- and $y$-flux, we first of all smear them over intervals $\I$,
\begin{align}
    \Fx{\I} &\coloneqq \int_{\I} \hat{E}^x \quad \text{and} \\ 
    \Fy{\I} &\coloneqq \int_{\I} \hat{E}^y , 
\end{align}
to finally obtain~\cite{Banerjee:quant}
\begin{align}
    \Fx{\I} |\graph,k,\mu,\nu\rangle &= \frac{\immirzi\lp{}^2}{2} \sum_{v \in V\lr{\graph \cap \I}} \mu_v |\graph,k,\mu,\nu\rangle \quad \text{and} \\
    \Fy{\I} |\graph,k,\mu,\nu\rangle &= \frac{\immirzi\lp{}^2}{2} \sum_{v \in V\lr{\graph \cap \I}} \nu_v |\graph,k,\mu,\nu\rangle.
\end{align}
Therein, we collected all contributions of vertices that lie in the union of $\I$ and $\graph$. We get a factor $\frac{1}{2}$ if an endpoint of $\I$ coincides with a vertex. We will later, however, use intervals that contain one vertex at most, as this simplifies the transition towards the AQG framework presented in Section \ref{Sec:AQGQuant}.

Before approaching the dynamics and the Hamiltonian constraint, we shortly illustrate how to deal with the Gauß constraint had it not been solved on the classical level already. Then, the pair $(\eta, \Peta)$ would still be part of the set of variables. Similar to the other variables, the point holonomy
\begin{align}
    h^{(\lambda_v)}_{v} \lr{\eta} \coloneqq \ex{\i\lambda_v\eta\lr{v}} \ , \ \lambda_v \in \mathbbm{Z}\ ,
\end{align}
as well as the flux
\begin{align}
    \Feta{\I} \coloneqq \int_{\I} \Peta
\end{align}
are defined. The corresponding holonomy operator $\hat{h}^{(\lambda_v)}_{v} \lr{\eta}$ acts multiplicatively on the basis states, whose composition \eqref{eq:state} is now additionally enriched with the holonomies $\ex{\i\lambda_v\eta\lr{v}}$ --- denoted as $|\graph,k,\mu,\nu,\lambda\rangle$. Accordingly, the corresponding flux operator $\hat{\cal F}_{\eta,\I} $ acts via differentiation:
\begin{align}
   \hat{\cal F}_{\eta,\I} |\graph,k,\mu,\nu,\lambda\rangle = \immirzi\lp{}^{2} \sum_{v\in V\lr{\graph \cap \I}} \lambda_v |\graph,k,\mu,\nu,\lambda\rangle.
\end{align}
We can then use these quantities to quantise the Gauß constraint \eqref{eq:GaussConstraint} by means of choosing  a suitable partition $\PT$ of $\Sd$ in terms of intervals $\I_n$ such that $\Sd=\cup_n\I_n$ with $\I_n:[\theta_n-\frac{\epsilon}{2},\theta_n+\frac{\epsilon}{2}]$. We can then obtain a regularisation of the Gau\ss{} constraint \`a la~\cite{Banerjee:quant}
\begin{align}
    \Gau^\epsilon & 
    =\frac{1}{\kappa' \immirzi} \sum\limits_{\I_n\in \PT}\int_{\I_n} \lr{\pt\E + \Peta} \d\theta \\
    & = \frac{1}{\kappa' \immirzi} \sum_{\I_n\in \PT}\lr{ \E\lr{\theta_n+\frac{\epsilon}{2}} - \E\lr{\theta_n-\frac{\epsilon}{2}} + \Feta{\I_n} } \\
\end{align}
and in the limit when we send the regulator to zero we rediscover the classical Gau\ss{} constraint, that is
\begin{align}
    \Gau &= \frac{1}{\kappa' \immirzi} \int_{\Sd} \lr{\pt\E + \Peta} \d\theta =
    \lim\limits_{\epsilon\to 0} \Gau^\epsilon.
\end{align}
The corresponding Gau\ss{} constraint operator is then obtained as
\begin{align}
\widehat{\Gau} & \coloneqq \lim\limits_{\epsilon\to 0} \frac{1}{\kappa' \immirzi}  \sum\limits_{\I_n\in\PT} \lr{ \hat\E\lr{\theta_n+\frac{\epsilon}{2}} - \hat\E\lr{\theta_n-\frac{\epsilon}{2}} + \hatFeta{\I_n} } . \label{eq:LQGGaußCstrt}    
\end{align}
Note that this Gau\ss{} operator agrees in its symmetric definition with the one used in \cite{Bojowald}, while it differs in this aspect to the one used in \cite{Banerjee:quant}.

In the limit where we send the regulator to zero, also known as the infinite refinement limit, we can choose the partition fine enough such that at most one vertex is contained in $\I_n$. 

Then, the action of this Gauß constraint operator on the basis states reads~\cite{Banerjee:quant}
\begin{align}
    \widehat{\Gau} |\graph,k,\mu,\nu,\lambda\rangle & = \frac{\lp{}^2}{\kappa'} \sum_{v\in V\lr{\graph}} \lr{ \frac{k_{\ev}-k_{\evm}}{2} + \lambda_v } |\graph,k,\mu,\nu,\lambda\rangle \label{eq:ActionGauß}.
\end{align}
Therein and from now on, we use as convention for the notation of the $k$-labels that we always work with outgoing edges and therefore run through the vertices with respective superscripts. This means that the $k$-label of the ingoing edge at vertex $v$ is the same as that of the outgoing edge at the left-neighbouring vertex $\vm$.

We can now solve the above Gauß constraint by imposing the following vertex-wise condition:
\begin{align}
    \lambda_v = - \frac{k_{\ev}-k_{\evm}}{2} \ , \quad \forall\,v\in V\lr{\graph}. \label{eq:GaußSolution}
\end{align}
As $\lambda_v \in \mathbbm{Z}$, the difference of the $k$-charges has to fulfil $ k_{\ev}-k_{\evm} \in 2\mathbbm{Z}$.

Note that from choosing an infinitely fine partition as above follows that if there is indeed a vertex within interval $\I_n$, there will be none in any of the two neighbouring intervals. Hence, in the action of the flux $\hat\E$ shown in \eqref{eq:ActionFlux}, the two terms add up the same charge as it is just one edge that gets split up by $\theta_n-\frac{\epsilon}{2}$, or $\theta_n+\frac{\epsilon}{2}$ respectively, and not two different ones from one in- and one outgoing edge. This then leads to \eqref{eq:ActionGauß}. We will later also show the implementation of a Gauß constraint operator in the AQG framework in Section \ref{Sec:AQGQuant}, but nevertheless also there stick to the strategy of solving the Gau\ss{} constraint already on the classical level. This is foremost due to the fact that it can be solved straightforwardly, eliminating also one pair of canonically conjugate variables $\lr{\eta,\Peta}$. So there really is no need to carry them along any further from this point onwards.

\subsection{Quantum dynamics in the reduced LQG model}
\label{sec:DynRedLQG}
While the quantisation of $\euclH $ and $\lorH$ is performed along the lines of \cite{Banerjee:quant}, the transition to the AQG formalism for the Brown--Kuchar model and the master constraint respectively can be found in \cite{MA:Boehm, MA:Alex}. Again, in terms of notation, we stay close to \cite{Banerjee:classical,Banerjee:quant} also used in \cite{MA:Andreas}. As before, we first regularise the classical expression for the physical Hamiltonian $\Hphys$ in order to be able to define the corresponding operator on ${\cal H}_{\rm phys}$.
First of all and following \cite{Bojowald:2005cb,Banerjee:quant}, we start with introducing the SU(2)-valued holonomies, which we can later use to reformulate $\euclH$ and $\lorH$:
\begin{align}
    h_\theta\lr{\I} &\coloneqq \exp\lr{\tau_3 k_0 \int_{\I} \A } = \cos\lr{\frac{k_0}{2}\int_{\I}\A} + 2\tau_3 \sin\lr{\frac{k_0}{2}\int_{\I}\A}, \label{eq:thetaSU2Holonomy} \\
    h_x\lr{\theta} &\coloneqq \exp\lr{\mu_0 \tau_x X} = \cos\lr{\frac{\mu_0}{2}X} + 2\tau_x \sin\lr{\frac{\mu_0}{2}X} \ \text{ and } \label{eq:xSU2Holonomy}\\
    h_y\lr{\theta} &\coloneqq \exp\lr{\nu_0 \tau_y Y} = \cos\lr{\frac{\nu_0}{2}Y} + 2\tau_y \sin\lr{\frac{\nu_0}{2}Y}.\label{eq:ySU2Holonomy}
\end{align}
Therein, we used
\begin{align}
\label{eq:tauxtauy}
    \tau_{x} \lr{\theta} \coloneqq \cos\beta\lr{\theta}\tau_1 + \sin\beta\lr{\theta}\tau_2 \quad \text{and} \quad \tau_y\lr{\theta} \coloneqq -\sin\beta\lr{\theta}\tau_1 + \cos\beta\lr{\theta}\tau_2 ,
\end{align}
where the $\mathfrak{su}(2)$ basis $\tau_i = -\frac{\i}{2}\sigma_i$, $i=1,2,3$, with the Pauli matrices $\sigma_i$ satisfies
\begin{equation}
    \tr \tau_i = 0 \quad\text{and}\quad \tau_i\tau_j = -\frac{1}{4}\delta_{ij}\mathbbm{1}_{\text{SU}(2)} + \frac{1}{2}\epsilon_{ijk}\tau_k.
\end{equation}
One can show the equality of the holonomies' splits into sine and cosine by using the easily verifiable identities
\begin{align}
    \tau_x{}^2 = \tau_y{}^2 = -\frac{1}{4}\mathbbm{1}_{\text{SU}(2)} . \label{eq:tauxtauysquared}
\end{align}

Having the action of the basic operators at hand, we can proceed towards the quantisation of the physical Hamiltonian operator. But not before we address the volume operator, which will serve as a crucial ingredient of the Hamiltonian constraint operator. We follow again closely~\cite{Banerjee:quant,MA:Andreas}.

As starting point, the volume of an arc $\I$ is classically given by the volume functional
\begin{align}
    V(\I) &\coloneqq 4\pi^2\int_{\I} \d\theta \sqrt{\lrabs{\det E}} =4\pi^2 \int_{\I} \d\theta \sqrt{\lrabs{\E\Ex\Ey}}. \label{eq:VolumeGeneral}
\end{align}
Now, similar to the discussion of the Gau\ss{} constraint above, we choose a partition $\PT_\I$ of $\I$ into intervals $\I_n$ such that we have $\I=\cup_{n}\I_n$. This allows us to rewrite the volume functional as
\begin{align}
    V(\I) &=4\pi^2\lim\limits_{\epsilon\to 0}  \sum_{\I_n\in\PT_\I} \int_{\I_n} \d\theta \sqrt{\lrabs{\E\Ex\Ey}\lr{\theta}} =4\pi^2 \lim\limits_{\epsilon\to 0}  \sum_{\I_n\in\PT_\I} \int_{\theta_n-\frac{\epsilon}{2}}^{\theta_n+\frac{\epsilon}{2}} \d\tilde{\theta}\sqrt{\lrabs{\E\Ex\Ey}(\tilde{\theta})},
\end{align}
where we choose the intervals $\I_n$ sufficiently small, that is $\I_n = [\theta_n-\frac{\epsilon}{2},\theta_n+\frac{\epsilon}{2}]$. The integral can then be replaced by a Riemann sum involving $\epsilon\sqrt{\lrabs{\E\Ex\Ey}(\theta_n)}$, yielding for the regularised volume functional $V^\epsilon(\I)$
\begin{align}
    V^\epsilon(\I) &=  4\pi^2\sum_{\I_n\in\PT_\I} \sqrt{\lrabs{\E}\lrabs{\epsilon\Ex}\lrabs{\epsilon\Ey}\lr{\theta_n}} 
    = 4\pi^2\sum\limits_{\I_n\in\PT_\I}
   \sqrt{ \lrabs{\E\lr{\theta_n}} \lrabs{\int_{\theta_n-\frac{\epsilon}{2}}^{\theta_n+\frac{\epsilon}{2}}\Ex} \lrabs{\int_{\theta_n-\frac{\epsilon}{2}}^{\theta_n+\frac{\epsilon}{2}}\Ey}}    
   \nonumber\\
   &=4\pi^2 \sum\limits_{\I_n\in\PT_\I} \sqrt{ \lrabs{\E\lr{\theta_n}} \lrabs{\mathcal{F}_{x,\I_n}} \lrabs{\mathcal{F}_{y,\I_n}}}.
\end{align}
From the first to the second line, we interpreted the two products including $\epsilon$ as approximations of infinitesimal integrals and then reintroduced the smeared fluxes $\mathcal{F}_{x,\I_n}, \mathcal{F}_{y,\I_n}$ --- now with intervals labelled by $n$. We then define the corresponding volume operator as
\begin{align}
    \V\lr{\I} &= 4\pi^2\lim\limits_{\epsilon\to 0} \sum_{\I_n\in\PT_\I} \sqrt{ \lrabs{\hat{\E}\lr{\theta_n}} \lrabs{\hat{\mathcal{F}}_{x,\I_n}} \lrabs{\hat{\mathcal{F}}_{y,\I_n}} } .
\end{align}
In the infinite refinement limit, we have at most one $\theta_n$ in each interval $\I_n$ and hence the action of $\V\lr{\I}$ on the basic states states~\eqref{eq:state} in the physical Hilbert space is given by~\cite{Banerjee:quant}
\begin{align}
    \V\lr{\I}|\graph,k,\mu,\nu\rangle &=\sum_{v\in V\lr{\graph\cap \I}}\V_v|\graph,k,\mu,\nu\rangle ,
\end{align}
where the sum involves all vertices of the graph $\graph$ that lie in the interval $\I$ and we have 
\begin{align}
\V_v|\graph,k,\mu,\nu\rangle & =
\frac{4\pi^2}{\sqrt{2}}\lr{\frac{\immirzi\lp{}^2}{2}}^{\frac{3}{2}} \sqrt{ \lrabs{k_{\ev}+k_{\evm}} \lrabs{\mu_v} \lrabs{\nu_v} } \;\! |\graph,k,\mu,\nu\rangle \label{eq:VvOperator}.
\end{align}

We can now turn to the regularisation and quantisation of the physical Hamiltonian $\Hphys$. For this, we construct the Hamilton constraint by first integrating $\Ham$ over the dust manifold $\dustmanifold$:
\begin{align}
    \HH_{\rm phys} &\coloneqq \int_{\Sd} \d\theta \Ham\lr{\theta} = \int_{\Sd} \d\theta \lr{\euclHam + \lorHam} \eqqcolon \euclH + \lorH.
\end{align}
We thereby introduced the convenient split into a so-called Euclidean and Lorentzian part
\begin{align}
    \euclHam &\coloneqq \euclHam^{(1)}+\euclHam^{(2)}+\euclHam^{(3)}  \nonumber \\
    \euclHam^{(1)} &\coloneqq -\frac{1}{\kappa'\immirzi^2}\frac{1}{\sqrt{\det E}}\lr{X\Ex Y\Ey}\label{eq:euclHam1}\\
    \euclHam^{(2)}&\coloneqq -\frac{1}{\kappa'\immirzi^2}\frac{1}{\sqrt{\det E}}\lr{\A\E X\Ex} \label{eq:euclHam2} \\
    \euclHam^{(3)}&\coloneqq -\frac{1}{\kappa'\immirzi^2}\frac{1}{\sqrt{\det E}}\lr{\A\E Y\Ey} \label{eq:euclHam3} \\
    \lorHam &\coloneqq \lorHam^{(1)} + \lorHam^{(2)} + \lorHam^{(3)} \nonumber\\
    \lorHam^{(1)} &\coloneqq -\frac{1}{4\kappa'} \frac{\lr{\pt\E}^2}{\sqrt{\det E}} \label{eq:lorHam1}\\
    \lorHam^{(2)} &\coloneqq \frac{1}{4\kappa'} \frac{\E^2}{\sqrt{\det E}}\lr{\frac{\pt\Ex}{\Ex}-\frac{\pt\Ey}{\Ey}}^2 \label{eq:lorHam2}\\
    \lorHam^{(3)} &\coloneqq \frac{1}{\kappa'} \pt\lr{\frac{\E\pt\E}{\sqrt{\det E}}}\label{eq:lorHam3},
    \end{align}
where we have, as in \cite{Banerjee:classical}, compared to \eqref{eq:Hamiltonconstraint} dropped contributions proportional to the Gau\ss{} constraint and those involving $\eta$ linearly.  We will now quantise $\Hphys$, starting with the Euclidean part and continuing with the Lorentzian one. The final physical Hamiltonian operator will then be taken to be the symmetric combination that is $\hatHphys = \frac{1}{2}\lr{\euclHhat + (\euclHhat)^\dagger+\lorHhat+(\lorHhat)^\dagger}$ as can be seen in \eqref{eq:FinalHphysLQG}.

Note that the $\tau_x, \tau_y$ can also be used to reformulate 
\begin{align}
    \Ex \tau_x = E^x_1 \tau_1 + E^x_2 \tau_2 \quad\text{and}\quad \Ey \tau_y = E^y_1 \tau_1 + E^y_2 \tau_2
\end{align}
of the $x$ and $y$ part of $E\lr{\theta} = \E\lr{\theta} \tau_3 \pt + E^x\lr{\theta}\tau_x\lr{\theta} \partial_x + E^y\lr{\theta}\tau_y\lr{\theta} \partial_y$ and as they just result from a rotation of $\tau_2$ and $\tau_3$ in the 2-3-plane --- which also explains \eqref{eq:tauxtauysquared} ---, it furthermore still holds that
\begin{equation}
    [\tau_x,\tau_y] = \tau_3, \quad [\tau_y,\tau_3] = \tau_x \quad\text{and}\quad [\tau_y,\tau_3] = -\tau_x.
\end{equation}

A difference to the already existing quantisations of the Hamiltonian constraint in \cite{Banerjee:quant,deBlas:2017goa}
is that here we consider the physical Hamiltonian that at the classical level is invariant under spatial diffeomorphisms. If we aim at carrying over these symmetries also to the corresponding physical Hamiltonian operator, then, as pointed out in \cite{AQG1,Giesel:2012rb} for the usual embedded LQG framework, we need to quantise $\Hphys$ in a graph-preserving way. Going back to the decomposition  of the physical Hilbert space ${\cal H}_{\rm phys}$ in terms of a direct sum of the individual graph Hilbert spaces ${\cal H}_\graph$ shown in \eqref{eq:DecHphys}, this means that the physical Hamiltonian operator $\hatHphys$ will preserve each ${\cal H}_\graph$ separately, similar to the situation in full reduced LQG \cite{AQG4}. This can be achieved by using the notion of minimal loops originally introduced in \cite{Sahlmann:2002qj,Sahlmann:2002qk} that we will adapt to the symmetry reduced case of the polarised Gowdy model here. As discussed in \cite{AQG4}, this has the consequence that the quantum theory involves infinitely many conserved charges that are absent in the classical theory and furthermore the physical Hilbert space is still non-separable in this model.

\subsubsection{Quantisation of the Euclidean part of the physical Hamiltonian} 
We notice that $\euclHam=\euclHam^{(1)}+\euclHam^{(2)}+\euclHam^{(3)}$  consists of three similarly structured terms. Hence, we illustrate the regularisation procedure and the quantisation in detail for the first contribution $\euclHam^{(1)}$ of \eqref{eq:euclHam1} only and then are more brief for the remaining two $\euclHam^{(2)}$ and $\euclHam^{(3)}$ since they can be obtained in a similar manner. As discussed in detail below, the regularisation chosen here is different from the one in \cite{Banerjee:quant} to ensure the graph-preserving property of the physical Hamiltonian operator $\hatHphys$. Such a choice of regularisation is, however, closer to the way how $\Hphys$ will be quantised in the AQG framework discussed in Section \ref{Sec:AQGQuant}.

We start with choosing a partition of $S_1$ and replacing the integral over $\dustmanifold$ by a corresponding Riemann sum of intervals  $\I_n=[\theta_n-\frac{\epsilon}{2},\theta_n+\frac{\epsilon}{2}]$ with $\Sd=\cup_n\I_n$ according to
\begin{align}
  \euclH^{(1)}&=\int_{\Sd}\d\theta\, \euclHam^{(1)}=-\frac{1}{\kappa'\immirzi^2} \int_{\Sd} \d \theta \frac{ X\lr{\theta} \Ex\lr{\theta} Y\lr{\theta} \Ey\lr{\theta} }{\sqrt{\det E\lr{\theta}}}   \nonumber\\ 
  &=-\frac{1}{\kappa'\immirzi^2}\lim\limits_{\epsilon\to 0} \sum_{\I_n\in\PT}
    \int_{\I_n} \d \theta \frac{ X\lr{\theta} \Ex\lr{\theta} Y\lr{\theta} \Ey\lr{\theta} }{\sqrt{\det E\lr{\theta}}}
    \nonumber\\
  &=  -\frac{1}{\kappa'\immirzi^2} \lim\limits_{\epsilon\to 0} \sum_{\I_n\in\PT} \epsilon \frac{ X\lr{\theta_n} \Ex\lr{\theta_n} Y\lr{\theta_n} \Ey\lr{\theta_n} }{\sqrt{\det E\lr{\theta_n}}},\nonumber\\ \label{eq:EuclideanStart}
\end{align}
where we used in the last step that the intervals of the partition have length $\epsilon$ and can be chosen to be sufficiently small. Note that we could restrain ourselves to the integral of $\theta$ over $\Sd$ since all quantities only depend on $\theta$. Also, we want to point out again our abuse of notation that is using $\dustcoords^{\theta} = \theta $ for the (cyclic) dust coordinate $\dustcoords^{\theta}$ on the dust manifold $\Sd$.

As the next step and following \cite{Banerjee:quant}, we use that we can regularise the summand on the RHS of \eqref{eq:EuclideanStart} \`a la
\begin{align}
  \frac{1}{4\pi^2} \tr\lr{\lr{ h_x h_y h_x\inv h_y\inv - h_y h_x h_y\inv h_x\inv } h_\theta\left\{ h_\theta\inv,V(\I_n) \right\} }\label{eq:EuclidPartIdentity1} \\ = \frac{\kappa'\immirzi}{2} k_0\mu_0\nu_0 \epsilon \cdot \frac{XY\Ex\Ey}{\sqrt{\det E}}\lr{\theta_n}+O(\epsilon^2,\mu_0^2,\nu^2_0), & \nonumber 
\end{align}
where $O(\epsilon^2,\mu_0^2,\nu^2_0)$ denotes all terms that involve at least second powers of either $\epsilon$, $\mu_0$ or $\nu_0$ respectively.  The expression in \eqref{eq:EuclidPartIdentity1} transforms the term  we started with into a straightforwardly quantisable expression of holonomies and the volume functional. This replacement neglects terms of second and higher orders in $\epsilon$ and holds for small $X,Y,\int_{\I}\A$ as we will see, where the smallness of the latter quantity corresponds to small intervals $\I$. Furthermore, $\tr$ denotes the SU(2) trace and the LHS depends of course on $\theta$ as well --- we just refrain from writing down this dependency when the formulae become more elongate. Along the path after \eqref{eq:VolumeGeneral}, restricting ourselves to infinitesimal intervals $\I_n$ around $\theta_n$ of length $\epsilon$ involved in the partition $\PT$ allows us to use the following form for the volume functional:
\begin{align}
    V\lr{\I_n} &\coloneqq 4\pi^2\int_{\I_n} \d \theta \sqrt{\lrabs{\E \Ex \Ey}}(\theta) =4\pi^2 \epsilon \sqrt{\lrabs{\E \Ex \Ey}}(\theta_n)\nonumber \\
&= 4\pi^2\sqrt{\lrabs{\E} \lrabs{\epsilon \Ex} \lrabs{\epsilon \Ey}}(\theta_n)
= 4\pi^2\sqrt{\lrabs{\E} \lrabs{\int_{\I_n} \Ex} \lrabs{\int_{\I_n} \Ey}}. \label{eq:infVolFunctional}
\end{align}
Then, we can compute the Poisson bracket of the $\theta$-holonomy and the (infinitesimal) volume functional, which implies the Thiemann identity:
\begin{align}
    h_\theta \pb{h_\theta\inv}{V(\I_n)} & = -\frac{\kappa'\immirzi}{2}k_0 \tau_3 \frac{\sqrt{ \lrabs{\int_{\I_n} \Ex } \lrabs{\int_{\I_n}\Ex}}}{\sqrt{\lrabs{\E}}} = -\frac{\kappa'\immirzi}{2}k_0 \tau_3 \frac{\lrabs{\int_{\I_n} \Ex } \lrabs{\int_{\I_n}\Ex}}{\frac{1}{4\pi^2}V\lr{\I_n}} \nonumber\\
    & = -\frac{\kappa\immirzi}{2}k_0 \tau_3 \epsilon \frac{\lrabs{\Ex } \lrabs{\Ex}}{\sqrt{\det E}}(\theta_n) +O(\epsilon^2). \label{eq:PBthetaHolV}
\end{align}
Note that we differ here from~\cite{Banerjee:quant} by a factor of $4\pi^2$, while it is in line with~\cite{Bojowald:2005cb}.
The formula above already provides $\Ex$ and $\Ey$ for the RHS of \eqref{eq:EuclidPartIdentity1}. Next, we apply the approximation of small $X, Y$ and $\int_{\I_n} \A$ to the sine and cosine formulation of the holonomies \eqref{eq:thetaSU2Holonomy}, \eqref{eq:xSU2Holonomy} and \eqref{eq:ySU2Holonomy}:
\begin{align}
    h_\theta\lr{\I_n} & = 1 + \tau_3 k_0 \int_{\I_n} \A +O(\epsilon^2)= 1 + \tau_3 k_0 \epsilon \A\lr{\theta_n} +O(\epsilon^2),\\
    h_x\lr{\theta_n} & = 1 + \tau_x(\theta_n) \mu_0 X\lr{\theta_n}+O(\mu_0^2) \ \text{ and } \\
    h_y\lr{\theta_n} & = 1 + \tau_y(\theta_n) \nu_0 Y\lr{\theta_n}+O(\nu_0^2) .
\end{align}
With this, we get
\begin{align}
    h_x h_y h_x\inv h_y\inv - h_y h_x h_y\inv h_x\inv = 2\tau_3 \mu_0 \nu_0 X\lr{\theta_n} Y\lr{\theta_n}+O(\mu_0^2,\nu^2_0), \label{eq:HolXHolY}
\end{align}
where $O(\mu_0^2,\nu^2_0)$ means terms that involve at least second powers of either $\mu_0$ and/or $\nu_0$ and taking the SU(2)-trace of this expression multiplied by \eqref{eq:PBthetaHolV} yields the result of \eqref{eq:EuclidPartIdentity1}. Note that it sufficed to expand the holonomies' trigonometric functions up to first order due to the multiplicative and subtractive structure of \eqref{eq:HolXHolY}'s LHS. Proceeding to the second order in the cosines only yields precisely these terms multiplied by the remaining holonomies' zeroth order terms as second order contribution. But these are then cancelled by the difference of the two products and hence there is no second order contribution other than the one above. Hence, the regularised expression for the first contribution denoted by $\euclH^{(1),\epsilon}$ is given by
\begin{align}
\euclH^{(1),\epsilon}  &=-\frac{2}{\kappa\kappa'\immirzi^3 k_0 \mu_0 \nu_0 }  \sum_{\I_n\in\PT}
\tr\lr{\lr{ h_x h_y h_x\inv h_y\inv - h_y h_x h_y\inv h_x\inv } h_\theta\left\{ h_\theta\inv,V(\I_n) \right\} }.
\end{align}
The corresponding operator $\hat{\rm H}_{\rm eucl}^{(1)}$ is obtained in the limit where the regulator is removed and where we also take into account that we can define the operator separately for each graph Hilbert space ${\cal H}_\graph$, yielding
\begin{align}
    \hat{\rm H}_{\rm eucl}^{(1)}&=
    \lim\limits_{\epsilon\to 0}\hat{\rm H}_{\rm eucl}^{(1),\epsilon} 
    = \lim\limits_{\epsilon\to 0}\bigoplus\limits_{\graph} \hat{\rm H}_{\rm eucl,\graph}^{(1),\epsilon} = \bigoplus\limits_{\graph} \hat{\rm H}_{\rm eucl,\graph}^{(1)},
\end{align}
with 
\begin{align}
    \hat{\rm H}_{\rm eucl,\graph}^{(1)}&\coloneqq 
    \frac{2\i\cdot4\pi^2}{\lp{}^2 \kappa k_0 \mu_0 \nu_0 \immirzi^3} \sum_{v\in V(\graph)} \tr\left(\lr{ \holOp_x \holOp_y \holOp_x\inv \holOp_y\inv - \holOp_y \holOp_x \holOp_y\inv \holOp_x\inv } \holOp_\theta \left[ \holOp_\theta\inv , \V_v \right] \right),\label{eq:euclHhatFirst}\nonumber\\
\end{align}
where the operator only acts on vertices due to the fact that the volume operator is involved --- $\V_v$ denotes the volume operator at vertex $v$ as given in \eqref{eq:VvOperator}.

Continuing with the remaining two terms of $\euclHam^{(2)}$ and $\euclHam^{(3)}$ in \eqref{eq:euclHam2} and \eqref{eq:euclHam3} respectively, we first of all state the corresponding Thiemann identities
\begin{align}
    h_x \pb{h_x\inv}{V(\I_n)} & = -\frac{\kappa\immirzi}{2} \mu_0 \tau_x(\theta_n) \frac{\E \lrabs{\Ey}}{\sqrt{\det E}}(\theta_n) +O(\mu_0^2)\quad\text{and} \label{eq:PBxHolV} \\
    h_y \pb{h_y\inv}{V(\I_n)} & = -\frac{\kappa\immirzi}{2} \nu_0 \tau_y(\theta_n) \frac{\E \lrabs{\Ex}}{\sqrt{\det E}}(\theta_n)+O(\nu_0^2), \label{eq:PByHolV}
\end{align}
which again constitute one part of the terms' regularisation. In analogy to \eqref{eq:HolXHolY}, we then find
\begin{align}
        h_\theta h_x h_\theta\inv h_x\inv - h_x h_\theta h_x\inv h_\theta\inv & = 2 k_0 \mu_0 \epsilon\tau_y(\theta_n) \A(\theta_n) X(\theta_n)+O(\epsilon^2,\mu_0^2)  \label{eq:HolXHolTheta} \quad\text{and}\\
        h_y h_\theta h_y\inv h_\theta\inv - h_\theta h_y h_\theta\inv h_y\inv & = 2 k_0 \nu_0 \epsilon \tau_x(\theta_n)\A)\theta_n Y(\theta_n0)+O(\epsilon^2,\nu_0^2) \label{eq:HolYHolTheta} .
\end{align}

Combining \eqref{eq:PBxHolV} with \eqref{eq:HolYHolTheta} and \eqref{eq:PByHolV} with \eqref{eq:HolXHolTheta}, we obtain
\begin{align}
    \frac{1}{4\pi^2}\tr\lr{\lr{ h_y h_\theta h_y\inv h_\theta\inv - h_\theta h_y h_\theta\inv h_y\inv } h_x\left\{ h_x\inv,V(\I_n) \right\} } &\nonumber\\
    = \frac{\kappa'\immirzi}{2} k_0\mu_0\nu_0 \epsilon \cdot \frac{\A Y \E \Ey}{\sqrt{\det E}}\lr{\theta_n} +O(\epsilon^2,\mu_0^2,\nu_0^2)\quad\text{and} \label{eq:EuclidPartIdentity2alt}& \\
    \frac{1}{4\pi^2}\tr\lr{\lr{ h_\theta h_x h_\theta\inv h_x\inv - h_x h_\theta h_x\inv h_\theta\inv } h_y\left\{ h_y\inv,V(\I_n) \right\} } &\nonumber \\ 
    = \frac{\kappa'\immirzi}{2} k_0\mu_0\nu_0 \epsilon \cdot \frac{\A X \E \Ex}{\sqrt{\det E}}\lr{\theta_n}+O(\epsilon^2,\mu_0^2,\nu^2_0).& \label{eq:EuclidPartIdentity3alt}
\end{align}

However, to stick closer to \cite{Banerjee:quant,MA:Boehm,MA:Alex,MA:Andreas}, we will use slightly different expressions. This is due to our choice of $\eta \approx 0$, which is not considered amongst the literature. Without fixing $\eta$, \eqref{eq:euclHam2} and \eqref{eq:euclHam3} are modified according to $\A \mapsto \A+\pt \eta$ and in order to be able to regularise the involved derivatives of $\eta$ one has to work with shifted holonomies of the form  $h_{x,\epsilon} = h_x\lr{\theta_{n}+\epsilon}$.\footnote{Note that the formulae involving shifted holonomies are furthermore geometrically motivated. They approximate the corresponding curvature within the loops described by the holonomies. Hence the evaluation on $\theta$ or $\theta+\epsilon$ --- depending on whether one travelled in $\theta$-direction before or not.} However, the latter create new vertices in a given graph and thus cannot be used if we require the final physical Hamiltonian operator to be graph-preserving. As discussed above, in the Gaussian dust model this requirement is dictated by the classical symmetries of the physical Hamiltonian that we would like to implement also in the quantum model. Hence, if we aim at regularising in terms of shifted holonomies as well, then we need to consider a shift to the next vertex, that is  $h_{x,\xi} = h_x\lr{\theta_{n+\xi}}$ where $\xi$ can be chosen to be $\xi=\pm 1$ depending on whether the shift goes into the left or right direction from $\theta_n$. That we only involve the next-neighbouring vertices corresponds to an analogue choice of a minimal loop that carries over to the choice of a minimal shift here. Now, in the former case where the shift involved the regularisation parameter $\epsilon$, it was ensured that in the limit where we send the regulator to zero the size of the shift can be assumed to be very tiny. This is no longer given if we associated the shift with the two neighbouring points that will be identified with the corresponding vertices of a given graph in the quantum theory. Then only for those graphs where the edge length between two neighbouring vertices can be assumed to be tiny will the regularised expression yield a good approximation of the corresponding classical expression. Note that this causes no severe issue here because we will follow the same strategy as used in the AQG framework \cite{AQG1}, although in a slightly different context. For the quantisation of the Euclidean part of the physical Hamiltonian part we do not require that the regularised expression reproduces the classical expression directly when we send the regulator to zero. Instead, we call an operator suitably quantised if for a chosen set of semiclassical states the corresponding expectation values reproduce in lowest order the correct classical expression. To judge this in detail, one needs to perform a semiclassical analysis of the relevant operators. However, even if we do not perform a detailed semiclassical computation here, using the existing results in \cite{Thiemann:2000bw,Thiemann:2000bx,Thiemann:2000ca} as well as \cite{Giesel:2021yop,Giesel:2020jkz} we can already draw some conclusions here if we restrict our discussions to the lowest order only.  A suitable choice of coherent states that we can consider here for each classical canonical pair are U$(1)$ complexifier coherent states that were introduced in \cite{Thiemann:2000bw}. Their expectation values as well as their peakedness property have been analysed in \cite{Thiemann:2000bx,Thiemann:2000ca}. From these results we know that in the lowest order of the semiclassical parameter, corresponding to the classical limit where $\hbar$ is sent to zero, the expectation value of the holonomy operator agrees with the classical holonomy and the same holds also for point holonomies. Furthermore, using the results of \cite{Giesel:2020jkz,Giesel:2021yop}, we also know that the expectation values of the operator  $\hat{h}[ \hat{h}\inv,\hat{V}]$ agrees in the lowest non-vanishing order of the semiclassical parameter with its classical counterpart. This motivates to define for the two remaining parts of the Euclidean physical Hamiltonian the following operators:
\begin{align}
    \hat{\rm H}_{\rm eucl}^{(2)} &= \bigoplus\limits_\graph  \hat{\rm H}_{\rm eucl,\graph}^{(2)}\quad{\rm and}\quad 
    \hat{\rm H}_{\rm eucl}^{(3)} = \bigoplus\limits_\graph  \hat{\rm H}_{\rm eucl,\graph}^{(3)},
\end{align}    
with     
\begin{align}
    \hat{\rm H}_{\rm eucl,\graph}^{(2)} &=    
    \frac{2\i\cdot4\pi^2}{\lp{}^2 \kappa k_0 \mu_0 \nu_0 \immirzi^3}\frac{1}{2} \sum_{\substack{v\in V(\graph) \\\xi=\pm 1}} \tr  \left( 
     \lr{ \holOp_y \holOp_\theta \holOp_{y,\xi}\inv \holOp_\theta\inv - \holOp_\theta \holOp_{y,\xi} \holOp_\theta\inv \holOp_y\inv } \holOp_x \left[ \holOp_x\inv , \V_v \right] \right)  \label{eq:euclHhatSecond}
\end{align}
and 
\begin{align}
    \hat{\rm H}_{\rm eucl,\graph}^{(3)} &=      
    \frac{2\i\cdot4\pi^2}{\lp{}^2 \kappa k_0 \mu_0 \nu_0 \immirzi^3}\frac{1}{2} \sum_{\substack{v\in V(\graph) \\\xi=\pm 1}} \tr  \left(  \lr{ \holOp_\theta \holOp_{x,\xi} \holOp_\theta\inv \holOp_x\inv - \holOp_x \holOp_\theta \holOp_{x,\xi}\inv \holOp_\theta\inv } \holOp_y \left[ \holOp_y\inv , \V_v \right] 
      \right)  \label{eq:euclHhatThird},
\end{align}
where again the sum runs over all vertices $v$, we used $\kappa'\hbar = \lp{}^2$ and we included an additional factor of $\frac{1}{2}$ because we considered $\xi=\pm 1$. Further, $\hat{h}_{x,\xi} = \hat{h}_x\lr{v_\xi}$ and $\hat{h}_{y, \xi} = \hat{h}_y\lr{v_\xi}$ where $v_+$ and $v_-$ denote the neighbouring vertices of $v$ to the right and left, respectively. 
Taking into account that these coherent states satisfy a resolution of identity together with their peakedness property \cite{Thiemann:2000bx,Thiemann:2000ca} as well as the results of semiclassical expectation values for U(1) coherent states of square root operators in terms of Kummer functions \cite{Giesel:2020jkz,Giesel:2021yop}, we can conclude that in the lowest order of the semiclassical parameter the operators $\hat{\rm H}_{\rm eucl}^{(2)}$ and $\hat{\rm H}_{\rm eucl}^{(3)}$ will reproduce the correct classical limit, that is
\begin{align*}
\langle & \Psi^t_{(\A,\E,X,\Ex,Y,\Ey)}\, | \, \hat{\rm H}_{\rm eucl}^{(I=2,3)}\, |\, \Psi^t_{(\A,\E,X,\Ex,Y,\Ey)}\rangle = \\
& =4\pi^2\int_{\Sd} \d\theta\, \euclHam^{(I)}\lr{\A(\theta),\E(\theta),X(\theta),\Ex(\theta),Y(\theta),\Ey(\theta)} +O(t,\epsilon,\mu_0,\nu_0),
\end{align*}
where one needs to choose a suitable set of coherent states such that the associated embedded graphs $\Gamma$ in $\Sd$ involved in the definition of $\Psi^t_{(\A,\E,X,\Ex,Y,\Ey)}$ approximate $\Sd$ well enough when the sum over all vertices of the graphs is considered. We would like to emphasise that we can use former results on semiclassical computations here only because we quantised the physical Hamiltonian in a graph-preserving manner --- suitable semiclassical states for graph-modifying operators in LQG are still an open and difficult question. Note that because we do not perform a detailed semiclassical computation here, we cannot make any statement about the terms involved in higher orders than the lowest order one. These can only be determined by computing the semiclassical expectation value in detail which, however, will not be part of this work.
As discussed above, working with shifted holonomies is motivated by the fact that one needs to regularise the derivative of $\eta$ being involved when the Gau\ss{} constraint is not solved at the classical level already. Since these formulae are still correct for $\eta \approx 0$, we will use these from now on, too, and thus enabling an easier transition between the two approaches.
Note that for $\eta \approx 0$ we have $\tau_x(\theta)=\tau_1$ and $\tau_y(\theta)=\tau_2$ because the $\theta$-dependent coefficients in \eqref{eq:tauxtauy} are either zero or one. Furthermore, this choice ensures that the operator obtained from following a Dirac quantisation procedure for the Gau\ss{} constraint and the one from the reduced quantisation considered here have the same regularisation. Because in the reduced case we could also choose a regularisation where the holonomy is located at the same vertices for all involved holonomies, we realise that such a choice is another example where Dirac and reduced quantisation would not yield the same final form of the operator similar to the situation discussed in \cite{Giesel:2016gxq}, although the latter shows a stronger difference between the two cases.

Altogether, this results in
\begin{align}
\label{eq:euclHhatAll}
    \euclHhat_{,\graph} = \frac{8\i\pi^2}{\lp{}^2 \kappa k_0 \mu_0 \nu_0 \immirzi^3} & \sum_{v\in V(\graph)} {\cal P}_\graph\tr  \left(  \lr{ \holOp_x \holOp_y \holOp_x\inv \holOp_y\inv - \holOp_y \holOp_x \holOp_y\inv \holOp_x\inv } \holOp_\theta \left[ \holOp_\theta\inv , \V_v \right] \right.  \nonumber  \\
    &  +\frac{1}{2}\sum\limits_{\xi=\pm 1}\lr{ \holOp_y \holOp_\theta \holOp_{y,\xi}\inv \holOp_\theta\inv - \holOp_\theta \holOp_{y,\xi} \holOp_\theta\inv \holOp_y\inv } \holOp_x \left[ \holOp_x\inv , \V_v \right]  \nonumber\\
    &  +\frac{1}{2}\sum\limits_{\xi=\pm 1} \left. \lr{ \holOp_\theta \holOp_{x,\xi} \holOp_\theta\inv \holOp_x\inv - \holOp_x \holOp_\theta \holOp_{x,\xi}\inv \holOp_\theta\inv } \holOp_y \left[ \holOp_y\inv , \V_v \right] \right){\cal P}_\graph , 
\end{align}
where we introduced in addition ${\cal P}_\graph:{\cal H}_{\rm phys}\to {\cal H}_{\graph}$, which are orthogonal projections that ensure that the operator is indeed graph-preserving if for instance two holonomies along a given edge combine to the identity. As before, $\V_v$ denotes the volume operator at vertex $v$, \eqref{eq:VvOperator}.
This finishes our discussion on the graph-preserving quantisation of the Euclidean part of the  physical Hamiltonian.
\subsubsection{Quantisation of the Lorentzian part of the physical Hamiltonian}
\label{sec:QuantLorPar}

Turning our attention to the quantisation of $\lorH$, we start again with the first part. According to \eqref{eq:lorHam1} and choosing again a partition of $\Sd$ such that $\Sd=\cup_n\I_n$ with $\I_n:[\theta_n-\frac{\epsilon}{2},\theta_n+\frac{\epsilon}{2}]$, we have
\begin{align}
    \lorH^{(1)} &= -\frac{1}{4\kappa'} \int_{\Sd}\d\theta \frac{\lr{\pt\E}^2}{\sqrt{\det E}}(\theta) \nonumber\\
    &=-\frac{1}{4\kappa'}\lim\limits_{\epsilon\to 0}\sum\limits_{\I_n\in \PT} \int_{\I_n}\d\theta \frac{\lr{\pt\E}^2}{\sqrt{\det E}}(\theta) \nonumber\\
    &= -\frac{1}{4\kappa'} \lim\limits_{\epsilon\to 0}\sum\limits_{\I_n\in \PT}\frac{\lr{\epsilon\pt\E\lr{\theta_n}}^2}{\epsilon\sqrt{\det E\lr{\theta_n}}} \nonumber \\
    &= -\frac{1}{4\kappa'} \lim\limits_{\epsilon\to 0}\sum\limits_{\I_n\in \PT}\frac{\lr{ \E\lr{\theta_n + \epsilon} - \E\lr{\theta_n} }^2}{\frac{1}{4\pi^2}V\lr{\mathcal{I}_n}}. \label{eq:LorentzianStart}
\end{align}
The last step then also used \eqref{eq:infVolFunctional} for the  volume of a tiny interval $\mathcal{I}_n$ around $\theta_n$ and $\E\lr{\theta_n+\epsilon} = \E\lr{\theta_n} + \epsilon\pt\E\lr{\theta_n}+O(\epsilon^2)$ for the derivative expression. The remaining task consists in dealing with the inverse volume involved in \eqref{eq:LorentzianStart}. For this purpose, we consider the Thiemann identity in \eqref{eq:PBthetaHolV} as well as the two analogue expressions in \eqref{eq:PBxHolV} and \eqref{eq:PByHolV} and use the quantity $\Z\ofI$ introduced already in \cite{Banerjee:quant} to obtain
\begin{align}
    \Z\ofIn & \coloneqq \epsilon^{abc} \tr\lr{ h_a \pb{h_a\inv}{V\ofIn} h_b\pb{h_b\inv}{V\ofIn} h_c\pb{h_c\inv}{V\ofIn}}\nonumber \\
    & = \frac{3}{2} \lr{\frac{\kappa\immirzi}{2}}^3 k_0\mu_0\nu_0 V\ofIn +O(\mu_0^2,\nu_0^2,\epsilon^2).
\end{align}
Following \cite{Banerjee:quant}, we can then derive
\begin{align}
    \Za\ofIn & \coloneqq \epsilon^{abc} \tr\lr{ h_a \pb{h_a\inv}{V^r\ofIn} h_b\pb{h_b\inv}{V^r\ofIn} h_c\pb{h_c\inv}{V^r\ofI}} \nonumber\\
    & = \frac{3}{2} \lr{\frac{\kappa\immirzi}{2}}^3 r^3 k_0\mu_0\nu_0 V^{3r-2}\ofIn +O(\mu_0^2,\nu_0^2,\epsilon^2) \nonumber \\
    &= r^3 V^{3r-3}\ofIn \Z\ofIn+O(\mu_0^2,\nu_0^2,\epsilon^2), \label{eq:defZalpha}
\end{align}
which allows us to introduce the following decomposition of unity:
\begin{align}
    \lr{1}^l = \lr{ \frac{16}{3\lr{\kappa\immirzi}^3 k_0\mu_0\nu_0} \frac{\Z\ofIn}{V\ofIn} }^l = \lr{\frac{16}{3\lr{\kappa\immirzi}^3 k_0\mu_0\nu_0}}^l \lr{\frac{\Za\ofI}{r^3 V^{3r-2}\ofIn}}^l, \label{eq:UnityDecomp}
\end{align}
with $l \in\mathbbm{R}$.
We can now use this identity to eliminate the inverse volume in \eqref{eq:LorentzianStart}. Setting
\begin{equation}
    \lr{3r-2}l = -1 \quad \Rightarrow \quad r=\frac{2}{3}-\frac{1}{3l}
\end{equation}
results in an inverse volume on the LHS of \eqref{eq:UnityDecomp}, which we then insert into \eqref{eq:LorentzianStart} to obtain a regularised expression $ \lorH^{(1),\epsilon}$ of the form:
\begin{equation}
    \lorH^{(1),\epsilon} \coloneqq  -\frac{4\pi^2}{4\kappa'}  \lr{ \frac{16}{3\lr{\kappa\immirzi}^3 r^3 k_0\mu_0\nu_0}}^l \sum_{\I_n\in \PT} \lr{ \E\lr{\theta_n+\epsilon} - \E\lr{\theta_n} }^2 \left. \Za{}^l\ofIn\right\rvert_{r=\frac{2}{3}-\frac{1}{3l} }.
\end{equation}
As for the Euclidean part, the  corresponding operator $\hat{\rm H}_{\rm lor}^{(1)}$ is obtained in the limit where the regulator is removed. We take again the infinite refinement limit where at most on vertex is in each $\I_n$ and where we also take into account that we can define the operator separately for each graph Hilbert space ${\cal H}_\graph$, yielding 
\begin{align}
    \hat{\rm H}_{\rm lor}^{(1)}&=
    \lim\limits_{\epsilon\to 0}\hat{\rm H}_{\rm lor}^{(1),\epsilon} 
    = \lim\limits_{\epsilon\to 0}\bigoplus\limits_{\graph} \hat{\rm H}_{\rm lor,\graph}^{(1),\epsilon} = \bigoplus\limits_{\graph} \hat{\rm H}_{\rm lor,\graph}^{(1)}
\end{align}
with
\begin{align}
\hat{\rm H}_{\rm lor,\graph}^{(1)} |\graph,k,\mu,\nu\rangle =&    
-\frac{4\pi^2}{4\kappa'} \lr{\frac{-\i}{\hbar}}^{3l} \lr{ \frac{16}{3\lr{\kappa\immirzi}^3 r^3 k_0\mu_0\nu_0}}^l \cdot \nonumber\\
&\cdot\sum_{v \in V(\graph)} \lr{\lr{k_{\evp} -k_{\evm}}}^2 \left.\ZaHatI{v}{}^l\right\rvert_{r=\frac{2}{3}-\frac{1}{3l} }  |\graph,k,\mu,\nu\rangle ,
\end{align}
where $k_{\evp}$ is the label attached to the outgoing edge of the vertex $\vp$ and $k_{\evm}$ is the label attached to the edge incoming at the vertex $v$ --- i.e.~outgoing from vertex $\vm$. Furthermore, we used that the operator $\ZaHatI{v}{}$ \cite{Banerjee:quant} given by
\begin{align}
   \ZaHatI{v}{}& \coloneqq 
  \epsilon^{abc} \tr\lr{ \holOp_a \comm{\holOp_a\inv}{\V_v^r} \holOp_b \comm{\holOp_b\inv}{\V_v^r} \holOp_c \comm{\holOp_c\inv}{\V_v^r}} \label{eq:ZhatLQG}
\end{align}
does not change the labels of the state $ |\graph,k,\mu,\nu\rangle$ that it acts on.

In similar ways, we obtain the respective expressions for the second and third Lorentzian part:
\begin{align}
\hat{\rm H}_{\rm lor,\graph}^{(2)} |\graph,k,\mu,\nu\rangle =&    
\frac{4\pi^2}{4\kappa'} \lr{\frac{-\i}{\hbar}}^{3l} \lr{ \frac{16}{3\lr{\kappa\immirzi}^3 r_2{}^3 k_0\mu_0\nu_0}}^l \nonumber\\
&\sum_{v \in V(\graph)} \lr{k_{\ev}+ k_{\evm}}^4 \lr{\mu_{v}\nu_{v_{+}}-\nu_{v}\mu_{v_{+}}}^2 \left.\ZbHatI{v}{}^l\right\rvert_{r_2=\frac{2}{3}-\frac{5}{3l} }  |\graph,k,\mu,\nu\rangle \\
\hat{\rm H}_{\rm lor,\graph}^{(3)} |\graph,k,\mu,\nu\rangle =&    
\frac{4\pi^2}{\kappa'} \lr{\frac{-\i}{\hbar}}^{3l} \lr{ \frac{16}{3\lr{\kappa\immirzi}^3 r^3 k_0\mu_0\nu_0}}^l \nonumber\\
&\sum_{v \in V(\graph)} \left( \lr{k_{\evp}+ k_{\ev}} \lr{k_{\evpp}-k_{\ev}}  \left.\ZaHatI{v_+}{}^l\right\rvert_{r=\frac{2}{3}-\frac{1}{3l} }\right. -\nonumber\\
& \qquad\quad - \lr{k_{\ev}+ k_{\evm}} \lr{ k_{\evp}-k_{\evm} } \left.\ZaHatI{v}{}^l\right\rvert_{r=\frac{2}{3}-\frac{1}{3l} } \bigg) |\graph,k,\mu,\nu\rangle
\end{align} 
where $\vpp$ denotes the subsequent vertex after $\vp$. Note that we evaluate $\hat{\operatorname{\Z}}$ within the second contribution $\hat{\rm H}_{\rm lor,\graph}^{(2)}$ on a different value of $r_2=\frac{2}{3}-\frac{5}{3l}$.

With that, the Lorentzian contribution to $\hatHphys$ reads
\begin{align}
\lorHhat&=\bigoplus\limits_{\graph}\hat{\rm H}_{\rm lor,\graph}=\bigoplus\limits_{\graph}\hat{\rm H}_{\rm lor,\graph}^{(1)}+\hat{\rm H}_{\rm lor,\graph}^{(2)}+\hat{\rm H}_{\rm lor,\graph}^{(3)}     \end{align}
and the physical Hamiltonian in the reduced LQG Gowdy model finally takes the following form:
\begin{align}
\hatHphys &=\bigoplus\limits_{\graph} \hat{\rm H}_{{\rm phys},\graph}
=\frac{1}{2}\bigoplus\limits_{\graph}\left(\euclHhat_{,\graph}+(\euclHhat_{,\graph})^\dagger+
\hat{\rm H}_{\rm lor,\graph}+(\hat{\rm H}_{\rm lor,\graph})^\dagger\right). \label{eq:FinalHphysLQG}
\end{align}
This finishes the discussion on the quantisation of the physical Hamiltonian of the Gowdy model in the reduced LQG framework.

\section{Quantisation of the model with polarised \texorpdfstring{$\mathbbm{T}^3$}{T3} Gowdy symmetry within Algebraic Quantum Gravity}
\label{Sec:AQGQuant}

As discussed in the previous Section \ref{sec:QuantRedLQG}, we needed to quantise the physical Hamiltonian in a graph-preserving way in order to carry over its classical symmetries to the quantum theory. In the case of reduced LQG, this corresponds to a quantisation that preserves each graph Hilbert space ${\cal H}_\gamma$ separately, yielding infinitely many conserved charges in the quantum theory that are absent in the classical theory. An alternative framework for the quantisation of these kind of operators where the graph-preserving feature of operators are implemented in a slightly different context is the Algebraic Quantum Gravity (AQG) approach introduced in \cite{AQG1} and combined with a reduced phase space quantisation for full LQG in \cite{AQG4}.

\subsection{The physical Hilbert space in AQG}
\label{sec:PhysHSAQG}
\begin{figure}
\begin{tikzpicture}
\def \n {5}
\def \firstvertex {-1.5}
\def \lastvertex {12.5}
\def \linelength {\lastvertex-\firstvertex}
\def \smallradius {0.2}
        \draw[densely dotted] (-2,0) -- (\firstvertex,0); 
        \draw[densely dotted] (\lastvertex,0) -- (13,0); 
        \fill[black!40] (\firstvertex,0) circle (2pt); 
        \fill[black!40] (\lastvertex,0) circle (2pt); 
        \draw[-latex] (\firstvertex,0) -- ({\firstvertex+(\lastvertex-\firstvertex)/(\n+1)},0) node[pos=0.5,below]{$k_{\n}$}; 
\foreach \s in {1,...,\n}
{
    \node at ({\firstvertex+(\lastvertex-\firstvertex)/(\n+1)+(\s-1)/(\n+1)*(\lastvertex-\firstvertex)},-0.3) {$v_{\s}$}; 
    \fill ({\firstvertex+(\lastvertex-\firstvertex)/(\n+1)+(\s-1)/(\n+1)*(\lastvertex-\firstvertex)},0) circle (2pt); 
    \draw[-latex] ({\firstvertex+(\s)*(\lastvertex-\firstvertex)/(\n+1)},0) -- ({\firstvertex+(\s+1)*(\lastvertex-\firstvertex)/(\n+1)},0) node[pos=0.5,above]{$k_{\s}$} node[pos=0.5,below]{$e_{\s}$}; 
    \draw [dash pattern=on 3pt off 1pt, black, rotate around x=-90, rotate around={-112.5:({\firstvertex+(\lastvertex-\firstvertex)/(\n+1)+(\s-1)/(\n+1)*(\lastvertex-\firstvertex)},0)}] ({\firstvertex+(\lastvertex-\firstvertex)/(\n+1)+(\s-1)/(\n+1)*(\lastvertex-\firstvertex)},0) arc (0:360:\smallradius) node[pos=0.65,right]{$\mu_{\s}$}; 
    \draw [dash pattern=on 3pt off 1pt, black!70, rotate around x=45, rotate around={-67.5:({\firstvertex+(\lastvertex-\firstvertex)/(\n+1)+(\s-1)/(\n+1)*(\lastvertex-\firstvertex)},0)}] ({\firstvertex+(\lastvertex-\firstvertex)/(\n+1)+(\s-1)/(\n+1)*(\lastvertex-\firstvertex)},0) arc (0:360:\smallradius) node[pos=0.35,left]{$\nu_{\s}$}; 
}
\end{tikzpicture}
    \caption{An abstract infinite AQG-graph, corresponding to the embedded one of Figure~\ref{fig:embeddedGraph}, with five vertices on which two point holonomies sit and six charged edges, where the most leftward one --- the copy of $k_5$ --- ensures the state's periodicity. Here, to keep the notation more compact, we used $k_{e_{v_I}}\eqqcolon k_I, \mu_{v_I}\eqqcolon\mu_I$ and $\nu_{v_I}\eqqcolon\nu_I$.}
    \label{fig:AQGgraph}
\end{figure}
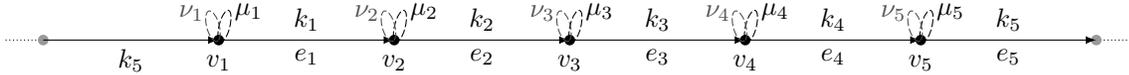

Here, we want to follow this quantisation approach in the symmetry reduced case of Gowdy models. One of the main difference is that AQG considers only one abstract infinite graph $\aqggraph$, whereas we had to include infinitely many finite embedded graphs $\graph$ for reduced LQG. The underlying Hilbert space in AQG is von-Neumann's infinite tensor product Hilbert space denoted by ${\cal H}_{\rm ITP}$ so that the physical Hilbert space in the AQG framework is ${\cal H}_{\rm phys} = {\cal H}_{\rm ITP}$. The topology of the abstract graph is chosen to define the corresponding AQG model and here this means that for the analogue of a Gowdy state's LQG-graph like the one of Figure~\ref{fig:embeddedGraph}, we need to rearrange it to a line with the same number of charged vertices $v_I$, see Figure~\ref{fig:AQGgraph}. To the right of every vertex $v$, the outgoing edge $\ev$ is attached and hence, the respective incoming edge at vertex $v$ is $\evm$. The charges $k_{\ev}$, $\mu_v$ and $\nu_v$ are then assigned correspondingly, with the only non-straightforward assignment being the one of edge $e_{v_0}$, incoming at $v_1$, which we have to charge with $k_{e_{v_N}}$, the charge at the edge $e_{v_N}$, in order to preserve the cyclic structure (cf. Fig.~\ref{fig:embeddedGraph}). All other edges and vertices are trivially charged and therefore do not contribute when operators such as the fluxes act on them. Note that Figure~\ref{fig:AQGgraph} shows again dashed miniature loops at the charged vertices representing the point holonomies in order to emphasise that they are not in fact holonomies along edges.

In general, dynamical operators can be carried over from the embedded LQG framework to AQG if they are spatially diffeomorphism invariant. In the case of the Gowdy model, this corresponds to operators that involve integrals over the dust manifold $\Sd$. 
All operators --- including the physical Hamiltonian --- will be implemented graph-preserving by construction, but one of the differences to the reduced LQG case is that in AQG we allow trivial representations on the edges of the infinite abstract graph $\aqggraph$.

Following the approach in \cite{AQG1}, the holonomies, states and fluxes are introduced as follows. To each edge of the abstract infinite graph $\aqggraph$ we associate an  U(1) element similar to the holonomy in~\eqref{eq:holonomy} for reduced LQG, but with the difference that here we do not express the U(1) element in terms of an integral along the edges but we associate to a given edge $e$ of the abstract graph $\aqggraph$ the U(1) element 
\begin{align}
    {h}^{(k_{e})}_{e}\lr{\A}&\coloneqq \ex{\i\frac{k_e}{2}\A_{e}}
    ,\quad{\rm with}\  \A_e\in\mathbbm{R} . \label{eq:AQGAholonomy}
\end{align}
Given this, we can define the analogue of the basis state in LQG  shown in~\eqref{eq:state} now in the AQG framework  as
\begin{align}
    |\aqggraph,k,\mu,\nu\rangle &\coloneqq \prod_{e\in E\lr{\aqggraph}} \ex{\i\frac{k_e}{2}\A_e} \prod_{v\in V\lr{\aqggraph}} \ex{\i\frac{\mu_v}{2}X_v} \ex{\i\frac{\nu_v}{2}Y_v}, \label{eq:AQGstate}
\end{align}
where for the point holonomies we introduced the notation $X_v \coloneqq X\lr{v}, Y_v \coloneqq Y\lr{v}$. The holonomy operators act on the AQG basis states $|\aqggraph,k,\mu,\nu\rangle$ in the following way:
\begin{align}
\hat{h}^{(k_{0})}_{e_I}\lr{\A}|\aqggraph,k,\mu,\nu\rangle &=
\ex{\i\frac{k_{0}}{2}\A_{e_I}} |\aqggraph,k,\mu,\nu\rangle 
 = |k_{e_I}+k_0,\mu,\nu\rangle,\quad k_0\in\mathbb{Z} \label{eq:actionholtheta} \\
\hat{h}^{(\mu_{0})}_{v_I} \lr{X} |\aqggraph,k,\mu,\nu\rangle&=  \ex{\i\frac{\mu_0}{2}X_{v_I}} |\aqggraph,k,\mu,\nu\rangle = |\aqggraph,k,\mu_{v_I}+\mu_0,\nu\rangle,\quad\mu_0\in\mathbb{R} \label{eq:actionholx}\\
  \hat{h}^{(\nu_{0})}_{v_I} \lr{Y}|\aqggraph,k,\mu,\nu\rangle &=   \ex{\i\frac{\nu_0}{2}Y_{v_I}} |\aqggraph,k,\mu,\nu\rangle =|\aqggraph,k,\mu,\nu_{v_I}+\nu_0\rangle,\quad\nu_0\in\mathbb{R}. \label{eq:actionholy}\end{align}
Because in the AQG model there is only one abstract graph $\aqggraph$, we will from now on neglect the label for the graph and denote the basis states just by $|k,\mu,\nu_v\rangle$. Note that we can recover the classical expression for the U(1) holonomy from the operator $\hat{h}^{(k_{0})}_{e_I}\lr{\A}$ by considering semiclassical states that encode in addition to their classical labels in the AQG framework also information about how the abstract graph $\aqggraph$ is embedded into a given spatial manifold from which an integral along the embedded edges involved in the classical holonomy can be rediscovered, see the results in \cite{AQG2,AQG3} for the case of the master constraint operator.

The main difference for the fluxes, in turn, is that they can now only act on vertices as there are no embedded edges at hand anymore. Therefore, the elementary flux operator within AQG read
\begin{align}
    \hat{\E}_v |k,\mu,\nu\rangle &= \frac{\immirzi\lp{}^2}{2}\frac{k_{\ev}+k_{\evm}}{2} |k,\mu,\nu\rangle, \label{eq:actionfluxtheta}\\
    \Fx{\I_v} |k,\mu,\nu\rangle &= \frac{\immirzi\lp{}^2}{2}  \mu_v |k,\mu,\nu\rangle \quad \text{and} \label{eq:actionfluxx}\\
    \Fy{\I_v} |k,\mu,\nu\rangle &= \frac{\immirzi\lp{}^2}{2} \nu_v |k,\mu,\nu\rangle, \label{eq:actionfluxy}
\end{align}
where $k_{\ev}$ is the at vertex $v$ outgoing edge's U(1)-charge and $k_{\evm}$ the incoming one's. Note that this trivial continuation from LQG to AQG is possible by choosing the occurring smearing intervals in such a way that they contain one vertex at most: the interval $\I_v$ of~\eqref{eq:actionfluxx} and~\eqref{eq:actionfluxy} includes solely vertex $v$.

The volume operator can be transferred to AQG as straightforwardly as for the basic operators themselves and we get
\begin{align}
    \V &\coloneqq \sum_{v\in V(\aqggraph)} \V_v \coloneqq 4\pi^2 \sum_{v\in V(\aqggraph)} \sqrt{\lrabs{\hat{\E}_v} \lrabs{\hat{\mathcal{F}}_{x,I_v}} \lrabs{\hat{\mathcal{F}}_{y,I_v}} }, \label{eq:VolAQG}
\end{align}
with its action on the AQG states~\eqref{eq:AQGstate}
\begin{align}
    \V_v|k,\mu,\nu\rangle = \frac{4\pi^2}{\sqrt{2}}\lr{\frac{\immirzi\lp{}^2}{2}}^{\frac{3}{2}} \sqrt{\lrabs{k_{\ev}+k_{\evm}} \lrabs{\mu_v} \lrabs{\nu_v}}\state{}{}{}. \label{eq:VolAQGaction}
\end{align}

To complete the discussion on the Gauß constraint, we also briefly present how to implement a Gauß constraint operator in AQG. This can be done quite directly as well by considering the LQG Gauß constraint \eqref{eq:LQGGaußCstrt} that had the form
\begin{align}
    \widehat{\Gau} & = \frac{1}{\kappa' \immirzi}\lim\limits_{\epsilon\to 0} \sum\limits_{\I_n\in\PT} \lr{ \hat\E\lr{\theta_n+\frac{\epsilon}{2}} - \hat\E\lr{\theta_n-\frac{\epsilon}{2}} + \hatFeta{\I_n} },\quad{\rm with}\ \theta_n\in\I_n. \tag{\ref{eq:LQGGaußCstrt}}
\end{align}
In AQG, we need to implement the operator such that it acts on the vertices of the abstract graph $\aqggraph$ only and thus we associate $\theta_n-\frac{\epsilon}{2}$ with vertex $\vm$, the left neighbouring vertex of $v$, and, accordingly, $\theta_n + \frac{\epsilon}{2}$ with vertex $\vp$, the right neighbouring vertex of the vertex $v$. This way, we obtain the following AQG version of the Gau\ss{} constraint operator:
\begin{align}
    \widehat{\Gau}|k,\mu,\nu\rangle &= \frac{1}{\kappa' \immirzi} \sum\limits_{v\in V(\aqggraph)} \lr{ \hat\E_{\vp} - \hat\E_{\vm} + \hatFeta{\I_v} }|k,\mu,\nu\rangle\nonumber \\
   & =\frac{\lp{}^2}{\kappa'} \sum\limits_{v\in V(\aqggraph)} \lr{ \frac{k_{\evp}+k_{\ev}-k_{\evm}-k_{\evmm}}{4} + \lambda_v }|k,\mu,\nu\rangle,
\end{align}
where $\vmm$ denotes the second vertex to the left of $v$ and  we also had to insert the AQG version of the flux conjugate to $\eta$:
\begin{align}
    \Feta{e_v} |k,\mu,\nu,\lambda\rangle = \immirzi\lp{}^{2} \lambda_v |k,\mu,\nu,\lambda\rangle.
\end{align}
In contrast to the interval $\I_n$ in LQG before, the edge $e_v$ in AQG can only contain one vertex at most, so there is only one contribution within the action of $\Feta{e_v}$. The action of the Gauß constraint on the basis states $|k,\mu,\nu,\lambda\rangle$, in turn, does differ more from its LQG counterpart because for the part involving the flux operators $\hat\E$  here the two neighbouring vertices of $v$ are involved. The solution of the AQG Gauss{} constraint being the equivalent to \eqref{eq:GaußSolution} in LQG then reads
\begin{align}
    \lambda_v = - \frac{k_{\evp}+k_{\ev}-k_{\evm}-k_{\evmm}}{4} \ , \quad \forall\,v.
\end{align}
We see that solving $\lambda_v$ to obtain solutions to the Gau\ss{} constraint does also constrain $k_{\ev}$. In contrast to the LQG case where only the charges of the neighbouring edges of $v$ were involved, we now have a condition depending on the two after next neighbouring ones as well. And without contributions from same edges adding up, as it was the case in LQG, the denominator remains to be 4. This finishes the considerations on the Gauß constraint and we close with the remark that for the work at hand, the Gauß constraint will be solved on the classical level.

\subsection{Dynamics of the model with polarised \texorpdfstring{$\mathbbm{T}^3$}{T3} Gowdy symmetry in AQG}
\label{sec:DynAQG}
In the following subsection, we will briefly discuss how the physical Hamiltonian operator that was so far quantised in reduced LQG can be implemented in the AQG Gowdy quantum model.

\subsubsection{Quantisation of the Euclidean part of the physical Hamiltonian within AQG}

We can now straightforwardly transfer the Euclidean part of the physical Hamiltonian operator in \eqref{eq:euclHhatAll} to AQG by means of the previously stated AQG holonomy operators \eqref{eq:actionholtheta}, \eqref{eq:actionholx} and \eqref{eq:actionholy} as well as the volume operator \eqref{eq:VolAQG}.
For this purpose, we define the following class of operators $\OOp{\theta/x/y}{r,v}$ for $r\in\mathbbm{R}, v\in V\lr{\aqggraph}$ according to
\begin{align}
    \OOp{\theta}{r,v} & \coloneqq \cos \frac{\A_{\ev}}{2}  \V^r_v  \sin \frac{\A_{\ev}}{2} - \sin \frac{\A_{\ev}}{2}  \V^r_v  \cos \frac{\A_{\ev}}{2} , \label{eq:defOOptheta} \\
    \OOp{x}{r,v} & \coloneqq \cos \frac{X_v}{2}  \V^r_v  \sin \frac{X_v}{2} - \sin \frac{X_v}{2}  \V^r_v  \cos \frac{X_v}{2} \quad \text{and}\label{eq:defOOpx} \\
    \OOp{y}{r,v} & \coloneqq \cos \frac{Y_v}{2}  \V^r_v  \sin \frac{Y_v}{2} - \sin \frac{Y_v}{2}  \V^r_v  \cos \frac{Y_v}{2}, \label{eq:defOOpy}
\end{align}
where we used the decomposition of the holonomies into sines and cosines (\eqref{eq:thetaSU2Holonomy}, \eqref{eq:xSU2Holonomy} and \eqref{eq:ySU2Holonomy}) offering an alternative, more concise description of the final operator. Now we can substitute the LQG expression involved in the first part of the Euclidean part in \eqref{eq:euclHhatFirst}
by the following AQG analogue:
\begin{align}
    \tr \lr{  \lr{ \holOp_x \holOp_y \holOp_x\inv \holOp_y\inv - \holOp_y \holOp_x \holOp_y\inv \holOp_x\inv } \holOp_\theta \left[ \holOp_\theta\inv , \V_v \right] } &\stackrel{\text{AQG}}{\longmapsto} -2 \sin X_v \sin Y_v \OOp{\theta}{1,v}.
\end{align}
The second and third part of the Euclidean part of the physical Hamiltonian shown in \eqref{eq:euclHhatSecond} and \eqref{eq:euclHhatThird} entail similar terms with shifted holonomies. Hence, their AQG expressions are of the following form
\begin{align}
    \tr \lr{  \lr{ \holOp_y \holOp_\theta \holOp_{y,\xi}\inv \holOp_\theta\inv - \holOp_\theta \holOp_{y,\xi} \holOp_\theta\inv \holOp_y\inv } \holOp_x \left[ \holOp_x\inv , \V_v \right] } &\stackrel{\text{AQG}}{\longmapsto} -4 \sin \frac{Y_{\vxi}}{2} \cos \frac{Y_{v}}{2} \sin \A_{e_v} \OOp{x}{1,v} \\
    \tr \lr{  \lr{ \holOp_\theta \holOp_{x,\xi} \holOp_\theta\inv \holOp_x\inv - \holOp_x \holOp_\theta \holOp_{x,\xi}\inv \holOp_\theta\inv } \holOp_y \left[ \holOp_y\inv , \V_v \right] } &\stackrel{\text{AQG}}{\longmapsto} -4 \sin \frac{X_{\vxi}}{2} \cos \frac{X_{v}}{2} \sin \A_{e_v} \OOp{y}{1,v}
\end{align}
Therein, $X_{\vxi} = X\lr{\vxi}$ and $Y_{\vxi} = Y\lr{\vxi}$, where $\vxi$ is $\vp$ for $\xi=1$ and $\vm$ for $\xi=-1$ denoting the two neighbouring vertices of $v$ to the right and left respectively.

With this, we can write the Euclidean part of the Hamilton operator in AQG in a form that is more concise and allows for a more compact evaluation of the corresponding action on the Gowdy states later:
\begin{align}
    \euclHhat &= 
    \sum_{v \in V(\aqggraph)} \euclHhatI,
\end{align}
\begin{align}
    \euclHhatI &\coloneqq
     -\frac{4\i}{\kappa'\lp{}^2k_0\mu_0\nu_0\immirzi^3} \bigg[ \sin X_v \sin Y_v \OOp{\theta}{1,v}  + \label{eq:euclH} \\
    & \qquad\quad +\frac{1}{2}\sum\limits_{\xi=\pm 1}\left( 2 \sin\frac{Y_{\vxi}}{2} \cos\frac{Y_v}{2} \sin \A_{e_v} \OOp{x}{1,v} + 2 \sin\frac{X_{\vxi}}{2} \cos\frac{X_v}{2} \sin\A_{e_v} \OOp{y}{1,v} \right)\bigg].\nonumber
\end{align}

\subsubsection{Quantisation of the Lorentzian part of the physical Hamiltonian within AQG}

With all the contained quantities depending on holonomies, fluxes or the volume, we can quantise the Lorentzian part of the physical Hamiltonian that was discussed for the case of reduced LQG in \ref{sec:QuantLorPar} straightforwardly and also directly in the AQG framework. We then obtain for the first part the following expression
\begin{align}
    \lorHhatpart{1} & = 
    \sum_{v \in V(\aqggraph)} \lorHhatIpart{1}\\
\end{align}
with 
\begin{align}
    \lorHhatIpart{1}& \eqqcolon 
    -\frac{4\pi^2}{4\kappa'} \lr{\frac{-\i}{\hbar}}^{3l} \lr{ \frac{16}{3\lr{\kappa\immirzi}^3 r^3 k_0\mu_0\nu_0}}^l  \lr{ \hat{\E}_{\vp} - \hat{\E}_v }^2 \left.\ZaHatI{v}{}^l\right\rvert_{ r=\frac{2}{3}-\frac{1}{3l} }, \nonumber\\
\end{align}
where we introduced
\begin{align}
    \ZaHatI{v} \coloneqq \epsilon^{abc} \tr\lr{ \holOp_a \comm{\holOp_a\inv}{\V_v} \holOp_b \comm{\holOp_b\inv}{\V_v} \holOp_c \comm{\holOp_c\inv}{\V_v} } = -12 \OOp{x}{r,v} \OOp{y}{r,v} \OOp{\theta}{r,v} \label{eq:ZintermsofOs}
\end{align}
as the equivalent of \eqref{eq:ZhatLQG}. In the semiclassical limit here the intervals corresponding to $\I_n$ in the reduced LQG case will be vertex-labelled intervals $\I_v$. 

This procedure can now be applied to the second and third part of the Lorentzian part of the Hamiltonian, \eqref{eq:lorHam2} and \eqref{eq:lorHam3}, that also act on the vertices only and thus we just present the operators for the individual vertices $v$. 
As the second part contains derivatives of $\Ex$ and $\Ey$, the according fluxes $\Fx{\I}$ and $\Fy{\I}$ will appear. Ultimately, the results read
\begin{align}
    \lorHhatIpart{2} &= \tfrac{4\pi^2}{4\kappa'} \lr{\tfrac{-\i}{\hbar}}^{3l} \lr{ \tfrac{16}{3\lr{\kappa\immirzi r_2}^3 k_0\mu_0\nu_0}}^l \hat{\E}_v{}^4 \lr{ \Fx{\I_v} \Fy{\I_{\vp}} - \Fy{\I_v} \Fx{\I_{\vp}} }^2 \left.\ZbHatI{v}{}^l\right\rvert_{ r_2=\frac{2}{3}-\frac{5}{3l} } \quad\text{and} \\
    \lorHhatIpart{3} &= \tfrac{4\pi^2}{\kappa'} \lr{\tfrac{-\i}{\hbar}}^{3l} \lr{ \tfrac{16}{3\lr{\kappa\immirzi r}^3 k_0\mu_0\nu_0}}^l \lr{ \hat{\E}_{\vp} \lr{ \hat{\E}_{\vpp}-\hat{\E}_{\vp} } \ZaHatI{\vp}{}^l -\hat{\E}_{v} \lr{ \hat{\E}_{\vp}-\hat{\E}_{v} } \ZaHatI{v}{}^l }\Big\rvert_{r=\frac{2}{3}-\frac{1}{3l} }.
\end{align}
In accordance with \cite{MA:Andreas,MA:Boehm,MA:Alex}, the second part was quantised in a different manner than in \cite{Banerjee:quant}. While the latter introduced the inverse flux to cope with the denominators $\Ex$ and $\Ey$ appearing in \eqref{eq:lorHam2}, the alternative route leading to the quantisation above uses
\begin{align}
    \frac{1}{\Ex} = \frac{\Ey\E}{\lr{\sqrt{\det E}}^2} \quad \text{and} \quad \frac{1}{\Ey} = \frac{\Ex\E}{\lr{\sqrt{\det E}}^2},
\end{align}
leading to the volume squared as denominator. Therefore, we can insert a new $\lr{1}^l$ for which
\begin{equation}
    r_2 = \frac{2}{3} - \frac{5}{3l}
\end{equation}
holds and resolve the inverse volume squared in the same manner as for the inverse volume before.
Lastly, we set $\mu_0=1=\nu_0$. 

Finally, the physical Hamiltonian operator now reads altogether
\begin{align}
  \hatHH_{\rm phys} &=  \sum_{v \in V(\aqggraph)}\hatHH_{\rm phys,v}= \frac{1}{2}\sum_{v \in V(\aqggraph)} \lr{\euclHhatI +(\euclHhatI)^\dagger + \lorHhatI + (\lorHhatI)^\dagger},
  \label{eq:FinalHphysAQG}
\end{align}
with
\begin{align}
   \euclHhat +& \lorHhat 
    = \euclHhat + \lorHhatpart{1} + \lorHhatpart{2} + \lorHhatpart{3} 
    = \sum_{v \in V(\aqggraph)} \lr{ \euclHhatI + \lorHhatIpart{1} + \lorHhatIpart{2} + \lorHhatIpart{3} }  \nonumber\\
    & = \sum_{v \in V(\aqggraph)} \Bigg\{ -\frac{4\i}{\kappa'\lp{}^2 k_0\mu_0\nu_0\immirzi^3} \bigg[ \sin X_v \sin Y_v \OOp{\theta}{1,v}+ \nonumber \\
    & \qquad\ +\frac{1}{2}\sum\limits_{\xi=\pm 1}\left( 2 \sin\tfrac{Y_{\vxi}}{2} \cos\tfrac{Y_v}{2} \sin \A_{e_v} \OOp{x}{1,v} + 2 \sin\tfrac{X_{\vxi}}{2} \cos\tfrac{X_v}{2} \sin\A_{e_v} \OOp{y}{1,v} \right) \bigg] \nonumber\\
    & \quad -\tfrac{4\pi^2}{4\kappa'}  \lr{ \tfrac{16\i}{3\lp{}^6\immirzi^3 r^3k_0\mu_0\nu_0}}^l  \lr{ \hat{\E}_{\vp}  -\hat{\E}_v }^2 \left.\ZaHatI{v}{}^l\right\rvert_{r=\frac{2}{3}-\frac{1}{3l} } \nonumber\\
    & \quad + \tfrac{4\pi^2}{4\kappa'}  \lr{ \tfrac{16\i}{3\lp{}^6\immirzi^3 r_2^3k_0\mu_0\nu_0}}^l \hat{\E}_v{}^4 \lr{ \Fx{\I_v} \Fy{\I_{\vp}} - \Fy{\I_v} \Fx{\I_{\vp}} }^2 \left.\ZbHatI{I}{}^l\right\rvert_{r_2=\frac{2}{3}-\frac{5}{3l} } \nonumber\\
    & \quad +\tfrac{4\pi^2}{\kappa'}  \lr{ \tfrac{16\i}{3\lp{}^6\immirzi^3 r^3k_0\mu_0\nu_0}}^l \lr{ \hat{\E}_{\vp} \lr{ \hat{\E}_{\vpp}-\hat{\E}_{\vp} } \ZaHatI{\vp}{}^l -\hat{\E}_{v} \lr{ \hat{\E}_{\vp}-\hat{\E}_{v} } \ZaHatI{v}{}^l }\Big\rvert_{r=\frac{2}{3}-\frac{1}{3l} } \Bigg\}. \label{eq:HhatphysAQG}
\end{align}
Comparing the results for the final physical Hamiltonian operator in reduced LQG in \eqref{eq:FinalHphysLQG} and for AQG in \eqref{eq:FinalHphysAQG}, they reflect again the underlying difference of the way graph-preserving operators are implemented. For reduced LQG, these involve a sum over all possible embedded finite graphs $\graph$ and the operator preserves each  graph Hilbert space ${\cal H}_\graph$ separately, whereas in AQG the operator involves a sum over the vertices of the abstract infinite graph $\aqggraph$. This finishes the discussion on the quantisation of the physical Hamiltonian of the Gowdy model in the AQG framework.

\section{First steps in applying the AQG Gowdy model}
\label{sec:FirstStepsAppl}
In this section, we present first steps in applying the AQG Gowdy model derived in the former sections of this article. In particular, we want to discuss the Schr\"odinger-like equation that encodes the dynamics of the quantum model. For this purpose, we compute the action of the physical Hamiltonian $\hatHphys$ on the basis states and due to its complexity we will discuss the individual parts of the Euclidean and Lorentzian contributions to $\hatHphys$ separately.

\subsection{The Schr\"odinger-like equation for the AQG Gowdy model}
Given the physical Hamiltonian operator in the AQG Gowdy model $\hatHphys$ in \eqref{eq:FinalHphysAQG}, we can take it as the starting point to derive the corresponding Schr\"odinger-like equation encoding the dynamics of the model. For simplicity, we will restrict our discussion to the case where we choose $\xi=1$ only and neglect the contribution coming from $\xi=-1$ in the sum in \eqref{eq:HhatphysAQG} in the Euclidean part because such a restriction will not be very relevant for the applications discussed in this section but simplifies the individual formulae. To ensure that the semiclassical limit is still correct, we need to add an additional factor of 2 here that cancels the factor of $\frac{1}{2}$ in front of the sum over $\xi$ in \eqref{eq:HhatphysAQG}. Carried over to the reduced LQG case, such a restriction can also be understood as a slightly different regularisation of the operator where the shifted holonomies involved act only to the right hand side of the vertex $v$ but not to the left. As discussed above, in the definition of $\hatHphys$ we choose the symmetric combination of the individual parts, i.e.~we have
\begin{align}
    \hatHphys = \frac{1}{2} \lr{ \euclHhat + \euclHhat^\dagger } + \lorHhat,
\end{align}
where we already used that $\lorHhat$ will turn out to be symmetric and thus we only need to consider the symmetric combination of $\euclHhat$. We can directly see this by calculating the adjoint version of $\euclHhat$. While the contained trigonometric functions $\sin X_v$, $\cos X_v$, $\sin Y_v$ etc. are self-adjoint due to
\begin{align}
    \lr{\sin X_v}^\dagger = \lr{ \frac{1}{2\i} \lr{ \e{\i X_v} - \e{-\i X_v} } }^\dagger = -\frac{1}{2\i} \lr{ \e{-\i X_v} - \e{\i X_v} } = \sin X_v, \label{eq:sinEvaluation}
\end{align}
the class of operators $\OOp{\theta/x/y}{r,v}$ is indeed not. Rewriting \eqref{eq:defOOpx} as
\begin{align}
    \OOp{x}{r,v} & = \cos \frac{X_v}{2}  \V^r_v  \sin \frac{X_v}{2} - \sin \frac{X_v}{2}  \V^r_v  \cos \frac{X_v}{2} = \frac{1}{2\i} \lr{ \e{-\frac{\i}{2}X_v} \V^r_v \e{\frac{i}{2}X_v} - \e{\frac{\i}{2}X_v} \V^r_v \e{-\frac{i}{2}X_v} },
\end{align}
which we will also later use to compute the action of the physical Hamilton operator on the basis states, we obtain
\begin{align}
    \lr{\OOp{x}{r,v}}^\dagger &= -\frac{1}{2\i} \lr{ \e{-\frac{\i}{2}X_v} \V^r_v \e{\frac{i}{2}X_v} - \e{\frac{\i}{2}X_v} \V^r_v \e{-\frac{i}{2}X_v} } = - \OOp{x}{r,v}.
\end{align}
Altogether, this results in
\begin{align}
    (\euclHhatI)^\dagger &= -\frac{4\i}{\kappa'\lp{}^2 k_0\mu_0\nu_0\immirzi^3} \left( \OOp{\theta}{1,v} \sin X_v \sin Y_v + 2\OOp{x}{1,v} \sin \frac{Y_{v+}}{2} \cos \frac{Y_v}{2} \sin \A_{e_v} + \right.\nonumber\\
    & \hphantom{-\frac{4\i}{\kappa'\lp{}^3\immirzi^3}}\qquad + \left. 2\OOp{y}{1,v}\sin \frac{X_{v+}}{2} \cos \frac{X_v}{2} \sin \A_{e_v} \right). \label{eq:euclHhatIdagger}
\end{align}
It is then straightforward to see that this acts differently on the states $|k,\mu,\nu\rangle$ than $\euclHhatI$ does (cf.~\eqref{eq:euclH}): The trigonometric functions, which act first in the adjoint version, increase or decrease the charges (cf.~\eqref{eq:sinEvaluation}) and the volume operator within the class of operators $\OOp{\theta/x/y}{1,v}$ then reads out different charges than the non-adjoint $\euclHhatI$ does.

The increasingly long expression for $\hatHphys$ can, however, also be reduced by one of its contributions. Namely, via the action\footnote{We describe the action of the main components on the basis states in more detail within the next subsection.} of $\lorHhatpart{3}$ on the basis states $|k,\mu,\nu\rangle$ 
\begin{align}
    \lorHhatpart{3} & |k,\mu,\nu\rangle = \sum_{v \in V(\aqggraph)} \lorHhatIpart{3} |k,\mu,\nu\rangle = \sum_{v \in V(\aqggraph)} \frac{4\pi^2\lp\sqrt{\immirzi}}{4\kappa'}\lr{\frac{1}{2r^3k_0\mu_0\nu_0}}^l \Big\rvert_{ r=\frac{2}{3}-\frac{1}{3l} } \cdot \nonumber\\
    & \!\!\!\!\!\!\! \cdot \Bigg\{ \lr{k_{\evp}+k_{\ev}}\lr{k_{\evpp}-k_{\ev}} \cdot \left[ \lr{ \lrabs{\mu_{\vp}+1}^{\frac{r}{2}} - \lrabs{\mu_{\vp}-1}^{\frac{r}{2}} }\lr{ \lrabs{\nu_{\vp}+1}^{\frac{r}{2}} - \lrabs{\nu_{\vp}-1}^{\frac{r}{2}} } \right. \nonumber\\
    & \cdot \left. \lr{\lrabs{k_{e_{\vp}}+k_{\ev}+1}^{\frac{r}{2}} - \lrabs{k_{e_{\vp}}+k_{\ev}-1}^{\frac{r}{2}}} \lrabs{\mu_{\vp}}^r \lrabs{\nu_{\vp}}^r \lrabs{k_{e_{\vp}}+k_{\ev}}^r \right]^l - \nonumber\\
    & - \lr{k_{\ev}+k_{\evm}}\lr{k_{e_{\vp}}-k_{\evm}} \cdot \left[ \lr{ \lrabs{\mu_{v}+1}^{\frac{r}{2}} - \lrabs{\mu_{v}-1}^{\frac{r}{2}} }\lr{ \lrabs{\nu_{v}+1}^{\frac{r}{2}} - \lrabs{\nu_{v}-1}^{\frac{r}{2}} } \right. \nonumber\\
    & \cdot \left. \lr{\lrabs{k_{\ev}+k_{\evm}+1}^{\frac{r}{2}} - \lrabs{k_{\ev}+k_{\evm}-1}^{\frac{r}{2}}} \lrabs{\mu_{v}}^r \lrabs{\nu_{v}}^r \lrabs{k_{\ev}+k_{\evm}}^r \right]^l \Bigg\} |k,\mu,\nu\rangle, \label{eq:lor3action}
\end{align}
we conclude that this expression vanishes: The minuend and the subtrahend of the difference within the curly brackets are structurally the same and only differ by the contained charges' indices via $\vpp\mapsto\vp,\vp\mapsto v$ and $v\mapsto \vm$, meaning that each vertex is mapped to its left neighbouring one. By taking the sum over all $v\in V(\aqggraph)$, i.e.~the sum over all vertices of the graph $\aqggraph$, this becomes a telescope series. Reminding ourselves that we implemented boundary conditions such that we mimic also in the AQG case the situation to sum along a closed circle as it is done in the reduced LQG model, we realise that the first and last contribution of the series then are the same which means as a result that the telescope series sums up to zero. Putting it into formulae, let us write this in compact form as
\begin{align}
    \eqref{eq:lor3action} &\eqqcolon \sum_{v \in V(\aqggraph)} \lr{ h_{\text{lor},\vp}^{(3)} - h_{\text{lor},v}^{(3)}} |k,\mu,\nu\rangle, \label{eq:telescopeStart}
\end{align}
where $h_{\text{lor},\vp}^{(3)}$ represents the respective minuends within the curly bracket of \eqref{eq:lor3action} and $h_{\text{lor},v}^{(3)}$ the corresponding subtrahends. Now, assuming the abstract graph has $N$ edges with non-trivial representations on the edges then via the difference within the summands of the series in \eqref{eq:telescopeStart}, all contributions but the ones for the  first vertex $v_1$ and the last one $v_{N+1}$  appear twice and in particular with different signs. Hence,
\begin{align}
    \eqref{eq:telescopeStart} = h_{\text{lor},v_{N+1}}^{(3)} - h_{\text{lor},v_1}^{(3)}
\end{align}
and as the definition of the abstract Gowdy graph involved periodic boundary conditions such that the vertices $v_{N+1}$ and $v_1$ are identified, this results in the contribution of $\lorHhatpart{3}$ to vanish. This means that the final physical Hamiltonian reads
\begin{align}
    \hatHphys = \frac{1}{2} \lr{ \euclHhat + (\euclHhat)^\dagger } + \lorHhatpart{1} +  \lorHhatpart{2}. \label{eq:physHam}
\end{align}

\subsubsection{Action of \texorpdfstring{$\hatHphys$}{Hphys}'s main components on the basis states \texorpdfstring{$|k,\mu,\nu\rangle$}{kmunu}} \label{sec:ActionOperatorsOnBasisStates}

In order to make future calculations easier and provide a concise overview, we will now state how the main components of $\hatHphys$ act on the basis states $|k,\mu,\nu\rangle$. We start with the trigonometric functions
\begin{align}
    \sin \A_{e_I} |k,\mu,\nu\rangle &= \frac{1}{2\i} \lr{ |k_{\ev}+2,\mu,\nu\rangle - |k_{\ev}-2,\mu,\nu\rangle} , \\
    \cos \A_{e_I} |k,\mu,\nu\rangle &= \frac{1}{2} \lr{ |k_{\ev}+2,\mu,\nu\rangle + |k_{\ev}-2,\mu,\nu\rangle} , \\
    \sin X_v |k,\mu,\nu\rangle &= \frac{1}{2\i} \lr{ |k,\mu_v+2,\nu\rangle - |k,\mu_v-2,\nu\rangle} , \\
    \cos X_v |k,\mu,\nu\rangle &= \frac{1}{2} \lr{ |k,\mu_v+2,\nu\rangle + |k,\mu_v-2,\nu\rangle} , \\
    \sin Y_v |k,\mu,\nu\rangle &= \frac{1}{2\i} \lr{ |k,\mu,\nu_v+2\rangle - |k,\mu,\nu_v-2\rangle} \quad\text{and} \\
    \cos Y_v |k,\mu,\nu\rangle &= \frac{1}{2} \lr{ |k,\mu,\nu_v+2\rangle + |k,\mu,\nu_v-2\rangle}
\end{align}
that we used instead of the actual holonomies. To have a complete list, we also recap the fluxes' actions
\begin{align}
    \hat{\E}_v |k,\mu,\nu\rangle &= \frac{\immirzi\lp{}^2}{2}\frac{k_{\ev}-k_{\evm}}{2} |k,\mu,\nu\rangle \tag{\ref{eq:actionfluxtheta}}\\
    \Fx{\I_v} |k,\mu,\nu\rangle &= \frac{\immirzi\lp{}^2}{2}  \mu_v |k,\mu,\nu\rangle \quad \text{and} \tag{\ref{eq:actionfluxx}}\\
    \Fy{\I_v} |k,\mu,\nu\rangle &= \frac{\immirzi\lp{}^2}{2} \nu_v |k,\mu,\nu\rangle. \tag{\ref{eq:actionfluxy}}
\end{align}
Then, for the class of the $\OOp{\theta/x/y}{r,v}$ operators, we get
\begin{align}
    \OOp{\theta}{r,v} |k,\mu,\nu\rangle & = \frac{1}{2\i}\lr{\frac{\lp{}^3\immirzi^{\frac{3}{2}}}{4}}^r \lr{ \lrabs{ k_{\ev}+1+k_{\evm} }^{\frac{r}{2}} - \lrabs{ k_{\ev}-1+k_{\evm} }^{\frac{r}{2}}} \lrabs{\mu_v}^{\frac{r}{2}}\lrabs{\nu_v}^{\frac{r}{2}}|k,\mu,\nu\rangle, \label{eq:OthetaAction}\\
    \OOp{x}{r,v} |k,\mu,\nu\rangle & = \frac{1}{2\i}\lr{\frac{\lp{}^3\immirzi^{\frac{3}{2}}}{4}}^r \lrabs{k_{\ev}+k_{\evm}}^\frac{r}{2} \lr{ \lrabs{ \mu_v+1 }^{\frac{r}{2}} - \lrabs{ \mu_v-1 }^{\frac{r}{2}}} \lrabs{\nu_v}^{\frac{r}{2}}|k,\mu,\nu\rangle \quad\text{and} \label{eq:OxAction}\\
    \OOp{y}{r,v} |k,\mu,\nu\rangle & = \frac{1}{2\i}\lr{\frac{\lp{}^3\immirzi^{\frac{3}{2}}}{4}}^r \lrabs{k_{\ev}+k_{\evm}}^\frac{r}{2} \lrabs{\mu_v}^{\frac{r}{2}} \lr{ \lrabs{ \nu_v+1 }^{\frac{r}{2}} - \lrabs{ \nu_v-1 }^{\frac{r}{2}}} |k,\mu,\nu\rangle ,\label{eq:OyAction}
\end{align}
where we used the action of the volume operator according to \eqref{eq:VolAQGaction}. And lastly,
\begin{align}
    \ZaHatI{v}{}^l & |k,\mu,\nu\rangle = \lr{-12}^l \lr{ \frac{1}{2\i}\lr{\frac{\lp{}^3\immirzi^{\nicefrac{3}{2}}}{4}}^{\!\! r}\, }^{\!\!3l} \left[  \lr{\lrabs{ k_{\ev}+1+k_{\evm} }^{\frac{r}{2}} - \lrabs{ k_{\ev}-1+k_{\evm} }^{\frac{r}{2}}} \cdot \right.\nonumber\\
    & \quad \cdot \left.  \lr{ \lrabs{ \mu_v+1 }^{\frac{r}{2}} - \lrabs{ \mu_v-1 }^{\frac{r}{2}}} \lr{ \lrabs{ \nu_v+1 }^{\frac{r}{2}} - \lrabs{ \nu_v-1 }^{\frac{r}{2}}} \lrabs{k_{\ev}+k_{\evm}}^r \lrabs{\mu_v}^r \lrabs{\nu_v}^r \right]^l |k,\mu,\nu\rangle . \label{eq:Zaction}
\end{align}
The individual action of these operators will be used in the next subsection where we discuss Gowdy states in the AQG framwork in more detail.

\subsubsection{Gowdy states in the AQG model}

We briefly discussed at the beginning of Section \ref{Sec:AQGQuant} how the symmetry reduced Gowdy model can be carried over to the AQG framework. Because the physical Hamiltonian operator $\hatHphys$ also involves the adjoint $(\euclHhat)^\dagger$, we need to discuss in more detail how we can perform an adaption of the AQG-graph we consider. Due to the appearance of $(\euclHhat)^\dagger$ within $\hatHphys$ and its action on $|k,\mu,\nu\rangle$, we allow only those states in the model that have only a finite number of the infinite number of edges with non-trivial U(1)-charges $k_{\ev}$, the remaining one carry trivial representations. The action of the physical Hamilton operator $\hatHphys$ of \eqref{eq:HhatphysAQG} on trivially charged vertices vanishes as there is always an operator of the class $\OOp{\theta/x/y}{r,v}$ acting first. Taking a look at their action on the basis states (\eqref{eq:OthetaAction}, \eqref{eq:OxAction} and \eqref{eq:OyAction}), we see that they vanish on trivially charged vertices --- cf. \eqref{eq:VolAQGaction} to see that it suffices that one of $k_{\ev}=-k_{\evm}$, $\mu_v=0$ or $\nu_v=0$ holds.

\begin{figure}
\begin{tikzpicture}
\def \n {5}
\def \firstvertex {-0.5}
\def \lastvertex {10.5}
\def \linelength {\lastvertex-\firstvertex}
\def \smallradius {0.2}
        \draw (-1.5,0) -- (\firstvertex,0) node[pos=0.5,below]{$-k_{\n}$}; 
        \draw (\lastvertex,0) -- (11.5,0) node[pos=0.5,below]{$-k_{\n}$}; 
        \draw (11.5,0) -- (12.5,0) node[pos=0.5,below]{$k_{\n}$}; 
        \fill[black!40] (\firstvertex,0) circle (2pt); 
        \fill[black!40] (11.5,0) circle (2pt); 
        \fill[black!40] (\lastvertex,0) circle (2pt); 
        \fill[black!40] (12.5,0) circle (2pt); 
        \fill[black!40] (-1.5,0) circle (2pt); 
        \draw[densely dotted] (12.5,0) -- (13,0); 
        \draw[densely dotted] (-2,0) -- (-1.5,0); 
        \draw[-latex] (\firstvertex,0) -- ({\firstvertex+(\lastvertex-\firstvertex)/(\n+1)},0) node[pos=0.5,below]{$k_{\n}$}; 
\foreach \s in {1,...,\n}
{
    \node at ({\firstvertex+(\lastvertex-\firstvertex)/(\n+1)+(\s-1)/(\n+1)*(\lastvertex-\firstvertex)},-0.3) {$v_{\s}$}; 
    \fill ({\firstvertex+(\lastvertex-\firstvertex)/(\n+1)+(\s-1)/(\n+1)*(\lastvertex-\firstvertex)},0) circle (2pt); 
    \draw[-latex] ({\firstvertex+(\s)*(\lastvertex-\firstvertex)/(\n+1)},0) -- ({\firstvertex+(\s+1)*(\lastvertex-\firstvertex)/(\n+1)},0) node[pos=0.5,above]{$k_{\s}$} node[pos=0.5,below]{$e_{\s}$}; 
    \draw [dash pattern=on 3pt off 1pt, black, rotate around x=-90, rotate around={-112.5:({\firstvertex+(\lastvertex-\firstvertex)/(\n+1)+(\s-1)/(\n+1)*(\lastvertex-\firstvertex)},0)}] ({\firstvertex+(\lastvertex-\firstvertex)/(\n+1)+(\s-1)/(\n+1)*(\lastvertex-\firstvertex)},0) arc (0:360:\smallradius) node[pos=0.65,right]{$\mu_{\s}$}; 
    \draw [dash pattern=on 3pt off 1pt, black!70, rotate around x=45, rotate around={-67.5:({\firstvertex+(\lastvertex-\firstvertex)/(\n+1)+(\s-1)/(\n+1)*(\lastvertex-\firstvertex)},0)}] ({\firstvertex+(\lastvertex-\firstvertex)/(\n+1)+(\s-1)/(\n+1)*(\lastvertex-\firstvertex)},0) arc (0:360:\smallradius) node[pos=0.35,left]{$\nu_{\s}$}; 
}
\end{tikzpicture}
    \caption{The abstract infinite AQG-graph of Figure~\ref{fig:AQGgraph}, now also ensuring trivially charged vertices not to contribute via the action of $(\euclHhat)^\dagger$ --- as guaranteed by alternating charges $\pm k_5$ on the previously uncharged edges $e_I,I>5$. To keep the notation more compact, we used $k_{e_{v_I}}\eqqcolon k_I, \mu_{v_I}\eqqcolon\mu_I$ and $\nu_{v_I}\eqqcolon\nu_I$.}
    \label{fig:GowdyAQGgraph}
\end{figure}
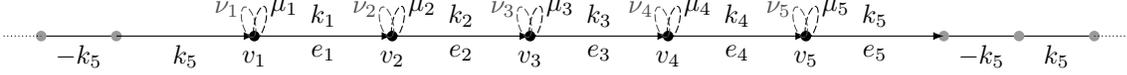

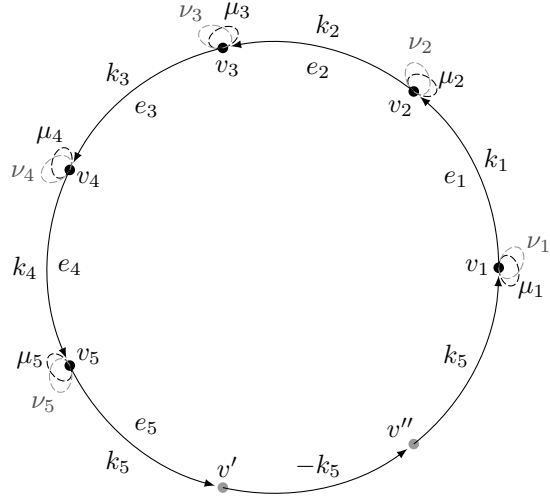
\begin{figure}
\begin{tikzpicture}
\pgfmathsetmacro \v {5} 
\pgfmathsetmacro \n {\v+2} 
\pgfmathsetmacro \firstnew {\v+1} 
\pgfmathsetmacro \secondnew {\v+2} 
\def \radius {3cm}
\def \smallradius {0.2}
\def \margin {0} 
\def \halfmargin {2} 

\foreach \s in {1,...,\n}
{
  \ifnum \s=\firstnew
  \fill[black!40] ({360/\n * (\s - 1)}:\radius) circle (2pt); 
  \node at ({360/\n * (\s - 1)}:\radius-8pt) {$v'$}; 
  \node at ({180/\n+(\s-1)*360/\n}:\radius*0.9) {$-k_{\v}$}; 
  \else
    \ifnum \s=\secondnew
        \fill[black!40] ({360/\n * (\s - 1)}:\radius) circle (2pt); 
        \node at ({360/\n * (\s - 1)}:\radius-8pt) {$v''$}; 
        \node at ({180/\n+(\s-1)*360/\n}:\radius*0.9) {$k_{\v}$}; 
    \else
        \fill ({360/\n * (\s - 1)}:\radius) circle (2pt); 
        \node at ({360/\n * (\s - 1)}:\radius-8pt) {$v_\s$}; 
        \node at ({180/\n+(\s-1)*360/\n}:\radius*1.1) {$k_\s$}; 
        \node at ({180/\n+(\s-1)*360/\n}:\radius*0.9) {$e_\s$};
        \draw  [dash pattern=on 3pt off 1pt, black, rotate around y=-30, rotate around z=180, rotate around={{360/\n * (\s - 1)}:({360/\n * (\s - 1)}:\radius)}] ({360/\n * (\s - 1)}:\radius) arc (0:360:\smallradius);
        \node at ({360/\n * (\s - 1)-5}:{\radius+0.45cm}) {$\mu_\s$}; 
        \draw  [dash pattern=on 3pt off 1pt, black!40, rotate around y=60, rotate around x=0, rotate around z=180, rotate around={{360/\n * (\s - 1)}:({360/\n * (\s - 1)}:\radius)}] ({360/\n * (\s - 1)}:\radius) arc (0:360:\smallradius);
        \node at ({360/\n * (\s - 1)+5}:{\radius+0.55cm}) {\color{black!70}$\nu_\s$}; 
    \fi
  \fi
  \draw[-latex] ({360/\n * (\s - 1)+\margin}:\radius) 
    arc ({360/\n * (\s - 1)+\margin}:{360/\n * (\s)-\halfmargin}:\radius);
}
\end{tikzpicture}
    \caption{An embedding of the abstract infinite AQG-graph of Figure~\ref{fig:GowdyAQGgraph}, which also ensures tri\-vially charged vertices not to contribute via the action of $(\euclHhat)^\dagger$ --- as guaranteed by the new uncharged vertices $v'$ and $v''$ and the new $k_5$- and $-k_5$-charged edges around them, compared to Fig.~\ref{fig:embeddedGraph}. To keep the notation more compact we used $k_{e_{v_I}}\eqqcolon k_I, \mu_{v_I}\eqqcolon\mu_I$ and $\nu_{v_I}\eqqcolon\nu_I$.}
    \label{fig:GowdyAQGgraphEmbedded}
\end{figure}

This changes for the action of the adjoint operator $(\euclHhat)^\dagger$, \eqref{eq:euclHhatIdagger}. 
Its first term acts with $\sin X_v \sin Y_v$ before $\OOp{\theta}{1,v}$, thereby charging the two previously neutral charges $\mu_v$ and $\nu_v$. Hence, $\OOp{\theta}{1,v}$ returns a non-zero value that is, i.a., $\sim \lrabs{ k_{\ev}+1+k_{\evm} }^{\frac{r}{2}} - \lrabs{ k_{\ev}-1+k_{\evm} }^{\frac{r}{2}}$. This still vanishes for vertices $v$ with trivially charged neighbouring edges $e_{v_{I-1}}$ and $e_{v_I}$. But taking a look at Figure~\ref{fig:AQGgraph}, we see that vertex $v_{N+1}$ --- which is trivially charged by means of $k_{e_{v_{N+1}}}=0$, $\mu_{v_{N+1}}=0$ and $\nu_{v_{N+1}}=0$ --- has also $e_{v_N}$ as neighbouring edge for which, i.a., $k_{e_{v_N}}\neq 0$ holds. Therefore, $(\euclHhatI)^\dagger|k,\mu,\nu\rangle$ does not vanish for $v_I=v_{N+1}$. However, we can fix this by fulfilling the condition $k_{\ev}=-k_{\evm}$ to obtain a zero eigenvalue and set $k_{e_{v_{N+1}}}=-k_{e_{v_N}}$. Consequently, to have all the following trivially charged vertices to have vanishing contributions as well, we need to set $k_{e_{v_{N+2}}}=-k_{e_{v_{N+1}}}=k_{e_{v_N}}$ and so forth, i.e.~all edges $e_I, I>N,$ are charged with $k_{\ev}\cdot\lr{-1}^{I-N}$. Figure \ref{fig:GowdyAQGgraph} illustrates these new states and Figure \ref{fig:GowdyAQGgraphEmbedded} does so for an embedded graph. The embedding is done by creating two additional, trivially charged vertices $v'$ and $v''$ between $v_N$ and $v_1$ and mapping all edges that are charged with $-k_N$ to the edge between $v'$ and $v''$, while all edges that are charged with $k_{e_{v_N}}$ are mapped to the edge between $v''$ and $v_1$. Note that the latter take over the role of the previously $k_{e_{v_N}}$-charged edge $e_0$ to the left of $v_1$ (cf. Fig.~\ref{fig:GowdyAQGgraph}).

Note that all the above is not the case for the other two contributions of $(\euclHhat)^\dagger$: Besides $\sin\A_{e_{v_I}}$, only trigonometric functions of either $X_v$ \textit{or} $Y_v$ act before $\OOp{y}{1,v}$ or $\OOp{x}{1,v}$ respectively. Hence, via their action according to \eqref{eq:OyAction} and \eqref{eq:OxAction}, $\OOp{y}{1,v}$ contributes with a value $\sim \lrabs{ \nu_v+1 }^{\frac{r}{2}} - \lrabs{ \nu_v-1 }^{\frac{r}{2}}$, where the charge $\nu_v=0$ is still the initial neutral one. The same holds for $\OOp{x}{1,v}$ and $\mu_v=0$ and we do not need to perform further adaptions.

Having found a suitable form for the basis states $|k,\mu,\nu\rangle$, we can now address the states $|\Psi\rangle\lr{\tau}$ that we will use for writing down an ansatz for the solution of the Schr\"odinger-like equation of the Gowdy model given by
\begin{equation}
i \hbar \frac{\partial}{\partial \tau}|\Psi(\tau)\rangle = \hatHphys |\Psi(\tau)\rangle
\end{equation}
later on in Subsection~\ref{subsec:Schrödinger}. For the state $|\Psi(\tau)\rangle$ we  use the following separation ansatz:
\begin{equation}
|\Psi\rangle\lr{\tau} = |\varphi\lr{k,\mu,\nu}\rangle |\chi\lr{\tau}\rangle, \label{eq:statesWithTime}    
\end{equation}
where we put the dependence on the physical time $\tau$ completely into $\chi\lr{\tau}$, while that quantity, in turn, does not depend on $k_{\ev}$, $\mu_v$ or $\nu_v$ and solely $|\varphi\rangle$ does. Our ansatz for $|\varphi\rangle$ then reads
\begin{align}
    |\varphi\rangle \coloneqq \sum_{k\in\mathbbm{Z}^N}\sum_{\mu_v\in m}\sum_{\nu\in n} \Coeff{}{}{} |k,\mu,\nu\rangle \label{eq:statesDef},
\end{align}
whose structure we illustrate in the following in more detail along the lines of~\cite{MA:Andreas}.

In the above, $N$ is the number of vertices and $k$, $\mu$ and $\nu$ are multi-labels:  
 $k\coloneqq (k_{e_{v_1}},...,k_{e_{v_N}})$, $\mu\coloneqq (\mu_{v_1},...,\mu_{v_N})$ and $\nu\coloneqq (\nu_{v_1},...,\nu_{v_N})$. 
Lastly,
 \begin{equation}
C _{k,\mu,\nu}\coloneqq C _{k_{e_{v_1}},...,k_{e_{v_N}},\mu_{v_1},...,\mu_{v_N},\nu_{v_1},...,\nu_{v_N}} 
\end{equation}
are coefficients that depend on all $k$-, $\mu$- and $\nu$-labels. While $k$ takes values in $\mathbbm{Z}^N$, the sets $m\coloneqq m_1\times ... \times m_N$ and $n\coloneqq n_1\times ... \times n_N$ allow more flexibility:
\begin{equation}
m_{v_j}\coloneqq \{\widetilde{\mu}_{v_j}+p\ |\ p\in\mathbb{Z}\} \;\; \text{and} \;\;  n_{v_j}\coloneqq \{\widetilde{\nu}_{v_j}+p\ |\ p\in\mathbb{Z}\} \;\; \text{with} \; \widetilde{\mu}_{v_j}, \widetilde{\nu}_{v_j} \in \mathbbm{R} \;\; \forall j\in\{1,...,N\}
\label{def_label_sets_mu_nu}
\end{equation}

To motivate this choice, we have a closer look at the point holonomies and take the one of $X$ as an example. Within its expression $\exp{(\frac{i}{2}\mu_{v_j} X_{v_j})}$, $\mu_{v_j}\in\mathbb{R}$ labels the specific irreducible representation of the Bohr compactification $\overline{\mathbb{R}}_{\text{Bohr}}$ for each vertex ${v_j}$. The corresponding Hilbert space $\mathcal{H}_{{v_j}}^{X}\coloneqq  L_2(\overline{\mathbb{R}}_{\text{Bohr}},d\mu_{\text{Bohr}})$ consists of square integrable functions $f$ over $\overline{\mathbb{R}}_{\text{Bohr}}$ with respect to its Haar measure $\d\mu_{\text{Bohr}}$. The inner product in this Hilbert space reads
\begin{equation}
\langle f\, | \, g\rangle\coloneqq \lim\limits_{R \to \infty}\frac{1}{2R} \int\limits_{-R}^{+R}\text{d}x f(x)^* g(x),
\end{equation}
wherein $f^*$ is the complex conjugate of $f$. Now, using $\langle x |{\mu_{v_j}}\rangle\coloneqq \exp{(\frac{i}{2}\mu_{v_j} x)}$, we find the inner product of two basis states to be
\begin{equation}
\langle \mu_{v_j}\, | \,\mu_{v_j}^\prime\rangle \coloneqq \lim\limits_{R \to \infty}\frac{1}{2R} \int\limits_{-R}^{+R}\text{d}x\ \e{\frac{i}{2}(\mu_{v_j}^\prime-\mu_{v_j})}=
\delta_{\mu_{v_j},\mu_{v_j}^\prime},
\end{equation}
with the Kronecker delta $\delta_{\mu_{v_j},\mu_{v_j}^\prime}$, and also deduce the completeness relation
\begin{equation}
\sum_{\mu_{v_j}\in m_j}|\mu_{v_j}\rangle\langle \mu_{v_j}\rangle =\mathbbm{1}_{\mathcal{H}_{v_j}^{X}}.
\end{equation}
Therein, $\mu_{v_j}\in m_j$, which is a finite subset of $\mathbbm{R}$. And it has to be a finite subset in order for a state 
\begin{equation}
\mathcal{H}_{v_j}^{X} \ni |\phi\rangle =\sum\limits_{\mu_{v_j}\in m_j} c_{\mu_{v_j}}|\mu_{v_j}\rangle, \label{general_element_bohr_HS}
\end{equation}
with arbitrary coefficients $c_{\mu_{v_j}}$, to be normalisable --- i.e.\ allowing it to be an element of the Hilbert space after all. When it comes to our applications of the Gowdy model, however, we need to fall back on formal states, whose labels can take values in infinite sets. The reason being that some of the operators we have to deal with will map out of an otherwise finite set: While the Lorentzian part of the physical Hamiltonian acts diagonally, we notice the Euclidean part~\eqref{eq:euclH} to contain the operators
\begin{equation}
\sin\left(X_{v_j}\right)=\frac{1}{2i}\left(\e{ i\frac{\mu_{v_j}}{2} \cdot 2X_{v_j}}-\e{- i\frac{\mu_{v_j}}{2} \cdot 2X_{v_j}}\right) \ \text{and}\ \sin\left(Y_{v_j}\right)=\frac{1}{2i}\left(\e{ i\frac{\nu_{v_j}}{2} \cdot 2Y_{v_j}}-\e{- i\frac{\nu_{v_j}}{2} \cdot 2Y_{v_j}}\right).
\label{sine_cosine_in_terms_of_holonomies}
\end{equation}
Clearly, those do not act diagonally and with the following operator $\OOp{\theta}{1,v}$ in~\eqref{eq:euclH} acting again diagonally, we do have an overall non-diagonal action. If we now choose the labels of the point holonomies to take values from a finite subset, say $m^\prime=n^\prime=u\times...\times u$ with $u\coloneqq \{1, 16, 25\}$, the above operators of~\eqref{sine_cosine_in_terms_of_holonomies} will map a value $16$ of one of the labels of a state i.a.\ to the new label $16+2=18 \notin u$. This problem clearly exists for any choice of finite $u$. Therefore, we have to choose infinite sets for the labels' values, where the individual elements are separated by steps of $\pm 1$ --- note that the physical Hamiltonian~\eqref{eq:euclH} also contains the above operators with their arguments divided by 2. Without having any constraint on ``where'' this sequence starts, we can choose an arbitrary value $\widetilde{\mu}_{v_j}\in\mathbbm{R}$ to construct the set $m_{v_j}\coloneqq \{\widetilde{\mu}_{v_j}+p\ |\ p\in\mathbb{Z}\} \ni \mu_{v_j}$ and similarly for all other vertices. For symmetry reasons, we get the same for the $Y$ point holonomy and its labels $\nu_{v_j}$, while for the $k$-label we have to choose full $\mathbbm{Z}$ itself. This leads us to the initially stated definition~\eqref{eq:statesDef} with~\eqref{def_label_sets_mu_nu}.

However, this infinite linear combination of the basis states is a rather formal ansatz as its norm
\begin{align}
   \langle\varphi\, |\,\varphi\rangle&= \norm{\varphi}^2  = \sum_{k\in\mathbbm{Z}^N}\sum_{\mu\in m}\sum_{\nu\in n} \sum_{k'\in\mathbbm{Z}^N}\sum_{\mu'\in m}\sum_{\nu'\in n} (\Coeff{}{}{} {})^\ast \cdot \Coeff{k'}{\mu'}{\nu'} \langle k,\mu,\nu|k',\mu',\nu'\rangle \\
    & = \sum_{k\in\mathbbm{Z}^N}\sum_{\mu\in m}\sum_{\nu\in n} \lrabs{\Coeff{}{}{}}^2
\end{align}
diverges. We will need these extensive states for the beginning, but when it comes to zero volume eigenstates, e.g., we will also find states with finite norm (cf.~Sec.~\ref{sec:ZeroVol}).
\subsubsection{Action of the physical Hamiltonian \texorpdfstring{$\hatHphys$}{Hphys} on the ansatz states}
We will now state the action of $\hatHphys$ on the state $|\varphi\rangle$ given in \eqref{eq:statesDef}. As the final result \eqref{eq:ActionHPhys} will be rather long, we start with presenting the actions of $\euclHhat$ and $(\euclHhat)^\dagger$ as well. This is furthermore convenient since we will later also use them individually. According to the action of the operators on the basis states as listed in Section \ref{sec:ActionOperatorsOnBasisStates}, we obtain
\vspace{-0.01cm}
\begin{align}
    & \euclHhat |\varphi\rangle = \sum_{v\in V\lr{\aqggraph}} \sum_{k\in\mathbbm{Z}^N}\sum_{\mu\in m}\sum_{\nu\in n} \Coeff{}{}{} \euclHhatI \state{}{}{} \nonumber\\
    & = \sum_{v\in V\lr{\aqggraph}} \sum_{k\in\mathbbm{Z}^N}\sum_{\mu\in m}\sum_{\nu\in n} \keucl  \Coeff{}{}{} \bigg\{ \lr{\sqrt{\lrabs{k_{\ev}+k_{\evm}+1}}-\sqrt{\lrabs{k_{\ev}+k_{\evm}-1}}} \sqrt{\lrabs{\mu_v}\lrabs{\nu_v}} \cdot \nonumber\\
    & \qquad \cdot \Big[ \state{}{\mu_v+2}{\nu_v+2} - \state{}{\mu_v+2}{\nu_v-2} - \state{}{\mu_v-2}{\nu_v+2} + \state{}{\mu_v-2}{\nu_v-2} \Big] + \nonumber\\
    & \quad + \sqrt{\lrabs{k_{\ev}+k_{\evm}}\lrabs{\nu_v}} \lr{\sqrt{\lrabs{\mu_v+1}}-\sqrt{\lrabs{\mu_v-1}}} \cdot \nonumber\\
    & \qquad \cdot \Big[ \state{k_{\ev}+2}{}{\nu_v+1,\nu_{\vp}+1} - \state{k_{\ev}+2}{}{\nu_v+1,\nu_{\vp}-1} + \nonumber\\
    & \qquad\quad + \state{k_{\ev}+2}{}{\nu_v-1,\nu_{\vp}+1} - \state{k_{\ev}+2}{}{\nu_v-1,\nu_{\vp}-1} - \nonumber\\
    & \qquad\quad - \state{k_{\ev}-2}{}{\nu_v+1,\nu_{\vp}+1} + \state{k_{\ev}-2}{}{\nu_v+1,\nu_{\vp}-1} -  \nonumber\\
    & \qquad\quad -\state{k_{\ev}-2}{}{\nu_v-1,\nu_{\vp}+1} + \state{k_{\ev}-2}{}{\nu_v-1,\nu_{\vp}-1} \Big] + \nonumber\\
    & \quad + \sqrt{\lrabs{k_{\ev}+k_{\evm}}\lrabs{\mu_v}} \lr{\sqrt{\lrabs{\nu_v+1}}-\sqrt{\lrabs{\nu_v-1}}} \cdot \nonumber\\
    & \qquad \cdot \Big[ \state{k_{\ev}+2}{\mu_v+1,\mu_{\vp}+1}{} - \state{k_{\ev}+2}{\mu_v+1,\mu_{\vp}-1}{} + \nonumber\\
    & \qquad\quad + \state{k_{\ev}+2}{\mu_v-1,\mu_{\vp}+1}{} - \state{k_{\ev}+2}{\mu_v-1,\mu_{\vp}-1}{} - \nonumber\\
    & \qquad\quad - \state{k_{\ev}-2}{\mu_v+1,\mu_{\vp}+1}{} + \state{k_{\ev}-2}{\mu_v+1,\mu_{\vp}-1}{} -  \nonumber\\
    & \qquad\quad - \state{k_{\ev}-2}{\mu_v-1,\mu_{\vp}+1}{} + \state{k_{\ev}-2}{\mu_v-1,\mu_{\vp}-1}{} \Big] \bigg\} \label{eq:ActionHeucl}
\end{align}
and likewise
\begin{align}
    & (\euclHhat)^\dagger |\varphi\rangle = \sum_{v\in V\lr{\aqggraph}} \sum_{k\in\mathbbm{Z}^N}\sum_{\mu\in m}\sum_{\nu\in n} \Coeff{}{}{} (\euclHhatI)^\dagger \state{}{}{} \nonumber\\
    & = \sum_{v\in V\lr{\aqggraph}} \sum_{k\in\mathbbm{Z}^N}\sum_{\mu\in m}\sum_{\nu\in n} \keucl  \Coeff{}{}{} \bigg\{ \lr{\sqrt{\lrabs{k_{\ev}+k_{\evm}+1}}-\sqrt{\lrabs{k_{\ev}+k_{\evm}-1}}} \cdot  \nonumber\\
    & \quad \cdot \Big[ \sqrt{\lrabs{\mu_v+2}\lrabs{\nu_v+2}} \state{}{\mu_v+2}{\nu_v+2} - \sqrt{\lrabs{\mu_v+2}\lrabs{\nu_v-2}} \state{}{\mu_v+2}{\nu_v-2} - \nonumber\\
    & \qquad - \sqrt{\lrabs{\mu_v-2}\lrabs{\nu_v+2}} \state{}{\mu_v-2}{\nu_v+2} + \sqrt{\lrabs{\mu_v-2}\lrabs{\nu_v-2}} \state{}{\mu_v-2}{\nu_v-2} \Big] + \nonumber\\
    & \quad + \lr{\sqrt{\lrabs{\mu_v+1}}-\sqrt{\lrabs{\mu_v-1}}} \Big[ \sqrt{\lrabs{k_{\ev}+k_{\evm}+2}\lrabs{\nu_v+1}}\state{k_{\ev}+2}{}{\nu_v+1,\nu_{\vp}+1} - \nonumber\\
    & \quad \hphantom{+ \lr{\sqrt{\lrabs{\mu_v+1}}-\sqrt{\lrabs{\mu_v-1}}} }\ - \sqrt{\lrabs{k_{\ev}+k_{\evm}+2}\lrabs{\nu_v+1}}\state{k_{\ev}+2}{}{\nu_v+1,\nu_{\vp}-1} + \nonumber\\
    & \quad \hphantom{+ \lr{\sqrt{\lrabs{\mu_v+1}}-\sqrt{\lrabs{\mu_v-1}}} }\ + \sqrt{\lrabs{k_{\ev}+k_{\evm}+2}\lrabs{\nu_v-1}}\state{k_{\ev}+2}{}{\nu_v-1,\nu_{\vp}+1} - \nonumber\\
    & \quad \hphantom{+ \lr{\sqrt{\lrabs{\mu_v+1}}-\sqrt{\lrabs{\mu_v-1}}} }\ - \sqrt{\lrabs{k_{\ev}+k_{\evm}+2}\lrabs{\nu_v-1}}\state{k_{\ev}+2}{}{\nu_v-1,\nu_{\vp}-1} - \nonumber\\
    & \quad \hphantom{+ \lr{\sqrt{\lrabs{\mu_v+1}}-\sqrt{\lrabs{\mu_v-1}}} }\ - \sqrt{\lrabs{k_{\ev}+k_{\evm}-2}\lrabs{\nu_v+1}}\state{k_{\ev}-2}{}{\nu_v+1,\nu_{\vp}+1} + \nonumber\\
    & \quad \hphantom{+ \lr{\sqrt{\lrabs{\mu_v+1}}-\sqrt{\lrabs{\mu_v-1}}} }\ + \sqrt{\lrabs{k_{\ev}+k_{\evm}-2}\lrabs{\nu_v+1}}\state{k_{\ev}-2}{}{\nu_v+1,\nu_{\vp}-1} - \nonumber\\
    & \quad \hphantom{+ \lr{\sqrt{\lrabs{\mu_v+1}}-\sqrt{\lrabs{\mu_v-1}}} }\ - \sqrt{\lrabs{k_{\ev}+k_{\evm}-2}\lrabs{\nu_v+1}}\state{k_{\ev}-2}{}{\nu_v-1,\nu_{\vp}+1} + \nonumber\\
    & \quad \hphantom{+ \lr{\sqrt{\lrabs{\mu_v+1}}-\sqrt{\lrabs{\mu_v-1}}} }\ + \sqrt{\lrabs{k_{\ev}+k_{\evm}-2}\lrabs{\nu_v+1}}\state{k_{\ev}-2}{}{\nu_v-1,\nu_{\vp}-1} \Big] + \nonumber\\
    & \quad + \lr{\sqrt{\lrabs{\nu_v+1}}-\sqrt{\lrabs{\nu_v-1}}} \Big[ \sqrt{\lrabs{k_{\ev}+k_{\evm}+2}\lrabs{\mu_v+1}}\state{k_{\ev}+2}{\mu_v+1,\mu_{\vp}+1}{} - \nonumber\\
    & \quad \hphantom{+ \lr{\sqrt{\lrabs{\mu_v+1}}-\sqrt{\lrabs{\mu_v-1}}} }\ - \sqrt{\lrabs{k_{\ev}+k_{\evm}+2}\lrabs{\mu_v+1}}\state{k_{\ev}+2}{\mu_v+1,\mu_{\vp}-1}{} + \nonumber\\
    & \quad \hphantom{+ \lr{\sqrt{\lrabs{\mu_v+1}}-\sqrt{\lrabs{\mu_v-1}}} }\ + \sqrt{\lrabs{k_{\ev}+k_{\evm}+2}\lrabs{\mu_v-1}}\state{k_{\ev}+2}{\mu_v-1,\mu_{\vp}+1}{} - \nonumber\\
    & \quad \hphantom{+ \lr{\sqrt{\lrabs{\mu_v+1}}-\sqrt{\lrabs{\mu_v-1}}} }\ - \sqrt{\lrabs{k_{\ev}+k_{\evm}+2}\lrabs{\mu_v-1}}\state{k_{\ev}+2}{\mu_v-1,\mu_{\vp}-1}{} - \nonumber\\
    & \quad \hphantom{+ \lr{\sqrt{\lrabs{\mu_v+1}}-\sqrt{\lrabs{\mu_v-1}}} }\ - \sqrt{\lrabs{k_{\ev}+k_{\evm}-2}\lrabs{\mu_v+1}}\state{k_{\ev}-2}{\mu_v+1,\mu_{\vp}+1}{} + \nonumber\\
    & \quad \hphantom{+ \lr{\sqrt{\lrabs{\mu_v+1}}-\sqrt{\lrabs{\mu_v-1}}} }\ + \sqrt{\lrabs{k_{\ev}+k_{\evm}-2}\lrabs{\mu_v+1}}\state{k_{\ev}-2}{\mu_v+1,\mu_{\vp}-1}{} - \nonumber\\
    & \quad \hphantom{+ \lr{\sqrt{\lrabs{\mu_v+1}}-\sqrt{\lrabs{\mu_v-1}}} }\ - \sqrt{\lrabs{k_{\ev}+k_{\evm}-2}\lrabs{\mu_v+1}}\state{k_{\ev}-2}{\mu_v-1,\mu_{\vp}+1}{} + \nonumber\\
    & \quad \hphantom{+ \lr{\sqrt{\lrabs{\mu_v+1}}-\sqrt{\lrabs{\mu_v-1}}} }\ + \sqrt{\lrabs{k_{\ev}+k_{\evm}-2}\lrabs{\mu_v+1}}\state{k_{\ev}-2}{\mu_v-1,\mu_{\vp}-1}{} \Big] \bigg\} \label{eq:ActionHeuclDagger}
\end{align}
Therein, we defined 
\begin{align}
    \keucl \coloneqq \frac{\lp}{8\kappa'\immirzi^{\nicefrac{3}{2}}k_0\mu_0\nu_0}.
\end{align}
Note that we always collected the generated states in squared brackets $[\ldots]$ to provide some clarity within the long formulae. Also, we could have structured the formula above differently and especially combined terms with the same charge-dependent prefactors --- like every pair within the series of eight states in \eqref{eq:ActionHeuclDagger}. We refrained from doing so as the structure, as it is, offers an easier overview of the created recharged states. Lastly, we want to point out the first collection of newly created states in \eqref{eq:ActionHeuclDagger} that all have prefactors $\sim\sqrt{\lrabs{\mu_v\pm2}\lrabs{\nu_v\pm2}}$. These are precisely the states that do not vanish if acting on trivially charged vertices and thus forced us to redefine the basis states by setting $k_{\ev}=-k_{\evm}$ for those in order to have their mutual prefactor $\sqrt{\lrabs{k_{\ev}+k_{\evm}+1}}-\sqrt{\lrabs{k_{\ev}+k_{\evm}-1}}$ vanish (cf.~Figure~\ref{fig:GowdyAQGgraph} and Figure~\ref{fig:GowdyAQGgraphEmbedded} and their discussions).
The corresponding actions of $\lorHhatpart{1}$ and $\lorHhatpart{2}$ are more compact and read
\begin{align}
    \lorHhatpart{1} |\varphi\rangle &= \sum_{v\in V\lr{\aqggraph}} \sum_{k\in\mathbbm{Z}^N}\sum_{\mu\in m}\sum_{\nu\in n} \Coeff{}{}{} \lorHhatIpart{1} \state{}{}{} \nonumber\\
    & = \sum_{v\in V\lr{\aqggraph}} \sum_{k\in\mathbbm{Z}^N}\sum_{\mu\in m}\sum_{\nu\in n} \klorOne \Coeff{}{}{} \bigg\{ - \lr{k_{e_{\vp}}-k_{\evm}}^2 \Big[ \lrabs{k_{\ev}+k_{\evm}}^r \lrabs{\mu_v}^r \lrabs{\nu_v}^r \cdot \nonumber\\
    & \qquad \cdot \lr{\lrabs{k_{\ev}+k_{\evm}+1}^\halfalpha - \lrabs{k_{\ev}+k_{\evm}-1}^\halfalpha} \lr{\lrabs{\mu_v+1}^\halfalpha-\lrabs{\mu_v-1}^\halfalpha} \cdot \nonumber\\
    & \qquad \cdot \left.\left.  \lr{\lrabs{\nu_v+1}^\halfalpha-\lrabs{\nu_v-1}^\halfalpha} \right]^l \bigg\}\right\rvert_{r=\frac{2}{3}-\frac{1}{3l}} \state{}{}{} \label{eq:ActionHlor1}
\end{align}
and
\begin{align}
    & \lorHhatpart{2} |\varphi\rangle = \sum_{v\in V\lr{\aqggraph}} \sum_{k\in\mathbbm{Z}^N}\sum_{\mu\in m}\sum_{\nu\in n} \Coeff{}{}{} \lorHhatIpart{2} \state{}{}{} \nonumber\\
    & = \sum_{v\in V\lr{\aqggraph}} \sum_{k\in\mathbbm{Z}^N}\sum_{\mu\in m}\sum_{\nu\in n} \klorTwo \Coeff{}{}{} \bigg\{ \lr{k_{\ev}+k_{\evm}}^4 \lr{\mu_v\nu_{\vp} - \mu_{\vp}\nu_v}^2 \left[ \lrabs{k_{\ev}+k_{\evm}}^{r_2} \lrabs{\mu_v}^{r_2} \cdot \right. \nonumber\\
    & \qquad \cdot \lrabs{\nu_v}^{r_2} \lr{\lrabs{k_{\ev}+k_{\evm}+1}^\halfbeta - \lrabs{k_{\ev}+k_{\evm}-1}^\halfbeta} \lr{\lrabs{\mu_v+1}^\halfbeta-\lrabs{\mu_v-1}^\halfbeta} \cdot \nonumber\\
    & \qquad \left.\left. \cdot \lr{\lrabs{\nu_v+1}^\halfbeta-\lrabs{\nu_v-1}^\halfbeta}\right]^l \bigg\}\right\rvert_{{r_2}=\frac{2}{3}-\frac{5}{3l}} \state{}{}{}, \label{eq:ActionHlor2}
\end{align}
where we collected all constants in 
\begin{equation}
\begin{aligned}
    \klorOne & \coloneqq \frac{4\pi^2\lp\sqrt{\immirzi}}{16\kappa'}\lr{\frac{1}{2r^3k_0\mu_0\nu_0}}^l\Bigg\rvert_{r=\frac{2}{3}-\frac{1}{3l}} \quad \text{and} \\
    \klorTwo & \coloneqq \frac{4\pi^2\lp\sqrt{\immirzi}}{16\kappa'}\lr{\frac{1}{2{r_2}^3k_0\mu_0\nu_0}}^l\Bigg\rvert_{{r_2}=\frac{2}{3}-\frac{5}{3l}}. \label{eq:klorConstants}
\end{aligned}
\end{equation}
We can now combine all these contributions to state the full action of $\hatHphys$:
\begin{align} 
    & \hatHphys |\varphi\rangle = \nonumber\\
    & = \sum_{v\in V\lr{\aqggraph}} \sum_{k\in\mathbbm{Z}^N}\sum_{\mu\in m}\sum_{\nu\in n} \Coeff{}{}{} \left\{ \frac{1}{2}\euclHhatI + \frac{1}{2}(\euclHhatI)^\dagger + \lorHhatIpart{1} + \lorHhatIpart{2} \right\} |k,\mu,\nu\rangle \\
    &= \sum_{v\in V\lr{\aqggraph}} \sum_{k\in\mathbbm{Z}^N}\sum_{\mu\in m}\sum_{\nu\in n} \Coeff{}{}{}  \left\{ \frac{\keucl}{2} \lr{\sqrt{\lrabs{k_{\ev}+k_{\evm}+1}}-\sqrt{\lrabs{k_{\ev}+k_{\evm}-1}}} \cdot \right.\nonumber\\
    & \qquad \cdot \left[ \lr{\sqrt{\lrabs{\mu_v+2}\lrabs{\nu_v+2}} + \sqrt{\lrabs{\mu_v}\lrabs{\nu_v}}} |k,\mu_v+2,\nu_v+2\rangle - \right.\nonumber\\
    & \quad\qquad - \lr{\sqrt{\lrabs{\mu_v+2}\lrabs{\nu_v-2}}+\sqrt{\lrabs{\mu_v}\lrabs{\nu_v}}} |k,\mu_v+2,\nu_v-2\rangle - \nonumber\\
    & \quad\qquad -\lr{\sqrt{\lrabs{\mu_v-2}\lrabs{\nu_v+2}}+\sqrt{\lrabs{\mu_v}\lrabs{\nu_v}}} |k,\mu_v-2,\nu_v+2\rangle + \nonumber\\
    & \,\quad\qquad \left. + \lr{\sqrt{\lrabs{\mu_-+2}\lrabs{\nu_v-2}}+\sqrt{\lrabs{\mu_v}\lrabs{\nu_v}}} |k,\mu_v-2,\nu_v-2\rangle |k,\mu_v-2,\nu_v-2\rangle \right] + \nonumber\\
    & \quad + \keucl \lr{\sqrt{\lrabs{\mu_v+1}}-\sqrt{\lrabs{\mu_v-1}}} \cdot \nonumber\\
    & \qquad \cdot \left[ \lr{\sqrt{\lrabs{k_{\ev}+k_{\evm}+2}\lrabs{\nu_v+1}}+\sqrt{\lrabs{k_{\ev}+k_{\evm}}\lrabs{\nu_v}}} |k_{\ev}+2,\mu,\nu_v+1,\nu_{\vp}+1\rangle -\right.\nonumber\\
    & \quad\qquad - \lr{\sqrt{\lrabs{k_{\ev}+k_{\evm}+2}\lrabs{\nu_v+1}}+\sqrt{\lrabs{k_{\ev}+k_{\evm}}\lrabs{\nu_v}}} |k_{\ev}+2,\mu,\nu_v+1,\nu_{\vp}-1\rangle +\nonumber\\
    & \quad\qquad + \lr{\sqrt{\lrabs{k_{\ev}+k_{\evm}+2}\lrabs{\nu_v-1}}+\sqrt{\lrabs{k_{\ev}+k_{\evm}}\lrabs{\nu_v}}} |k_{\ev}+2,\mu,\nu_v-1,\nu_{\vp}+1\rangle -\nonumber\\
    & \quad\qquad - \lr{\sqrt{\lrabs{k_{\ev}+k_{\evm}+2}\lrabs{\nu_v-1}}+\sqrt{\lrabs{k_{\ev}+k_{\evm}}\lrabs{\nu_v}}} |k_{\ev}+2,\mu,\nu_v-1,\nu_{\vp}-1\rangle - \nonumber\\
    & \quad\qquad -\lr{\sqrt{\lrabs{k_{\ev}+k_{\evm}-2}\lrabs{\nu_v+1}}+\sqrt{\lrabs{k_{\ev}+k_{\evm}}\lrabs{\nu_v}}} |k_{\ev}-2,\mu,\nu_v+1,\nu_{\vp}+1\rangle + \nonumber\\
    & \quad\qquad +\lr{\sqrt{\lrabs{k_{\ev}+k_{\evm}-2}\lrabs{\nu_v+1}}+\sqrt{\lrabs{k_{\ev}+k_{\evm}}\lrabs{\nu_v}}} |k_{\ev}-2,\mu,\nu_v+1,\nu_{\vp}-1\rangle - \nonumber\\
    & \quad\qquad - \lr{\sqrt{\lrabs{k_{\ev}+k_{\evm}-2}\lrabs{\nu_v-1}}+\sqrt{\lrabs{k_{\ev}+k_{\evm}}\lrabs{\nu_v}}} |k_{\ev}-2,\mu,\nu_v-1,\nu_{\vp}+1\rangle +\nonumber\\
    & \quad\qquad \left. + \lr{\sqrt{\lrabs{k_{\ev}+k_{\evm}-2}\lrabs{\nu_v-1}}+\sqrt{\lrabs{k_{\ev}+k_{\evm}}\lrabs{\nu_v}}} |k_{\ev}-2,\mu,\nu_v-1,\nu_{\vp}-1\rangle  \right] + \nonumber\\
    & \quad + \keucl \lr{\sqrt{\lrabs{\nu_v+1}}-\sqrt{\lrabs{\nu_v-1}}} \cdot \nonumber\\
    & \qquad \cdot \left[ \lr{\sqrt{\lrabs{k_{\ev}+k_{\evm}+2}\lrabs{\mu_v+1}}+\sqrt{\lrabs{k_{\ev}+k_{\evm}}\lrabs{\mu_v}}} |k_{\ev}+2,\mu_v+1,\mu_{\vp}+1,\nu\rangle -\right.\nonumber\\
    & \quad\qquad - \lr{\sqrt{\lrabs{k_{\ev}+k_{\evm}+2}\lrabs{\mu_v+1}}+\sqrt{\lrabs{k_{\ev}+k_{\evm}}\lrabs{\mu_v}}} |k_{\ev}+2,\mu_v+1,\mu_{\vp}-1,\nu\rangle +\nonumber\\
    & \quad\qquad + \lr{\sqrt{\lrabs{k_{\ev}+k_{\evm}+2}\lrabs{\mu_v-1}}+\sqrt{\lrabs{k_{\ev}+k_{\evm}}\lrabs{\mu_v}}} |k_{\ev}+2,\mu_v-1,\mu_{\vp}+1,\nu\rangle -\nonumber\\
    & \quad\qquad - \lr{\sqrt{\lrabs{k_{\ev}+k_{\evm}+2}\lrabs{\mu_v-1}}+\sqrt{\lrabs{k_{\ev}+k_{\evm}}\lrabs{\mu_v}}} |k_{\ev}+2,\mu_v-1,\mu_{\vp}-1,\nu\rangle - \nonumber\\
    & \quad\qquad -\lr{\sqrt{\lrabs{k_{\ev}+k_{\evm}-2}\lrabs{\mu_v+1}}+\sqrt{\lrabs{k_{\ev}+k_{\evm}}\lrabs{\mu_v}}} |k_{\ev}-2,\mu_v+1,\mu_{\vp}+1,\nu\rangle + \nonumber\\ 
    & \quad\qquad +\lr{\sqrt{\lrabs{k_{\ev}+k_{\evm}-2}\lrabs{\mu_v+1}}+\sqrt{\lrabs{k_{\ev}+k_{\evm}}\lrabs{\mu_v}}} |k_{\ev}-2,\mu_v+1,\mu_{\vp}-1,\nu\rangle - \nonumber\\
    & \quad\qquad - \lr{\sqrt{\lrabs{k_{\ev}+k_{\evm}-2}\lrabs{\mu_v-1}}+\sqrt{\lrabs{k_{\ev}+k_{\evm}}\lrabs{\mu_v}}} |k_{\ev}-2,\mu_v-1,\mu_{\vp}+1,\nu\rangle +\nonumber\\
    & \quad\qquad \left. + \lr{\sqrt{\lrabs{k_{\ev}+k_{\evm}-2}\lrabs{\mu_v-1}}+\sqrt{\lrabs{k_{\ev}+k_{\evm}}\lrabs{\mu_v}}} |k_{\ev}-2,\mu_v-1,\mu_{\vp}-1,\nu\rangle  \right] - \nonumber\\
    & \quad - \klorOne \lr{k_{\evp}-k_{\evm}}^2 \left[ \lrabs{k_{\ev}+k_{\evm}}^r \lrabs{\mu_v}^r\lrabs{\nu_v}^r \lr{\lrabs{k_{\ev}+k_{\evm}+1}^{\halfalpha} - \lrabs{k_{\ev}+k_{\evm}-1}^{\halfalpha}} \cdot \right. \nonumber\\
    &\quad\qquad \cdot\left. \lr{\lrabs{\mu_v+1}^{\halfalpha}-\lrabs{\mu_v-1}^{\halfalpha}} \lr{\lrabs{\nu_v+1}^{\halfalpha}-\lrabs{\nu_v-1}^{\halfalpha}} \right]^l |k,\mu,\nu\rangle + \nonumber\\
    & \quad + \klorTwo \lr{k_{\ev}+k_{\evm}}^4 \lr{\mu_v\nu_{\vp}-\mu_{\vp}\nu_v}^2 \left[ \lrabs{k_{\ev}+k_{\evm}}^{r_2} \lrabs{\mu_v}^{r_2} \lrabs{\nu_v}^{r_2}  \right. \nonumber\\
    &\quad\qquad \cdot\lr{\lrabs{k_{\ev}+k_{\evm}+1}^{\halfbeta} - \lrabs{k_{\ev}+k_{\evm}-1}^{\halfbeta}} \lr{\lrabs{\mu_v+1}^{\halfbeta}-\lrabs{\mu_v-1}^{\halfbeta}} \cdot\nonumber\\
    & \quad\qquad\cdot \left.\left.\left. \lr{\lrabs{\nu_v+1}^{\halfbeta}-\lrabs{\nu_v-1}^{\halfbeta}} \right]^l |k,\mu,\nu\rangle \right\}\right\rvert_{r=\frac{2}{3}-\frac{1}{3l}\,\land\, {r_2}=\frac{2}{3}-\frac{5}{3l}} \label{eq:ActionHPhys}
\end{align}
Note that some contributions from $\euclHhat$ and $(\euclHhat)^\dagger$ were combined, while some identical numerical charge-dependent prefactors were not factored out in order to keep a form that allows for an easy read-out of the newly created states.

\subsection{On specific solutions of the Schrödinger equation} \label{subsec:Schrödinger}
Having found an appropriate physical Hamiltonian \eqref{eq:physHam} and states \eqref{eq:statesWithTime}, we can approach solving the Schrödinger-like equation
\begin{align}
    \i\hbar\,\partial_\tau |\Psi\lr{\tau}\rangle = \hatHphys |\Psi\lr{\tau}\rangle .
\end{align}
We already introduced the well-known separation ansatz
\begin{equation}
    |\Psi\lr{\tau}\rangle = |\varphi\lr{k,\mu,\nu}\rangle |\chi\lr{\tau}\rangle \tag{\ref{eq:statesWithTime}}
\end{equation}
for the states and we will later see that additional ansätze of this kind allow us to better understand the action of the physical Hamiltonian.

With this partitioning of the state, we can proceed to the time-independent version of the Schrödinger-like equation just like in standard quantum mechanics and obtain
\begin{align}
    \hatHphys |\varphi\rangle = E |\varphi\rangle,
\end{align}
where we used $E$ as the constant that arises due to the separation of the variables.

This eigenvalue equation is now easier to solve, yet the involved action of the physical Hamiltonian makes it still very complicated to find general solutions. For this reason, we will first search for zero-volume eigenstates in the next subsection, as all terms of $\hatHphys$ do indeed contain the volume operator this corresponds to the case $E=0$. This is also an illustrative introduction in how to handle the action of operators on Gowdy states because determining the spectrum of $\hatHphys$ is far beyond the scope of this article. Furthermore, the special case of choosing $E=0$ corresponds at the classical level to the limiting case where the dust energy density vanishes and thus should in some formal sense make contact to the vacuum Gowdy case. A more rigorous understanding of taking this limit in the quantum theory will be necessary in future work as well as analysing the question whether zero is involved in the spectrum of $\hatHphys$ at all; both questions will not be addressed in this work. Here, considering this specific choice should rather be understood as a toy example in which we can obtain some first intuition about the action of the physical Hamiltonian operator on Gowdy states. Since the physical Hamiltonian consists of a sum of individual operators that splits into an Euclidean and a Lorentzian part, knowing the spectrum of the individual parts does not give us insight into the spectrum of the total operator in general. However, as we will discuss in the conclusions for the Gowdy model presented in this paper, a perturbative approach along the lines of \cite{Assanioussi:2017tql}, where the Euclidean contribution is treated as a perturbation of the Lorentzian part, may be an option for analysing the model. From this perspective, it is therefore useful to discuss the actions of the Euclidean and Lorentzian parts separately. 

\subsubsection{Zero-volume eigenstates} \label{sec:ZeroVol}

This section is about finding states $|\varphi\rangle$ for which the volume vanishes. While this certainly holds for trivially charged states $k_{\ev}=\mu_v=\nu_v=0, \forall v\in V\lr{\aqggraph},$ there are also ones with less rigid restrictions. We will use this chapter about finding those zero-volume states also as an introduction for what comes afterwards, as the technique of finding constraints for the coefficients $\Coeff{}{}{}$ such that the corresponding state $|\varphi\rangle$ features a desired property is the basis of our treatment of the Schrödinger-like equation, too.

We can derive the action of the volume operator on the states $|\varphi\rangle$ from \eqref{eq:VolAQGaction}:
\begin{align}
    \V|\varphi\rangle & = \sum_{v\in V\lr{\aqggraph}} \sum_{k\in\mathbbm{Z}^N}\sum_{\mu\in m}\sum_{\nu\in n} \Coeff{}{}{} \V_v \state{}{}{} \nonumber\\
    & = \sum_{v\in V\lr{\aqggraph}} \sum_{k\in\mathbbm{Z}^N}\sum_{\mu\in m}\sum_{\nu\in n} \Coeff{}{}{} \frac{1}{\sqrt{2}}\lr{\frac{\immirzi\lp{}^2}{2}}^{\frac{3}{2}} \sqrt{\lrabs{k_{\ev}+k_{\evm}} \lrabs{\mu_v} \lrabs{\nu_v}} \state{}{}{}. 
\end{align}
If we now want to find solutions for which the above eigenvalue vanishes, we deduce the condition
\begin{align}
    \sqrt{\lrabs{k_{\ev}+k_{\evm}} \lrabs{\mu_v} \lrabs{\nu_v}} = 0 \ \forall v\in V\lr{\aqggraph}. \label{eq:zeroVolume}
\end{align}
Note that contributions from different vertices can not sum up to zero as there are no negative eigenvalues, i.e.~no negative volume contribution. This leads to the following basic conditions:
\begin{enumerate}
    \item $k_{\ev}=0=k_{\evm}$ and $\mu_v, \nu_v$ arbitrary
    \item $k_{\ev}=-k_{\evm}$ and $\mu_v, \nu_v$ arbitrary
    \item $\mu_v=0$ and $k_{\ev}, \nu_v$ arbitrary
    \item $\nu_v=0$ and $k_{\ev}, \mu_v$ arbitrary
\end{enumerate}
The charges $k_{\ev}$ play a special role as neighbouring $k_{\ev}$ are coupled via $ \sqrt{\lrabs{k_{\ev}+k_{\evm}}}$. This is the reason why setting $k_{\ev}=0$ is not sufficient for fulfilling \eqref{eq:zeroVolume}, but instead at least $k_{\evm}=0$ has to be chosen, too. Now, for having the total volume of $|\varphi\rangle$ to vanish, one may combine vertex-wise any of the above conditions.
However, if one wants to construct zero-volume states, one rather works with the coefficients $\Coeff{}{}{}$. The above conditions for the charges then translate into conditions for the coefficients by assigning only to all those $\Coeff{}{}{}$ a non-zero value such that the thereby non-suppressed states $\state{}{}{}$ fulfil, as a set, for all vertices at least one of the above conditions. As an example, if we wish to have a state that has zero volume through $\mu_v=0 \ \forall v \in V\lr{\aqggraph},$ we set all those $\Coeff{}{}{}$ to zero whose set $\mu$ contains at least one $\mu_v\neq0$:
\begin{align}
    \mu_v=0 \ \forall v \in V\lr{\aqggraph} \Longleftrightarrow \Coeff{}{}{}=0 \ \text{if}\ \exists v \in V\lr{\aqggraph} \colon 0\neq \mu_v \in \mu.
\end{align}
This way, by assigning specific values to certain coefficients $\Coeff{}{}{}$, we can construct states that fulfil a desired property such as having zero volume.

\subsubsection{Vanishing action states for \texorpdfstring{$\euclHhat$}{Heucl}} 

The procedure of carrying over to conditions the coefficients $\Coeff{}{}{}$ have to satisfy can also be used to construct states that cause a vanishing action of $\euclHhat$. We exemplarily show this for the three-vertex graph --- with basis states denoted by $|\varphi_3\rangle = \state{}{}{}_{3}$ --- as it allows for clearer formulae. To keep our notation more compact, we use the abbreviations
\begin{align}
 v_I&\eqqcolon I, & & & & \\
 k_{e_{v_I}}&\eqqcolon k_I,&  k_{e_{v_I^-}}&\eqqcolon k_{I-1}, & k_{e_{v_I^+}}&\eqqcolon k_{I+1}, \nonumber\\
 \mu_{v_I}&\eqqcolon \mu_I,& \mu_{\vm}&\eqqcolon \mu_{I-1}, & \mu_{\vp}&\eqqcolon \mu_{I+1}, \nonumber \\
 \nu_{v_I}&\eqqcolon \nu_I,& \nu_{\vm}&\eqqcolon \nu_{I-1},& \nu_{\vp}&\eqqcolon \nu_{I+1}.
\end{align}
With this notation, we deduce from~\eqref{eq:ActionHeucl} the following form of the action of $\euclHhat$ on $|\varphi_3\rangle$ after having performed substitutions of the charges so there is no shift in the basis states anymore:
\begin{align}
    & \euclHhat |\varphi_3\rangle = \sum_{I=1}^3 \sum_{k\in\mathbbm{Z}^3}\sum_{\mu\in m}\sum_{\nu\in n} \Coeff{}{}{} \euclHhatI \state{}{}{}_3 \nonumber\\
    & = \sum_{I=1}^3 \sum_{k\in\mathbbm{Z}^3}\sum_{\mu\in m}\sum_{\nu\in n} \keucl \bigg\{ \lr{\sqrt{\lrabs{k_I+k_{I-1}+1}}-\sqrt{\lrabs{k_I+k_{I-1}-1}}} \cdot \nonumber\\
    & \qquad \cdot \left( \Coeff{}{\mu_I-2}{\nu_I-2}\sqrt{\lrabs{\mu_I-2}\lrabs{\nu_I-2}} + \Coeff{}{\mu_I+2}{\nu_I+2} \sqrt{\lrabs{\mu_I+2}\lrabs{\nu_I+2}} \right. - \nonumber\\
    & \qquad \quad\ \left. - \Coeff{}{\mu_I+2}{\nu_I-2}\sqrt{\lrabs{\mu_I+2}\lrabs{\nu_I-2}} - \Coeff{}{\mu_I-2}{\nu_I+2} \sqrt{\lrabs{\mu_I-2}\lrabs{\nu_I+2}} \right) + \nonumber\\
    & \quad + \lr{\sqrt{\lrabs{\mu_I+1}}-\sqrt{\lrabs{\mu_I-1}}} \cdot \nonumber\\
    & \qquad \cdot \Big[ \lr{\Coeff{k_I+2}{}{\nu_I+1,\nu_{I+1}+1} - \Coeff{k_I+2}{}{\nu_I+1,\nu_{I+1}-1}} \sqrt{\lrabs{k_I+k_{I-1}+2}\lrabs{\nu_I+1}} + \nonumber\\
    & \qquad\quad + \lr{\Coeff{k_I+2}{}{\nu_I-1,\nu_{I+1}+1} - \Coeff{k_I+2}{}{\nu_I-1,\nu_{I+1}-1}} \sqrt{\lrabs{k_I+k_{I-1}+2}\lrabs{\nu_I-1}} + \nonumber\\
    & \qquad\quad + \lr{\Coeff{k_I-2}{}{\nu_I+1,\nu_{I+1}-1} - \Coeff{k_I-2}{}{\nu_I+1,\nu_{I+1}+1}} \sqrt{\lrabs{k_I+k_{I-1}-2}\lrabs{\nu_I+1}} +  \nonumber\\
    & \qquad\quad + \lr{\Coeff{k_I-2}{}{\nu_I-1,\nu_{I+1}-1} - \Coeff{k_I-2}{}{\nu_I-1,\nu_{I+1}+1}} \sqrt{\lrabs{k_I+k_{I-1}-2}\lrabs{\nu_I-1}} \Big]_\sharp + \nonumber\\
    & \quad + \lr{\sqrt{\lrabs{\nu_I+1}}-\sqrt{\lrabs{\nu_I-1}}} \cdot \Big[ \ldots \lr{\mu\longleftrightarrow\nu} \ldots \Big]_\sharp \bigg\} \state{}{}{}_3 \label{eq:euclAction3state}.
\end{align}
Therein, \guillemotright\ $\big[ \ldots \lr{\mu\longleftrightarrow\nu} \ldots \big]_\sharp$ \guillemotleft\ stands for the square bracket with subscript $\sharp$ of the four lines before just with the roles of $\mu$ and $\nu$ interchanged. The notation for the coefficients $\Coeff{}{}{}{}$ is similar to the one of the states that we already used: Only the charges that were de- or increased are specifically denoted.
The above formula is, of course, only possible since we sum over all charges from $-\infty$ to $\infty$ and the substitution therefore does not change the solution space.

Noticing that there is no mixing of the three classes of charges $k$, $\mu$ and $\nu$, we introduce the separation ansatz
\begin{equation}
    \Coeff{}{}{} = \kCoeff{} \cdot \muCoeff{} \cdot \nuCoeff{} .
\end{equation}
With this, the above action becomes
\begin{align}
    & \euclHhat |\varphi_3\rangle = \sum_{I=1}^3 \sum_{k\in\mathbbm{Z}^3}\sum_{\mu\in m}\sum_{\nu\in n} \keucl \Bigg\{ \kCoeff{} \lr{\sqrt{\lrabs{k_I+k_{I-1}+1}}-\sqrt{\lrabs{k_I+k_{I-1}-1}}} \cdot \nonumber\\
    & \qquad \cdot \lr{ \muCoeff{\mu_I+2}\sqrt{\lrabs{\mu_I+2}} - \muCoeff{\mu_I-2}\sqrt{\lrabs{\mu_I-2}} } \cdot \lr{ \nuCoeff{\nu_I+2}\sqrt{\lrabs{\nu_I+2}} - \nuCoeff{\nu_I-2}\sqrt{\lrabs{\nu_I-2}} } + \nonumber\\
    & + \lr{\kCoeff{k_I+2}\sqrt{\lrabs{k_I+k_{I-1}+2}}-\kCoeff{k_I-2}\sqrt{\lrabs{k_I+k_{I-1}-2}} }\cdot \nonumber\\
    & \cdot \bigg[ \muCoeff{}\lr{\sqrt{\lrabs{\mu_I+1}}-\sqrt{\lrabs{\mu_I-1}}}  \lr{ \lr{\nuCoeff{\nu \stackpp 1}-\nuCoeff{\nu\stackpm 1}}\sqrt{\lrabs{\nu_I+1}} + \lr{\nuCoeff{\nu\stackmp1}-\nuCoeff{\nu\stackmm1}}\sqrt{\lrabs{\nu_I-1}} }  \nonumber\\
    & \quad\ + \qquad \text{\dittostraight}_{\stackrel{}{\lr{\mu\longleftrightarrow\nu}}}  \qquad  \bigg] \Bigg\} \state{}{}{}_3, \label{eq:euclAction3stateSeparation}
\end{align}
where the last line's \guillemotright\ $\text{\dittostraight}_{\stackrel{}{\lr{\mu\longleftrightarrow\nu}}}$ \guillemotleft\ denotes the ditto mark of the line before with the roles of $\mu$ and $\nu$ interchanged. As before, while $\kCoeff{}$ stands for the coefficient representing all unshifted $k$-charges, $\kCoeff{k_I+2}$ means that all but $k_I$ are unshifted and $k_I$ is increased by two. We then introduced a new abbreviation for coefficients that feature shifts in the charges of both $v_I$ and $v_{I+1}$: $\nuCoeff{\nu\stackpp1} \coloneqq \nuCoeff{\nu_I+1,\nu_{I+1}+1},\ \nuCoeff{\nu\stackpm1} \coloneqq \nuCoeff{\nu_I+1,\nu_{I+1}-1},\ \nuCoeff{\nu\stackmp1} \coloneqq \nuCoeff{\nu_I-1,\nu_{I+1}+1} \text{ and } \nuCoeff{\nu\stackmm1} \coloneqq \nuCoeff{\nu_I-1,\nu_{I+1}-1}$.

To get an intuition of the formula above, we may consider graphs for which the action vanishes. We can then state two basic principles to achieve this --- or, in fact, any other degeneracy, too:
\begin{enumerate}
    \item the sum over the vertices causes the individual contributions to cancel each other / equal the desired value, or
    \item each individual contribution vanishes / amounts for the same contribution to the desired value.
\end{enumerate}
While for the last one we can ignore the sum over the vertices and just need to find coefficients that make up for $\nicefrac{1}{\sharp\lr{\text{vertices}}}$-th of the final result's value, the first one is more complicated and we may not find general solutions that reflect, e.g., the symmetries of the Gowdy models. For that reason, we concentrate on solutions that fulfil the chosen constraint vertex-wise.

\begin{figure}
    \begin{tikzpicture}
    \node (c111) [] {$\kCoeff{111}$}; 
    \node (2ndCentre) [below of=c111,yshift=-1cm] {}; 
    \node (3rdCentre) [below of=2ndCentre,yshift=-1cm] {}; 
    \node (4thCentre) [below of=3rdCentre, yshift=-1cm] {$\leftarrow \text{\Lightning} \rightarrow$}; 
    \node (2ndLeft) [left of=2ndCentre, xshift=-1cm] {$\sqrt{\frac{2}{6}}\kCoeff{111}=\kCoeff{115}$}; 
    \node (3rdLeft) [below =of 2ndLeft.east, anchor=east, yshift=-1cm] {$\sqrt{\frac{2}{6}}\kCoeff{115}=\kCoeff{155}$}; 
    \node (4thLeft) [below = of 3rdLeft.east, anchor=east, yshift=-1cm] {$\kCoeff{155}=\sqrt{\frac{1}{9}}\kCoeff{111}$}; 
    \node (2ndRight) [right of=2ndCentre,xshift=1cm] {$\kCoeff{151}=\sqrt{\frac{2}{6}}\kCoeff{111}$}; 
    \node (3rdRight) [below = of 2ndRight.west,anchor=west, yshift=-1cm] {$\kCoeff{155} = \sqrt{\frac{6}{10}}\kCoeff{151}$}; 
    \node (4thRight) [below = of 3rdRight.west,anchor=west, yshift=-1cm] {$\kCoeff{155}=\sqrt{\frac{1}{5}}\kCoeff{111}$}; 
    \tikzstyle{arrow} = [thick,->,>=stealth]
    \draw [arrow] (c111) --node[anchor=south east] {$ k_3\!\uparrow$} (2ndLeft.north east); 
    \draw [arrow,transform canvas={xshift=-.5cm}] (2ndLeft.south east) --node[anchor=east] {$k_2\!\uparrow$} (3rdLeft.north east);
    \draw [arrow,transform canvas={xshift=-.5cm}] (3rdLeft.south east) --node[anchor=east] {i.e.} (4thLeft.north east);
    \draw [arrow] (c111) --node[anchor=south west] {$k_2\!\uparrow$} (2ndRight.north west); 
    \draw [arrow,transform canvas={xshift=.5cm}] (2ndRight.south west) --node[anchor=west] {$k_3\!\uparrow$} (3rdRight.north west);
    \draw [arrow,transform canvas={xshift=.5cm}] (3rdRight.south west) --node[anchor=west] {i.e.} (4thRight.north west);
    \end{tikzpicture}
    \caption{On the condition \eqref{eq:kCondition} on $k$}
    \label{fig:kConditionGraph}
\end{figure}
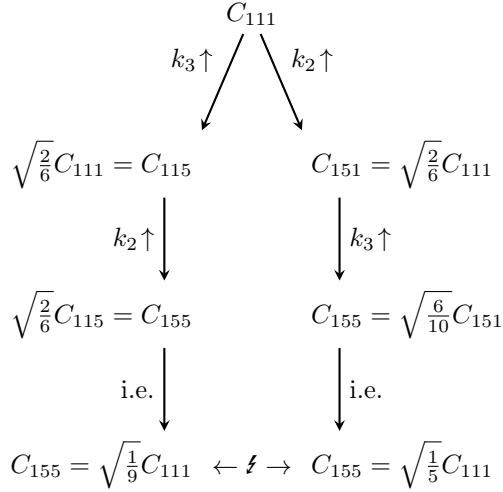

If we want \eqref{eq:euclAction3stateSeparation} to vanish, we first of all notice that we face a sum of two contributions within the curly brackets. As the two are products and sums/differences of square roots, we may conclude that it is unlikely for the two to annihilate, although that can be achieved for specific choices. Hence, we want both summands to vanish on their own, which can be fulfilled, e.g., by
\begin{enumerate}
    \item $\muCoeff{\mu_I+2}\sqrt{\lrabs{\mu_I+2}} = \muCoeff{\mu_I-2}\sqrt{\lrabs{\mu_I-2}} $
    \item $\kCoeff{k_I+2}\sqrt{\lrabs{k_I+k_{I-1}+2}} = \kCoeff{k_I-2}\sqrt{\lrabs{k_I+k_{I-1}-2}}$.
\end{enumerate}
The first choice is, of course, equivalently replaceable by the corresponding condition with $\mu_I\mapsto\nu_I$ due to the multiplicative structure of the first summand within \eqref{eq:euclAction3stateSeparation}. Likewise, one could replace the above second condition with one that makes the second summand of \eqref{eq:euclAction3stateSeparation} vanish via a zero contribution from the square brackets. Considering this bracket's additive structure, in turn, more than one such condition would be required. Therefore, the above conditions can be considered as the most basic ones. The chart in Figure~\ref{fig:kConditionGraph} illustrates what this condition means in terms of the three $k_I$-charges, after having gone over to the equivalent form
\begin{equation}
    \kCoeff{k_I+4} = \sqrt{\frac{\lrabs{k_I+k_{I-1}}}{\lrabs{k_I+k_{I-1}+4}}}\kCoeff{}: \label{eq:kCondition}
\end{equation}
Starting with a specific value for $\kCoeff{111}$, we can determine the value of, say, $\kCoeff{115}$ afterwards. This means that there are two paths leading to the new value of $\kCoeff{155}$: Either via $\kCoeff{111}\rightarrow\kCoeff{115}\rightarrow\kCoeff{155}$ or $\kCoeff{111}\rightarrow\kCoeff{151}\rightarrow\kCoeff{155}$. Doing so, we get contradictory results for the two paths (cf.~Figure~\ref{fig:kConditionGraph}). This is due to the condition depending also on $k_{I-1}$. During the path via $\kCoeff{151}$, this charge was increased by four before being evaluated, while it was increased only after being evaluated during the other path --- which is also why there the prefactor $\sqrt{\nicefrac{2}{6}}$ appears in both steps. Lastly, note that the trivial solution is of course not excluded and marks the only scenario where the above contradiction does not apply.
Even though this condition turned out to be inapplicable, it showed the general strategy we pursue --- and where one has to be cautious. Proceeding to the condition for $\mu$ (or $\nu$), we first of all notice that there is no link between the respective charges of different vertices. Therefore, we may separate the coefficients once more into
\begin{align}
    \muCoeff{} &= \singleCoeff{\mu_1}\cdot \singleCoeff{\mu_2} \cdot \singleCoeff{\mu_3} \text{ and}\\
    \nuCoeff{} &= \singleCoeff{\nu_1}\cdot \singleCoeff{\nu_2} \cdot \singleCoeff{\nu_3} .
\end{align}
With this and moving the shift again into one of the coefficients only, the condition now reads
\begin{equation}
    \muCoeff{\mu_I+2}\sqrt{\lrabs{\mu_I+2}} = \muCoeff{\mu_I-2}\sqrt{\lrabs{\mu_I-2}} \ \Rightarrow\ \singleCoeff{\mu_I+4} = \sqrt{\frac{\lrabs{\mu_I}}{\lrabs{\mu_I+4}}}\singleCoeff{\mu_I} . \label{cond:muConditionFirstPart}
\end{equation}
This condition is, e.g., fulfilled by
\begin{align}
    \singleCoeff{\mu_I} = \begin{cases}
    \text{arbitrary}, & \text{for}\ \mu_I=0 \\
    0, & \forall \mu_I \in 4\mathbbm{Z} \\
    \sqrt{\frac{1}{\lrabs{\mu_I}}}, &\text{rest}  \end{cases} . \label{sol:muConditionFirstPart}
\end{align}
Due to the denominator diverging for $\mu_I=-4$, the first two cases of the solution above are rather formal. As the condition \eqref{cond:muConditionFirstPart} is a recurrence relation of order 4, we chose the remaining three initial conditions as $\singleCoeff{1}=1, \singleCoeff{2} = \sqrt{\frac{1}{2}}, \singleCoeff{3}=\sqrt{\frac{1}{3}}$ to obtain the intuitive solution \eqref{sol:muConditionFirstPart}. With this, we can now set the first part of the action to zero for all vertices. However, as shown before, we need a different procedure than the initially discussed one for the second part. Starting with the relevant part after the last separation, the square brackets of \eqref{eq:euclAction3stateSeparation}
\begin{align}
    & \Big[\ldots\Big]\big(\eqref{eq:euclAction3stateSeparation}\big) =  \nonumber\\
    & = \muCoeff{}\big(\sqrt{\lrabs{\mu_I+1}}-\sqrt{\lrabs{\mu_I-1}}\big) \singleCoeff{\nu_{I+2}} \lr{\singleCoeff{\nu_{I+1}+1}-\singleCoeff{\nu_{I+1}-1}} \big( \singleCoeff{\nu_I+1}\sqrt{\lrabs{\nu_I+1}} + \singleCoeff{\nu_I-1}\sqrt{\lrabs{\nu_I-1}} \big) \nonumber\\
    & \quad + \qquad \text{\dittostraight}_{\stackrel{}{\lr{\mu\longleftrightarrow\nu}}} , \label{eq:euclSeparationSecondPart}
\end{align}
we realise we can choose any of the following four options for making this expression vanish:
\begin{subequations}
  \begin{align}
    \singleCoeff{\nu_{I+1}+1}-\singleCoeff{\nu_{I+1}-1}&=0 \qquad \text{or} \label{cond:nuEquality}\\
    \singleCoeff{\nu_I+1}\sqrt{\lrabs{\nu_I+1}} + \singleCoeff{\nu_I-1}\sqrt{\lrabs{\nu_I-1}} &=0 \label{cond:nuFraction}
  \end{align}
\end{subequations}
\noindent and
\begin{subequations}
  \begin{align}
    \!\!\singleCoeff{\mu_{I+1}+1}-\singleCoeff{\mu_{I+1}-1} &=0 \qquad \text{or} \label{cond:muEquality} \\
    \!\!\singleCoeff{\mu_I+1}\sqrt{\lrabs{\mu_I+1}} + \singleCoeff{\mu_I-1}\sqrt{\lrabs{\mu_I-1}} &=0 . \label{cond:muFraction}
  \end{align}
\end{subequations}
With the second line representing conditions on the $\singleCoeff{\mu_I}$, we have to guarantee compatibility with the previously obtained solution \eqref{sol:muConditionFirstPart}, or rather the constraint \eqref{cond:muConditionFirstPart} behind it. However, it is easy to show that solutions of \eqref{cond:muFraction} automatically fulfil \eqref{cond:muConditionFirstPart}:
\begin{align}
    \eqref{cond:muFraction} \Rightarrow
    \begin{cases}
    \singleCoeff{\mu_I+2}\sqrt{\lrabs{\mu_I+2}} = -\singleCoeff{\mu_I}\sqrt{\lrabs{\mu_I}} \searrow \\
    \singleCoeff{\mu_I}\sqrt{\lrabs{\mu_I}} = -\singleCoeff{\mu_I-2}\sqrt{\lrabs{\mu_I-2}} \nearrow
    \end{cases} \hspace{-15.13pt} \ & \stackrel{\!\!\circleddash}{\Rightarrow}  \singleCoeff{\mu_I+2}\sqrt{\lrabs{\mu_I+2}} = \nonumber \\ & = \singleCoeff{\mu_I-2}\sqrt{\lrabs{\mu_I-2}} = \eqref{cond:muConditionFirstPart}.  \label{eq:proofSimulSol} 
\end{align}
Note that the inverse does not hold --- solutions to \eqref{cond:muConditionFirstPart} do not automatically also solve \eqref{cond:muFraction}. The proof thereof is analogous to \eqref{eq:proofSimulSol}.

In contrast to the condition \eqref{cond:muConditionFirstPart} that we solved before, we now have its equivalent difference equation of second order and with alternating sign at hand. Hence, the solution to \eqref{cond:muFraction} can be derived from \eqref{sol:muConditionFirstPart}:
\begin{align}
    \singleCoeff{\mu_I}\Big\vert_{\eqref{cond:muFraction}} = \begin{cases}
    \text{arbitrary}, & \text{for}\ \mu_I=0 \\
    0, & \forall \mu_I \in 2\mathbbm{Z} \\
    \lr{-1}^{\frac{\mu_I-1}{2}}\sqrt{\frac{1}{\lrabs{\mu_I}}}, &\text{rest}  \end{cases} . \label{sol:muFraction}
\end{align}
I.e., while it remains structurally the same, it considers the alternating minus sign and the necessity of setting every second coefficient zero --- compared to every fourth one before.

With \eqref{sol:muFraction} and the adapted form for $\nu_I$, we can achieve a vanishing action according to \eqref{eq:euclAction3stateSeparation}. Alternatively, we can replace one of the two by \eqref{cond:muEquality} or \eqref{cond:nuEquality}, whose solutions are obtained straightforwardly as
\begin{align}
    \singleCoeff{\mu_I}\Big\vert_{\eqref{cond:muEquality}} = \begin{cases}
    \text{arbitrary}, & \text{for}\ \mu_I\in\{0,1\} \\
    \singleCoeff{0}, & \forall \mu_I \in 2\mathbbm{Z}\setminus\{0\} \\
    \singleCoeff{1}, & \forall \mu_I \in 2\mathbbm{Z}+1\setminus\{1\} \end{cases} \label{sol:muEquality}
\end{align}
and the corresponding expression for $\nu_I$. Note that \eqref{cond:nuEquality} and \eqref{cond:muEquality} are conditions for vertices $v_{I+1}$, but as we sum over all vertices and try to find conditions that hold already vertex-wise, we can neglect that shift.

Summarising, we can make the action of $\euclHhat$ vanish by
\begin{align}
    \eqref{sol:muFraction}\Big\vert_{\mu_I} \land \lr{ \eqref{sol:muEquality}\Big\vert_{\nu} \lor \eqref{sol:muFraction}\Big\vert_{\nu}} \tag{S1} \label{sol:euclVanish}
\end{align}
or, equivalently, with $\mu$ and $\nu$ interchanged. Note that these turn out to be the only two combinations of the solutions stated above: \eqref{eq:euclSeparationSecondPart} showed that we need one condition for $\mu$ and one for $\nu$ as the solutions to \eqref{cond:nuEquality} and \eqref{cond:nuFraction} are not compatible (and likewise those to \eqref{cond:muEquality} and \eqref{cond:muFraction}). Furthermore, the solution \eqref{sol:muConditionFirstPart} for either $\mu$ or $\nu$, which makes the first part of the Euclidean action vanish, does not so for any contribution of the second part, \eqref{eq:euclSeparationSecondPart}. In turn, the condition \eqref{cond:muConditionFirstPart} behind that solution is automatically fulfilled by solutions to \eqref{cond:muFraction} (and, again, the same holds for the respective $\nu$ equivalents). Hence, we have to choose \eqref{sol:muFraction} for at least one of the charges $\mu$ and $\nu$, which then makes one part of the Euclidean action's second contribution vanish as well as the first contribution. The remaining contribution then vanishes by setting \eqref{sol:muFraction} or \eqref{sol:muEquality} for the respective other set of charges.

We therefore found states $|\varphi\rangle \coloneqq \sum_{k\in\mathbbm{Z}^N}\sum_{\mu\in m}\sum_{\nu\in n} \Coeff{}{}{} |k,\mu,\nu\rangle$ that experience a vanishing action of $\euclHhat$ by fulfilling constraints for the separated coefficients $\Coeff{}{}{} = \kCoeff{} \prod_I \singleCoeff{\mu_I}\singleCoeff{\nu_I}$. The solution we stated above, however, poses restrictions for the $\mu$ and $\nu$ coefficients only.
\subsubsection{Degeneracies of the action of the Lorentzian part \texorpdfstring{$\lorHhat$}{Hlor}}

While the diagonal action \eqref{eq:ActionHlor} of $\lorHhat$ makes a discussion as the one of the previous section irrelevant --- note that a diagonal action includes that there are no shifted coefficients ---, it in turn allows for a discussion of degeneracies.

Recall the action of $\lorHhat$ on the state $|\varphi\rangle$ via \eqref{eq:ActionHlor1} and \eqref{eq:ActionHlor2}
\begin{align}
    \lorHhat |\varphi\rangle =& \sum_{v\in V\lr{\aqggraph}} \sum_{k\in\mathbbm{Z}^N}\sum_{\mu\in m}\sum_{\nu\in n} \Coeff{}{}{} \Bigg\{ \bigg\{ -\klorOne \lr{k_{\ev}-k_{\evm}}^2 \Big[ \lrabs{k_{\ev}+k_{\evm}}^r \lrabs{\mu_v}^r \lrabs{\nu_v}^r \cdot \nonumber\\
    & \qquad \cdot \lr{\lrabs{k_{\ev}+k_{\evm}+1}^\halfalpha - \lrabs{k_{\ev}+k_{\evm}-1}^\halfalpha} \lr{\lrabs{\mu_v+1}^\halfalpha-\lrabs{\mu_v-1}^\halfalpha} \cdot \nonumber\\
    & \qquad \cdot \left.\left.  \lr{\lrabs{\nu_v+1}^\halfalpha-\lrabs{\nu_v-1}^\halfalpha} \right]^l \bigg\}\right\rvert_{r=\frac{2}{3}-\frac{1}{3l}} + \nonumber\\
    &  + \bigg\{ \klorTwo \lr{k_{\ev}+k_{\evm}}^4 \lr{\mu_v\nu_{\vp} - \mu_{\vp}\nu_v}^2 \left[ \lrabs{k_{\ev}+k_{\evm}}^{r_2} \lrabs{\mu_v}^{r_2} \cdot \right. \nonumber\\
    & \qquad \cdot \lrabs{\nu_v}^{r_2} \lr{\lrabs{k_{\ev}+k_{\evm}+1}^\halfbeta - \lrabs{k_{\ev}+k_{\evm}-1}^\halfbeta} \lr{\lrabs{\mu_v+1}^\halfbeta-\lrabs{\mu_v-1}^\halfbeta} \cdot \nonumber\\
    & \qquad \left.\left. \cdot \lr{\lrabs{\nu_v+1}^\halfbeta-\lrabs{\nu_v-1}^\halfbeta}\right]^l \bigg\}\right\rvert_{{r_2}=\frac{2}{3}-\frac{5}{3l}} \Bigg\} \state{}{}{} . \label{eq:ActionHlor}
\end{align}
We notice that i.a.~due to the appearance of $r$ and ${r_2}$ as different exponents in the two contributions (and in $\klorOne$ and $\klorTwo$, cf. \eqref{eq:klorConstants}), it is very unlikely for the two contributions to result in the same value or annihilate each other. Therefore, we consider both summands alone.

From the vertex-wise composition follow immediately two of the most basic degeneracies, namely rotations and flips of the graph and its vertices. Rotations and their corresponding rearrangement of the vertices are described by
\begin{equation}
    \forall v_I \in V\lr{\aqggraph} \colon v_I \mapsto v_{I+n}, \text{ for some } n\in\mathbbm{N}, \tag{S2}
\end{equation}
while flips of a graph $\aqggraph$ with $N=\lrabs{V\lr{\aqggraph}}$ many vertices can be represented as
\begin{equation}
    \forall v_I \in V\lr{\aqggraph} \colon v_I\mapsto v_{N+1-I+n}, \text{ for some } n\in\mathbbm{N} .  \tag{S3}
\end{equation}
With this, all charges $\mu_v$ and $\nu_v$ change their indices the same way. Hence, these two mappings do not change the value of \eqref{eq:ActionHlor} when the summation over all vertices is considered.

The other basic degeneracy is the interchange of all $\mu$ and $\nu$ charges,
\begin{equation}
    \forall v \in V\lr{\aqggraph} \colon \mu_v \mapsto \nu'_v \land \nu_v \mapsto \mu'_v \tag{S4} .
\end{equation}
As for all considerations before, $\mu$ and $\nu$ play the same role in \eqref{eq:ActionHlor}. The only factor that does not reflect this behaviour immediately is $\lr{\mu_v\nu_{\vp}-\mu_{\vp}\nu_v}^2$, but due to the even exponent, the minus within the bracket that arises via the interchange of the $\mu$ and $\nu$ charges does not change its value as well.

After having specified these three basic degeneracies, an aspect of interest may be whether shifts created by the action of $\euclHhat$ result in a new degenerate state with the same eigenvalue of $\lorHhat$. Addressing this question, we recapitulate that $\euclHhat$ acts only vertex-wise. Changing only the contribution of one vertex in \eqref{eq:ActionHlor}, however, is in general not preserving its value, but we can indeed find configurations that do fulfil this connection of $\euclHhat$ and $\lorHhat$. From \eqref{eq:ActionHeucl}, we recapitulate that $\euclHhat \state{}{}{}$ generates the shifted states
\begin{equation}
    \state{}{}{} \overset{\euclHhat}{\mapsto} \state{k_{\ev}\pm 2}{\mu_v\pm 1, \mu_{\vp}\pm 1}{}, \state{k_{\ev}\pm 2}{}{\nu_v\pm 1, \nu_{\vp}\pm 1}, \state{}{\mu_v\pm 2}{\nu_v\pm 2}. \label{eq:euclHhatGeneratedStates}
\end{equation}
We then see that individual states of the set above do preserve some factors' values within $\eqref{eq:ActionHlor}$ for specific values of the charges, but not the whole expression. It is in particular the link between the $k$-charges of different vertices that causes trouble: $\lr{k_{\ev}+k_{\evp}}^4$ can not be preserved when only $k_{\ev}$ is de- or increased by two. The only way that's possible is for $k_{\ev}=0$ and $k_{\ev}=\pm 1$. But as we would have to fulfil it for every vertex, this choice leads to a contradiction. While this excludes the first two sets of states of \eqref{eq:euclHhatGeneratedStates}, we can indeed find states of the third set that have the same Lorentz energy\footnote{We call the eigenvalues of $\lorHhat$ ``Lorentz energies'', even though they are, of course, not necessarily one part/summand of the (proper) energy eigenvalues of $\hatHphys$.} as the initial $\state{}{}{}$. They are shown in Figure \ref{fig:degeneracies} and represent states where, at one vertex $v$, the charges $\mu_v$ and $\nu_v$ happen to be $\pm1$. Acting with $\euclHhat$ on such a state produces i.a.~states of the third set of states of \eqref{eq:euclHhatGeneratedStates} that only change the signs of these charges $\mu_v$ and $\nu_v$: from -1 to +1 or vice versa. This clearly preserves all factors $\lrabs{\mu_v,\nu_v}^{r,{r_2}}$ of the Lorentz action \eqref{eq:ActionHlor} and also the products $\lr{\lrabs{\mu_v+1}^{\halfalpha,\halfbeta}-\lrabs{\mu_v-1}^{\halfalpha,\halfbeta}}\lr{\lrabs{\nu_v+1}^{\halfalpha,\halfbeta}-\lrabs{\nu_v-1}^{\halfalpha,\halfbeta}}$ due to the double change of the sign. Lastly, $\lr{\mu_v\nu_{\vp} - \mu_{\vp}\nu_v}^2$ experiences a change of the sign in both its subtrahend and its minuend, hence preserves its value because of the square. Note that this factor has to be preserved for the vertex $v_{I-1}$ as well, due to the mixture of the charges of the current and the next vertex. All other $\mu$- and $\nu$-charges as well as all $k_{\ev}$ can, however, be chosen arbitrary as the symmetry of -1 and +1 suffices to conserve the Lorentz energy.

\begin{figure}
\begin{subfigure}{\textwidth}
    \begin{subfigure}{.4\textwidth}
    \begin{tikzpicture}
    \draw[-latex] (230:3) arc (230:268:3) node[pos=.6,anchor=south]{$k_{e-}$};
    \draw[-latex] (270:3) arc (270:308:3) node[pos=.45, anchor=south]{$k_{\ev}$};
    \draw[densely dotted] (200:3) arc (200:230:3);
    \draw[densely dotted] (310:3) arc (310:340:3);
    \fill (0,-3) circle (2pt);
    \fill (230:3) circle (2pt);
    \fill (310:3) circle (2pt);
    \node[anchor=south] at (0,-3) {$v$};
    \node[anchor=south west] at (230:3) {$v_{I-1}$};
    \node[anchor=south east] at (310:3) {$v_{I+1}$};
    \node[anchor=north, align=center] at (0,-3.1) {$\mu_v=-1$ \\ $\nu_v=-1$};
    \node[anchor=north east, align=center] at (230:3) {$\mu_{\vm}$ \\ $\nu_{\vm}$};
    \node[anchor=north west, align=center] at (310:3) {$\mu_{v+}$ \\ $\nu_{v+}$};
    \end{tikzpicture}
    \end{subfigure}
    \begin{subfigure}{.1\textwidth}
    $\underset{\lower.2em\hbox{\footnotesize $\nu_v \pm 2$}}{\overset{\raise.4em\hbox{\footnotesize $\mu_v \pm 2$}}{\Longleftrightarrow}}$
    \end{subfigure}
    \begin{subfigure}{.4\textwidth}
    \begin{tikzpicture}
    \draw[-latex] (230:3) arc (230:268:3) node[pos=.6,anchor=south]{$k_{\evm}$};
    \draw[-latex] (270:3) arc (270:308:3) node[pos=.45, anchor=south]{$k_{\ev}$};
    \draw[densely dotted] (200:3) arc (200:230:3);
    \draw[densely dotted] (310:3) arc (310:340:3);
    \fill (0,-3) circle (2pt);
    \fill (230:3) circle (2pt);
    \fill (310:3) circle (2pt);
    \node[anchor=south] at (0,-3) {$v$};
    \node[anchor=south west] at (230:3) {$v_{I-1}$};
    \node[anchor=south east] at (310:3) {$v_{I+1}$};
    \node[anchor=north, align=center] at (0,-3.1) {$\mu_v=1$ \\ $\nu_v=1$};
    \node[anchor=north east, align=center] at (230:3) {$\mu_{\vm}$ \\ $\nu_{\vm}$};
    \node[anchor=north west, align=center] at (310:3) {$\mu_{v+}$\\$\nu_{v+}$};
    \end{tikzpicture}
    \end{subfigure}
\caption{parallel degeneracy}
\end{subfigure}
\begin{subfigure}{\textwidth}
    \vspace{.6cm}
    \begin{subfigure}{.4\textwidth}
    \begin{tikzpicture}
    \draw[-latex] (230:3) arc (230:268:3) node[pos=.6,anchor=south]{$k_{\evm}$};
    \draw[-latex] (270:3) arc (270:308:3) node[pos=.45, anchor=south]{$k_{\ev}$};
    \draw[densely dotted] (200:3) arc (200:230:3);
    \draw[densely dotted] (310:3) arc (310:340:3);
    \fill (0,-3) circle (2pt);
    \fill (230:3) circle (2pt);
    \fill (310:3) circle (2pt);
    \node[anchor=south] at (0,-3) {$v$};
    \node[anchor=south west] at (230:3) {$v_{I-1}$};
    \node[anchor=south east] at (310:3) {$v_{I+1}$};
    \node[anchor=north, align=center] at (0,-3.1) {$\mu_v=-1$ \\ $\nu_v=1$};
    \node[anchor=north east, align=center] at (230:3) {$\mu_{\vm}$ \\ $\nu_{\vm}$};
    \node[anchor=north west, align=center] at (310:3) {$\mu_{v+}$ \\ $\nu_{v+}$};
    \end{tikzpicture}
    \end{subfigure}
    \begin{subfigure}{.1\textwidth}
    $\underset{\lower.2em\hbox{\footnotesize $\nu_v \mp 2$}}{\overset{\raise.4em\hbox{\footnotesize $\mu_v \pm 2$}}{\Longleftrightarrow}}$
    \end{subfigure}
    \begin{subfigure}{.4\textwidth}
    \begin{tikzpicture}
    \draw[-latex] (230:3) arc (230:268:3) node[pos=.6,anchor=south]{$k_{\evm}$};
    \draw[-latex] (270:3) arc (270:308:3) node[pos=.45, anchor=south]{$k_{\ev}$};
    \draw[densely dotted] (200:3) arc (200:230:3);
    \draw[densely dotted] (310:3) arc (310:340:3);
    \fill (0,-3) circle (2pt);
    \fill (230:3) circle (2pt);
    \fill (310:3) circle (2pt);
    \node[anchor=south] at (0,-3) {$v$};
    \node[anchor=south west] at (230:3) {$v_{I-1}$};
    \node[anchor=south east] at (310:3) {$v_{I+1}$};
    \node[anchor=north, align=center] at (0,-3.1) {$\mu_v=1$ \\ $\nu_v=-1$};
    \node[anchor=north east, align=center] at (230:3) {$\mu_{\vm}$ \\ $\nu_{\vm}$};
    \node[anchor=north west, align=center] at (310:3) {$\mu_{v+}$\\$\nu_{v+}$};
    \end{tikzpicture}
    \end{subfigure}
\caption{cross degeneracy}
\end{subfigure}
\caption{Specific degeneracies of states linked by the action of $\euclHhat$. As before in the figures also here to keep the notation more compact we used $k_{e_{v_I}}\coloneqq k_I, \mu_{v_I}\coloneqq\mu_I$ and $\nu_{v_I}\coloneqq\nu_I$ etc.}
\label{fig:degeneracies}
\end{figure}

To give a further example of a more special degeneracy, we deduce from \eqref{eq:ActionHlor} that the charges $k_{\ev}$ always appear in the combination $k_{\ev}+k_{\evp}$ or $k_{\evp}-k_{\evm}$. This allows to set
\begin{equation}
    k_{e_{v_I}}\mapsto \begin{cases}
    k_{e_{v_I}} \pm a & \text{ for } I\in2\mathbbm{Z}, a \in \mathbbm{R} \\
    k_{e_{v_I}} \mp a & \text{ for } I\in 2\mathbbm{Z}+1, a \in \mathbbm{R}
    \end{cases} \tag{S5}
\end{equation}
for graphs with an even number of vertices and yet get the same Lorentz energy.

\section{Conclusions}
\label{sec:Conclusion}
In this article, we considered a reduced model with a polarised \texorpdfstring{$\mathbbm{T}^3$}{T3} Gowdy symmetry resulting from Gaussian dust coupled to general relativity and afterwards applying the symmetry reduction. The corresponding physical phase space has three independent canonical pairs consisting of Dirac observables associated with the connection and triad variables and describe an unconstrained  U(1) gauge field theory. The evolution of these Dirac observables is generated by a physical Hamiltonian that itself is a Dirac observable. This classical model was taken as a starting point and quantised in the reduced LQG as well as the AQG framework in this work. In both cases, due to the symmetry of the classical physical Hamiltonian a graph-preserving quantisation was chosen in order to implement these symmetries also at the quantum level. The results presented here extend the ones in the literature in the following aspects: On the one hand, the models existing so far that use a loop but not hybrid quantisation \cite{Banerjee:quant,deBlas:2017goa} have all applied a Dirac quantisation where a kinematical Hilbert space is chosen as an intermediate step on which the Hamiltonian, spatial diffeomorphism and Gau\ss{} constraints of the Gowdy model are implemented as operators. The physical Hilbert space then involves those physical states that are annihilated by all constraint operators. The model discussed in \cite{Martin-Benito:2008eza,Martin-Benito:2010dge,Garay:2010sk} considers a hybrid quantisation where the homogeneous modes are quantised using loop quantum gravity techniques whereas for the quantisation of the inhomogeneous modes a Fock quantisation has been chosen. They are thus not easy to relate to those models where no Fock quantisation has been used such as, e.g., full LQG. The model in \cite{deBlas:2017goa} derives the full physical Hilbert space in a simpler setup where vacuum Gowdy spacetimes have been considered with an additional rotational symmetry. The model that comes most closely to the one discussed here is the one in \cite{Banerjee:quant} where similarities but also differences exist. The main difference is that \cite{Banerjee:quant} also follows a Dirac quantisation for the individual constraints with partly a different regularisation. Since also for their chosen regularisation the structure of the constraint operators is similarly complicated as the Schr\"odinger-like equation we obtain here, the physical Hilbert space of that model has not yet been derived. Furthermore, because one works with the Hamiltonian constraint instead of a physical Hamiltonian, the properties of the constraint algebra, as in the full theory, favour a graph-modifying quantisation of the Hamiltonian constraints. This yields a setup where the construction of semiclassical states as well as solutions of the constraint operator equations become more complicated compared to the model presented in this work due to the fact that operators modify the underlying graph they are acting on. In contrast, in the models presented here, the graph-preserving property comes in accordance with the requirement to implement classical symmetries also at the quantum level in the case of the reduced LQG model where the usual Ashtekar-Lewandowski representation is chosen for the physical Hilbert space. We further discussed the differences in the implementation of graph-preserving operators in the reduced LQG and AQG framework. 
Furthermore, because we couple Gaussian dust to gravity, the number of physical degrees of freedom differ in the two models. The one of \cite{Banerjee:quant} has just one independent degree of freedom, whereas here we have three. This is reflected in the fact that all geometric degrees of freedom encoded in the Dirac observables of the model presented here are unconstrained, while for the corresponding quantities on the kinematical Hilbert space described in \cite{Banerjee:quant} constraints still exist. We also derived the explicit form of the Schr\"odinger-like equation of the model in the AQG framework.  This result provides an option for future work in which one can analyse this Schr\"odinger-like equation numerically or perform a semiclassical analysis of this equation in order to derive the corresponding effective model. As far as these future computations are considered, the model with polarised \texorpdfstring{$\mathbbm{T}^3$}{T3} Gowdy symmetry introduced here has --- due to its symmetry reduction --- the advantage that the volume operator acts diagonally on the basis states and hence the spectrum of the volume operator is known in the quantum theory. For semiclassical computations we therefore do not need to apply semiclassical perturbation theory along the lines of \cite{AQG3}, as it is necessary for full LQG. As we do not analyse the Schr\"odinger-like equation in full detail here but just derive it for the model and then discuss some very specific zero volume solutions in order to obtain a first intuition on how the physical Hamiltonian operator acts, it will be an interesting question for future work to better understand whether the model in \cite{Banerjee:quant} can in some sense be embedded in the model presented here at the quantum level, when we extend our model by additional constraints that reduce the dust degrees of freedom and allow to go back to the vacuum case --- and how this might be reflected in the solutions of the Schr\"odinger-like equation we obtain in this work.

Another scenario where we can get a notion of how the physical Hamiltonian and especially its Euclidean and Lorentzian parts (inter-)act is degenerate perturbation theory. The idea that the (symmetrised) Eucledian part of the physical Hamiltonian can be considered as a perturbation of the Lorentzian part in case the Immirzi parameter is chosen large enough was already introduced in \cite{Assanioussi:2017tql} in the full LQG setup using a different graph-modifying regularisation for the corresponding physical Hamiltonian operators in \cite{Assanioussi:2017tql}. 
In the case of this Gowdy model presented in this work first steps were already performed in~\cite{Refik}, where the action of the symmetrised Euclidean part $\frac{1}{2}\lr{\euclHhat + (\euclHhat)^\dagger}$ is treated as a perturbation on top of the action of the Lorentzian part. However, as the complete set of degeneracies of $\lorHhat$ is not known, a comprehensive treatment of that ansatz is not possible. The special cases considered in~\cite{Refik} still illustrate nicely the interplay of the actions of the two parts of the physical Hamiltonian and how one can in general approach degenerate perturbation theory for Gowdy models like the one considered herein.

\tocless{\section*{Acknowledgements}}
The work of D.W. was supported by a stipend provided from the FAU Erlangen-Nürnberg. Further, D.W. thanks the Studienstiftung des deutschen Volkes (German Academic Scholarship Foundation) for financial support at an earlier stage of the project.
\bibliography{Gowdy}
\bibliographystyle{unsrtnat} 

\end{document}